\documentclass{report}
\usepackage[utf8]{inputenc}
\usepackage[T1]{fontenc}
\usepackage[french]{babel}
\usepackage{graphicx}
\usepackage{subcaption}
\usepackage{amsmath,amsfonts,amssymb}
\usepackage{dsfont}
\usepackage{appendix}
\usepackage{aligned-overset}
\usepackage{bbold}
\usepackage{enumitem}

\DeclareMathOperator{\rank}{rank}

\DeclareMathOperator{\sign}{sign}

\usepackage{geometry}
 \usepackage{hyperref}

\usepackage{nomencl}
\makenomenclature

\renewcommand{\nompreamble}{\noindent La liste suivante décrit l'intégralité des notations utilisées dans ce document.\\ \vspace{0.5cm}}
\newcommand{\nomunit}[1]{%
\renewcommand{\nomentryend}{\hspace*{\fill}#1}}

\usepackage[
backend=biber,
style=ieee,
uniquename=init,
citestyle=numeric
]{biblatex}
\usepackage{csquotes}
\usepackage{siunitx}

\newcommand{\skewsym}[1]{[#1]_{\times}}

\begin{document}

\begin{titlepage}
    \begin{center}
        \vspace*{1cm}
            
        \Huge
        \textbf{Commande hybride d’un drone convertible pour des déplacements sous-optimaux}
            
        \vspace{0.5cm}
        \LARGE

        \vspace{1.5cm}
            
        \textbf{SANSOU Florian}
            
        \vfill
            
        Rapport de stage présenté pour l'obtention d'un diplôme d'ingénieur aéronautique à l'École Nationale de l'Aviation Civile 
            
        \vspace{0.8cm}
        \begin{figure}[!h]
            \centering
            \begin{subfigure}[b]{0.49\textwidth}
                \centering
                \includegraphics[width=\textwidth]{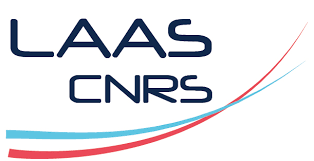}
            \end{subfigure}
            \begin{subfigure}[b]{0.49\textwidth}
                \centering
                 \includegraphics[width=\textwidth]{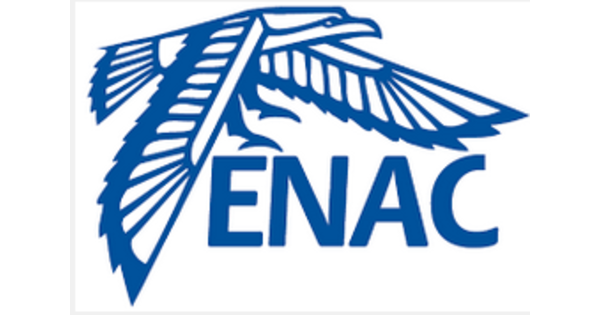}
            \end{subfigure}

        \end{figure}

        \Large
        Encadrants : Luca Zaccarian (LAAS) \& Gautier Hattenberger (ENAC)\\
        Équipe MAC\\
        LAAS - CNRS\\
        France\\
        Février - Août 2021
            
    \end{center}
\end{titlepage}
\setcounter{figure}{0}
\setcounter{page}{2}

\clearpage\null\newpage

\vspace*{2cm}

\section*{Remerciements}\label{sec:Remerciements}
\addcontentsline{toc}{chapter}{\nameref{sec:Remerciements}}
\vspace{1cm}
Je tiens à remercier toutes les personnes qui ont contribué au succès de mon stage et qui m’ont aidé lors de la rédaction de ce mémoire.

Je remercie tout particulièrement M. Luca Zaccarian, mon référent au LAAS, qui m'a soumis un sujet, l'a guidé d'un point de vue scientifique et m'a accompagné tout au long de mon travail.

Ma reconnaissance va également à M. Gautier Hattenberger, mon tuteur ENAC, qui a encadré mon expérience et m'a épaulé lors de la préparation de l'implémentation. Je n'oublie pas les membres de l'équipe drone de l'ENAC, Jean Philippe Condomines et Murat Bronz, qui ont bien voulu m'accorder un peu de leur temps. 

Aussi, je témoigne toute ma sympathie à Mme Lucie Baudouin, responsable de l'équipe MAC, qui a transmis ma candidature et m'a ainsi offert l'opportunité de ce stage.

Je remercie également toute l’équipe pédagogique de l’École Nationale de l'Aviation Civile et les intervenants professionnels responsables de ma formation, pour avoir assuré la partie théorique de celle-ci.

Enfin, je tiens à témoigner toute ma reconnaissance à ma compagne et à mes proches, pour leur soutien constant et leurs encouragements dans la réalisation de ce mémoire.

\clearpage\null\newpage

\vspace*{2cm}
\section*{Résumé}\label{sec:Résumé}
\vspace{1cm}
\addcontentsline{toc}{chapter}{\nameref{sec:Résumé}}

Que ce soit pour leur maniabilité, leur autonomie ou leur ergonomie, les drones convertibles, c'est-à-dire ceux ayant la capacité de décoller ou d’atterrir à la verticale et de voler comme un avion, présentent de nombreux intérêts.

Cependant, leurs dynamiques étant complexifiées par divers phénomènes comme les non-linéarités, les couplages aérodynamiques ou encore le nombre important de degrés de liberté, il est nécessaire d'utiliser de nouveaux outils de commande. Ainsi, devient centrale la commande hybride, utile lors de choix binaires.

 Dès lors, le drone étant en mesure de se déplacer de deux manières, en vol pseudo-stationnaire et en vol horizontal, se pose la question de la pertinence d'une commande vis-à-vis de l'autre, en fonction du contexte opérationnel. Cette problématique trouve ses réponses dans la conception de deux lois de commande non-linéaires, basées sur des fonctions de Lyapunov, qui permettent de valider les exigences. Donc, en fonction de la configuration, il s’agira pour le mécanisme hybride de sélectionner le mode de vol opportun. Enfin, dans l’objectif d’un vol en conditions réelles, une implémentation a été initiée.

\clearpage\null\newpage
\vspace*{2cm}
\section*{Abstract}\label{sec:Abstract}
\vspace{1cm}
\addcontentsline{toc}{chapter}{\nameref{sec:Abstract}}

Whether for their maneuverability, autonomy or ergonomics, convertible UAVs, meaning those with the ability to take off or land vertically and fly like an airplane, present many interests. 

However, their dynamics are complicated by various phenomena such as non-linearities, aerodynamic couplings or the large number of degrees of freedom, and it is necessary to use new control tools. Thus, hybrid control becomes crucial, useful when making binary choices.

Since the drone is able to move in two ways, in pseudo-stationary flight and in horizontal flight, the question of the relevance of a command with respect to the other, according to the operational context, arises. This problem is answered by the design of two nonlinear control laws, based on Lyapunov functions, which allow to validate the requirements. Therefore, depending on the configuration, the hybrid mechanism will have to select the appropriate flight mode. Finally, with the objective of a flight in real conditions, an implementation has been initiated.

\clearpage\null\newpage
\addcontentsline{toc}{chapter}{Table des matières}
\tableofcontents

\newpage
\addcontentsline{toc}{chapter}{Table des figures}
\listoffigures

\clearpage\null\newpage
\vspace*{2cm}
\section*{Sujet et objectifs}\label{sec:sujet}
\vspace{1cm}
\addcontentsline{toc}{chapter}{\nameref{sec:sujet}}
Lors de l’utilisation d’un drone convertible (capable de passer d’un mode avion à un mode de vol stationnaire), se pose la question de la planification des déplacements locaux. Effectivement, lors de grands trajets, il est préférable d'effectuer une transition pour voler comme un avion, afin d’être le moins consommateur en énergie. Cependant, quand le mouvement intervient sur des distances plus courtes, on peut se demander quel mode de pilotage est le plus opportun ?

L’objectif de ce stage est de déterminer une couche de commande en guidage permettant de définir, de manière formelle et stable, la meilleure commande pour effectuer ce déplacement. Cet asservissement sera optimisé notamment sur des critères de consommation énergétique. Il s’agira de définir un modèle de drone, dans un premier temps relativement simple, de manière à pouvoir choisir les outils nécessaires à l’élaboration de cette couche de guidage. Il est possible, en fonction des résultats de simulation, d'effectuer des tests en situation réelle sur des maquettes de drone, en collaboration avec l’ENAC.

En rapport avec le contexte défini ci-dessus, le stage suivra une approche mathématique rigoureuse, basée sur la description moderne des systèmes dynamiques hybrides \cite{65}. La loi de commande proposée sera caractérisée par un mécanisme de commutation basé sur des variables logiques. Ces dernières font partie de l’état interne de la dynamique hybride. Dès lors que les conditions de saut sont validées, il y a une commutation. Cependant, cette dernière doit garantir, par sa structure, l’absence de phénomènes Zeno et avoir un comportement approprié issu des propriétés comme l’hystérésis de commutation entre les différents modes de vol. On axera le stage sur différentes manœuvres de vol telles que des manœuvres d’évitement d’obstacles (décrites dans \cite{9064903}) ou des manœuvres de récupération d’urgence (présentées dans \cite{DBLP:conf/IEEEcca/AndreettoFZ16}).

\clearpage\null\newpage

\setcounter{section}{0}
\section*{Introduction}\label{sec:Introduction}
\addcontentsline{toc}{chapter}{\nameref{sec:Introduction}}
\vspace{1cm}

L'essor des drones est sans appel. Aujourd'hui, nous les utilisons dans de plus en plus de domaines tels que la surveillance, la météorologie, l'agriculture ou la topographie. Chaque contexte opérationnel ayant ses caractéristiques et ses exigences propres, il s'agit de concevoir un appareil qui soit adapté à sa mission. Ainsi, on peut distinguer trois grandes architectures : les aéronefs à rotor, les aéronefs à ailes fixes et les convertibles, chacun présentant des avantages et des inconvénients. Les aéronefs à rotor possèdent une très grande agilité et la capacité d'effectuer un vol stationnaire. Cependant, leur sustentation étant réalisée uniquement par les hélices, ils sont très énergivores et ne permettent pas des vols de longues durées. Les aéronefs à ailes fixes ont des temps de vol très importants car ils génèrent leur portance grâce à un profil aérodynamique. Toutefois, ils n'ont pas la possibilité d'effectuer de vol stationnaire, ce qui engendre un domaine de vol relativement réduit. Enfin, les convertibles, sujet de l'étude de notre stage, sont un compromis entre les deux architectures précédentes. Étant équipés d'une aile fixe, ils ont la possibilité de voler comme un avion, avec les avantages énergétiques que cela leur confère, mais aussi de se stabiliser en hovering pour maintenir une position. Ainsi, ils ont une enveloppe de vol très importante.

\paragraph{}
La motivation opérationnelle de ce travail réside dans la situation qui suit : supposons que le drone soit stable en hovering (vol stationnaire vertical) en un point de l'espace. On souhaite déplacer ce drone en un autre point de l'espace, distinct du premier. Dès lors, la problématique autour de laquelle s'articule notre réflexion est la suivante : est-il préférable de maintenir une orientation relativement proche de l'équilibre, drone à la verticale, et de se déplacer vers la cible ou bien faut-il basculer en vol horizontal pour rejoindre ce point ?  
\paragraph{}
Notre objectif est d'obtenir une loi de commande, basée modèle, permettant d'assurer le contrôle du drone sur l'intégralité de son domaine de vol. Dans chacun des modes de vol, les manœuvres sont identiques mais réalisées de manières différentes, en fonction de la configuration. Par exemple, pour effectuer un changement de niveau de vol, il est possible d'effectuer un vol en montée rectiligne uniforme, utilisant la portance générée par la vitesse du drone pour sustenter ce dernier, ou de voler en hovering, compensant la gravité avec une force opposée, générée par les hélices, tout surplus de puissance étant converti en énergie potentielle et donc en hauteur.  De cette possibilité qu'est le vol dans plusieurs configurations, résulte la question du choix de la configuration en fonction de la mission et des critères tels que la consommation énergétique. Ainsi, le contrôleur devant être en mesure de maintenir un vol en hovering et un vol horizontal, un mécanisme hybride est nécessaire.  
Concrètement, le vol en hovering doit permettre d'orienter le drone dans une situation qui lui permette de rejoindre un point cible et stabiliser sa position sur ce point. Le vol horizontal doit, quant à lui, permettre d'orienter le drone de manière à maintenir un vol stabilisé, à une vitesse constante, tout en rejoignant un point cible à une distance relativement élevée. Ainsi, notre choix d'asservissement réside dans des commandes par retour d'état, sur modèle linéarisé et non-linéaire, basées sur des fonctions de Lyapunov, ce qui assure ainsi leur convergence et leur attractivité.
Aussi, il est nécessaire d'étudier le modèle du drone pour déterminer une inversion de modèle, ce qui rend possible la définition des commandes sur les actionneurs, pour obtenir les moments voulus.
\\
\indent
Dans l'idée de couvrir l'intégralité des capacités du drone, nous nous tournons vers un modèle unique, sans singularité. C'est donc naturellement que, pour décrire l'orientation du drone, nous utilisons une représentation complexe, à base de quaternions unitaires.
De plus, la complexité de la commande provient des divers phénomènes qui entrent en jeu. On peut notamment penser aux non-linéarités, très présentes dans la mesure où nous atteignons des angles d'incidence importants et, par conséquent, le décrochage de la partie non-soufflée de l'aile. Citons aussi les couplages aérodynamiques qui nécessitent des commandes particulières ainsi que des efforts supérieurs sur les gouvernes pour obtenir le découplage. De plus, les modèles reposent  sur l'identification, parfois fastidieuse et imprécise, des coefficients aérodynamiques. De même, le drone possédant six degrés de liberté et seulement quatre actionneurs indépendants, cela engendre un sous-actionnement.

\paragraph{}
Par ailleurs, soulignons qu'une commande basée modèle permet d'envisager une certification EASA, à laquelle sont soumis les drones civils actuels. En cela, il est intéressant de constater que si d'autres commandes ont été étudiées dans le même objectif et donnent de bons résultats, notamment la commande sans modèle \cite{olszaneckibarth:hal-02542982}, elles ne permettent pas la réglementation, en raison de l'absence de preuve démontrant leur convergence. Un autre type de commande existe, basé sur de l'inversion numérique de la dynamique du drone (INDI : Incremental Nonlinear Dynamic Inversion) \cite{smeur:hal-02293693}. Dans celle-ci, on intègre les mesures des gyromètres pour obtenir le comportement de l'aéronef ; cependant la forte densité de bruit complexifie l'intégration, ce qui a tendance à faire diverger la commande.

\paragraph{}
Notre travail a été pensé en trois phases : la mise en place d'un modèle, l'étude des lois de commande et leurs simulations, et les expérimentations.
L'étude du modèle est le point de départ, en ce qu'il nous permet de comprendre le comportement du drone. Pour cela, nous avons choisi de nous appuyer sur les travaux de thèse de M. Leandro Ribeiro Lustosa \cite{lustosa:hal-03035938}.
Celle-ci envisage une modélisation des drones convertibles (\textit{tail-sitter}) sans singularité, avec des quaternions. La représentation matricielle des équations envisagées permet un calcul plus simple des équations, en faisant abstraction des angles de dérapage et d'incidence du drone. Cependant, cette modélisation s'appuie sur des coefficients aérodynamiques pour lesquels il est nécessaire de faire une identification. Il en est de même avec la loi de commande proposée dans cet article \cite{DBLP:journals/automatica/KaiHS19}.\\ \indent 
Une fois l'étude du modèle aboutie, nous nous sommes intéressés au vol en mode vertical, l'objectif étant de réaliser de petits trajets avec une loi de commande stabilisante non-linéaire. Cependant, nous observons certaines limites lors de déplacements plus importants, ce qui justifie un passage en mode horizontal.
Ainsi, en mode hovering, une loi de commande non-linéaire sera conçue sur la base d'une commande \textit{zero-moment} \cite{2020e-MicCenZacFra}.\\ \indent  Par la suite, nous avons analysé le comportement du drone en vol horizontal, notamment au travers d'une linéarisation autour d'un point d'équilibre. Puis, nous avons commencé à étudier la possibilité de concevoir une loi de commande non-linéaire permettant de stabiliser le drone vers une direction fixée, à une vitesse de vol choisie. \\ \indent  Ces études ont été suivies d'une implémentation, pour effectuer des simulations, dans la plate-forme Simulink.
\paragraph{}
La suite de ce mémoire présente les études réalisées, les résultats obtenus et les enseignements qui en découlent.

\newpage
\chapter{Présentation et déroulement du stage}
\section{Présentation du LAAS \texorpdfstring{\cite{LAAS}}{}}

\subsection{Présentation générale}
J'ai effectué mon stage de fin d'études au sein du Laboratoire d'Analyse et d'Architecture des Systèmes (LAAS), structure rattachée au Centre National de la Recherche Scientifique (CNRS). Plus précisément, il s'est déroulé dans l'équipe Méthodes et Algorithmes en Commande (MAC), au sein du département "Décision et Optimisation".\\

Le Laboratoire d'Analyse et d'Architecture des Systèmes (LAAS-CNRS) est une unité propre du CNRS, en lien avec l'Institut des sciences de l'ingénierie et des systèmes (INSIS) et l'Institut des sciences de l'information et de leurs interactions (INS2I). Le laboratoire, implanté à Toulouse, au sein du complexe scientifique de Rangueil, développe des démarches cohérentes pour comprendre, concevoir et maîtriser des systèmes complexes.

\subsection{Le département Décision et Optimisation}

Ce département du LAAS mène des activités de recherche théoriques et méthodologiques pour la conception de lois mathématiques et de techniques algorithmiques servant à la commande et à la décision. 
Les trois équipes (DISCO - Diagnostic, Supervision et Conduite, MAC - Méthodes et Algorithmes en Commande, ROC - Recherche Opérationnelle, Optimisation Combinatoire et Contraintes) composant le département couvrent une variété de champs disciplinéaires de l’automatique et de l’informatique. Elles partagent certaines particularités comme : être centrées sur des classes de modèles représentant des réalités physiques, fonctionnelles ou organisationnelles que l’on souhaite piloter ; proposer des outils théoriques pour l’analyse des propriétés et performances atteignables ou atteintes ; adosser ces résultats à des méthodes de conception de lois de commande, de diagnostic ou d’optimisation ; illustrer les résultats sur des exemples d’applications fournis par des partenaires extérieurs qui, dans l’échange, alimentent les équipes en problématiques nouvelles.
\subsection{Équipe MAC}
Les domaines de recherche dans lesquels s'inscrit le groupe MAC concernent les méthodes et les techniques de commande pour les systèmes à états continus. Autrement dit, ses activités sont ancrées dans le domaine de l'automatique et de la théorie des systèmes. Les thématiques principales autour desquelles s'articulent les travaux de l'équipe sont
la commande robuste, la prise en compte de non-linéarités et l'optimisation. Les travaux menés sont principalement orientés sur les problèmes de modélisation, d'estimation, d'analyse et de commande de systèmes dynamiques.
\newpage
\begin{figure}[h]
    \centering
    \includegraphics[width=0.5\textwidth]{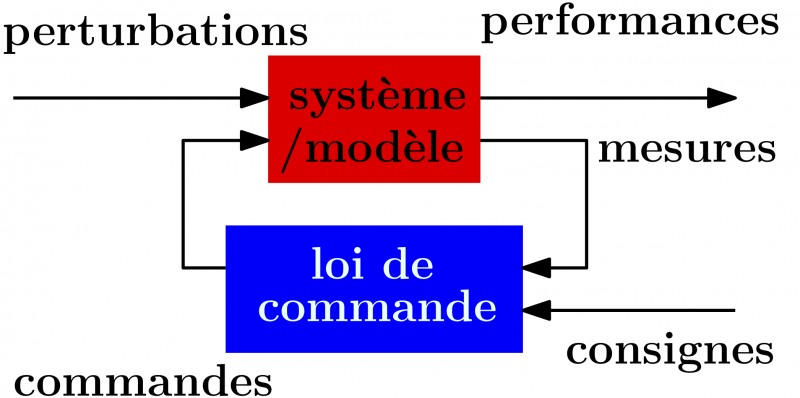}
    \caption{Vision de la modélisation de l'équipe MAC}
    \label{fig:mac_vision}
\end{figure}

\section{Formations au cours du stage}
 Avant toutes choses, soulignons que ce stage a été l'occasion de mobiliser des connaissances acquises lors de ma formation à l'ENAC, notamment sur la linéarisation d'un modèle. Celles-ci ont pu être complétées par des échanges réguliers avec mon maître de stage et des recherches personnelles.
\paragraph{}
La première partie du stage a été consacrée à l'étude d'une nouvelle méthode de commande : la commande hybride. Celle-ci permet de modéliser la partie continue et la partie discrète d'un système et de voir leurs interactions. Cet apprentissage s'est déroulé en deux étapes importantes. Premièrement, j'ai participé aux cours donnés par Sophie Tarbouriech et Isabelle Queinnec,  qui m'ont permis d'acquérir les notions fondamentales. En effet, m'ont été proposés les aspects théoriques de la commande hybride et divers exemples et exercices. Puis, j'ai travaillé sur la simulation d'un article \cite{DBLP:conf/IEEEcca/AndreettoFZ16}, dans l'objectif d'obtenir les mêmes résultats que ceux présentés dans celui-ci et ainsi m'initier aux fonctionnements et problématiques hybrides. J'ai observé que l'utilisation de mécanismes hybrides dans le contexte de ce stage est très intéressante, en ce qu'elle nous permet la conception de contrôleurs avec des performances prouvées mathématiquement. La principale difficulté rencontrée dans cette tâche a été l'assimilation de la notion de temps généralisé, composante de cette loi de commande. \\
\indent
Ce papier \cite{DBLP:conf/IEEEcca/AndreettoFZ16}, présentant une loi de commande basée sur un contrôleur local et un contrôleur global, a aussi été le point de départ de notre travail, dans la mesure où il a inspiré la création de deux contrôleurs permettant de stabiliser le vol en hovering et le vol horizontal. Parallèlement, le contrôleur temps quasi optimal (QTO), non-linéaire saturé, présent dans la loi de commande globale, a nécessité une étude approfondie de sa forme et de son fonctionnement, via des travaux récents \cite{journals/csysl/InvernizziLZ20}. 
\\
\indent
Ce stage visant l'étude d'un drone convertible tel que DarkO, c'est naturellement que j'ai étudié son processus de développement, et notamment ses différentes formes. L'impact de ces dernières sur la commande du drone, c'est-à-dire les changements dans la dynamique de DarkO en fonction de la forme de l'aile, est développé dans l'annexe \ref{ann:comp}.
\\
\indent

\begin{figure}[h]
    \centering
    \includegraphics[width=0.4\textwidth]{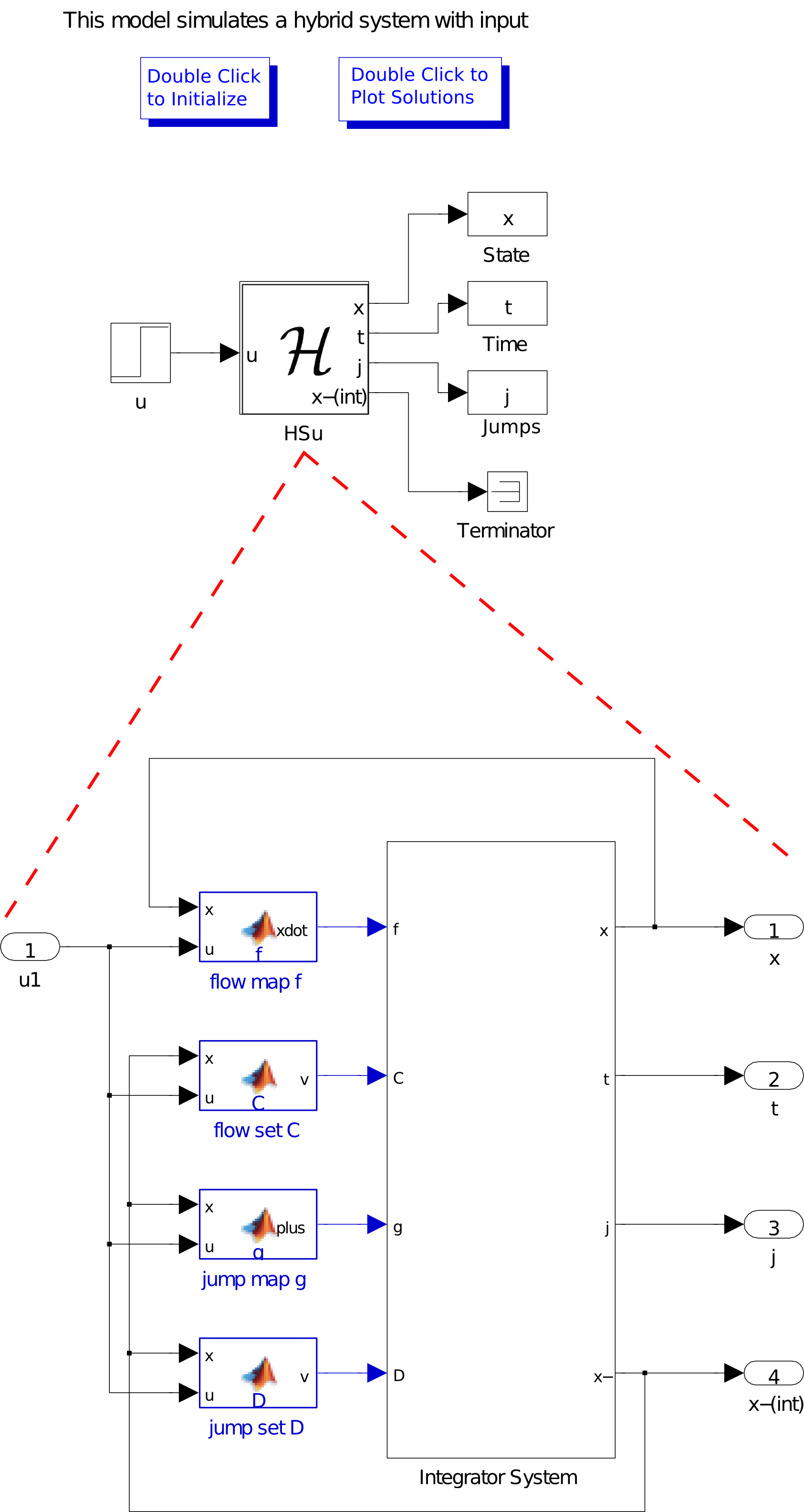}
    \caption{Modèle Simulink de simulation hybride}
    \label{fig:sim_hybrid}
\end{figure}

Aussi, la simulation du mécanisme hybride a demandé un apprentissage car bien qu'elle s'opère sur la plateforme Simulink (Matlab), elle nécessite une \textit{toolbox, Hybrid Equations Toolbox (HyEQ)}, développée par Ricardo Sanfelice
\cite{sanfelice_2017} (figure \ref{fig:sim_hybrid}).

\newpage
Cet outil permet de décrire l'ensemble des espaces définis dans la figure \ref{fig:sim_hybrid} (\textit{flow map, flow set, jump map, jump set} et l'espace état) et de représenter graphiquement les résultats.
\nomenclature{Flow map}{Fonction décrivant l'évolution continue de la dynamique.}
\nomenclature{Flow set}{Espace décrivant les conditions de saut de la dynamique continue.}
\nomenclature{Jump map}{Fonction décrivant l'évolution discrète de la dynamique.}
\nomenclature{Jump set}{Espace décrivant les conditions de saut de la dynamique discrète.}

\begin{figure}[h]
    \centering
    \includegraphics[width=0.7\textwidth]{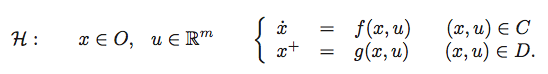}
    \caption{Description mathématique d'un système hybride}
    \label{fig:math_hybrid}
\end{figure}
\nomenclature{$\dot x$}{Évolution continue de la dynamique.}
\nomenclature{$x^{+}$}{Évolution discrète de la dynamique.}
Pour se familiariser avec la \textit{toolbox}, plusieurs systèmes ont été étudiés tels qu'un pendule (simple et rebondissant sur un mur), un réservoir d'eau et un système de chauffage dans une pièce.    
\paragraph{}
Un soin particulier a été apporté à l'apprentissage des LMIs, outil puissant de commande moderne. Pour cela, j'ai suivi le cours de Riccardo Bertollo, qui propose plusieurs exemples permettant de mettre en contexte les notions théoriques liées à l'utilisation des LMIs. Bien que nous regrettons, faute de temps, de n'avoir pu utiliser cet outil, nous avons entrevu leurs intérêts et imaginé leurs utilisations dans l'optimisation de certaines lois de commande, proposées par la suite.   
\nomenclature{LMI}{Linear Matrix Inequality ou Inégalité matricielle linéaire.}

\clearpage\null\newpage
\chapter{Description du drone}
\section{Modélisations}
La dynamique des drones de type \textit{tail-sitter} est particulièrement difficile à appréhender, notamment à cause des effets aérodynamiques et des couplages tridimensionnels. Dans cette section, nous allons décrire le modèle mathématique utilisé pour modéliser le comportement du drone, inspiré d'un travail de thèse \cite{Lustosa2017LaP}.
\paragraph{}
Dans un premier temps, nous allons décrire les référentiels utilisés, les systèmes d'axes, les paramètres cinématiques et le vecteur de commande.
Puis, on décrira les équations du modèle ainsi que les forces et moments qui sont en action.
Pour finir, nous appliquerons des hypothèses simplificatrices pour permettre le contrôle du drone.
\paragraph{}
Pour toute la suite de ce rapport, nous utiliserons la notation suivante: tous les vecteurs seront notés avec une lettre minuscule, en gras, alors que les scalaires seront en lettre minuscule, simple. Par exemple, $\boldsymbol{x} \in \mathbb{R}^n$ est un vecteur de dimension n, contrairement à $a \in \mathbb{R}$, qui est un scalaire.

\subsection{Référentiels et systèmes d'axes}
\subsubsection{Les repères}
\noindent
Les référentiels utilisés sont les référentiels usuels de la mécanique du vol.
\renewcommand \labelitemi{$\blacksquare$}
\begin{itemize}
    \item $\mathcal{I}$ est le repère inertiel. Il est lié à un point O de la surface terrestre. On le suppose galiléen, donc le théorème fondamental de la dynamique s'applique. On associe à $\mathcal{I}$ le système d'axes $\left[ x_{0},y_{0},z_{0}\right]$, où $x_{0}$ pointe vers le nord, $y_{0}$ pointe vers l'ouest et $z_{0}$ forme un trièdre direct. 
    \item $\mathcal{B}$ est le repère du corps. Il est lié au centre de gravité G du drone.  On associe à $\mathcal{B}$ le système d'axes $\left[ x_{b},y_{b},z_{b}\right]$, où $x_{b}$ est l'axe de roulis du drone, porté par l'axe des moteurs, $y_{b}$ est l'axe de tangage du drone, porté par l'axe des ailes et $z_{b}$ est l'axe de lacet, formant un trièdre direct.
    \item $\mathcal{A}$ est le repère aérodynamique. Il est lié au centre de gravité G du drone. On associe à $\mathcal{A}$ le système d'axes $\left[ x_{a},y_{a},z_{a}\right]$. Ce système d'axes est particulièrement bien adapté pour l'expression des efforts aérodynamiques. 
\end{itemize}
\newpage
Le schéma ci-dessous illustre quelques-uns de ces repères. 
\begin{figure}[h]
    \centering
    \includegraphics[width=0.7\textwidth]{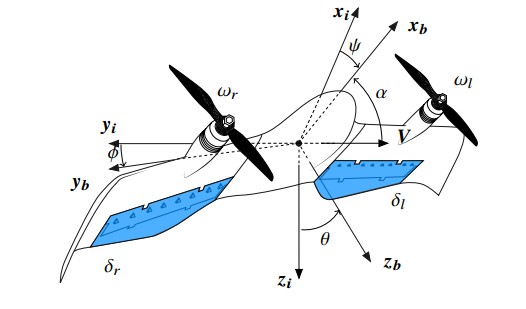}
    \caption{Référentiels du drone}
    \label{fig:ref_drone}
\end{figure}
\subsubsection{Quaternions et matrices de changement de base entre les repères}
\nomenclature{Angles d'Euler}{Technique permettant de représenter l'orientation d'un solide par rapport à un repère, avec des singularités.}
Dans ce travail, nous adoptons une représentation de l'attitude du drone avec le formalisme des quaternions unitaires. Cette représentation permet d'éviter les singularités (présentent avec les angles d'Euler) et de simplifier les équations. Soit un quaternion $q$, couramment représenté par un vecteur de dimension quatre, avec une norme unitaire, composé d'une partie scalaire $\eta \in \mathbb{R}$ et d'une partie vectorielle $\boldsymbol{\epsilon} \in \mathbb{R}^{3}$ tel que $q = \left [ \eta ~ \epsilon^\top \right]^\top$. Les matrices de changement de base permettent de passer librement d'un repère à un autre. Nous travaillons dans les bases orthonormées, donc les matrices de changement de base appartiennent au groupe spécial orthogonal SO(3). 
\begin{subequations}
    \begin{align}
        R(q) &= I_{3} + 2\eta \skewsym{\epsilon} + 2\skewsym{\epsilon}^{2}\\
        &= I_{3} + 2\eta \skewsym{\epsilon} + 2(\epsilon\epsilon^\top - \epsilon^\top\epsilon I_{3})
    \end{align}
\end{subequations}

\subsection{Paramètres du système}
Dans cette section, nous allons exprimer les paramètres cinématiques permettant de décrire le mouvement du drone dans l'espace, en rotation comme en translation. Par la suite, nous nous intéresserons aux propriétés cinétiques telles que la masse, le centrage et l'inertie du véhicule. Ces quantités permettent de faire le lien entre les forces et les dérivées des paramètres cinématiques.

\subsubsection{Paramètres cinématiques}
Les paramètres cinématiques qui permettent de décrire le mouvement du drone dans l'espace sont la position, la vitesse, l'attitude et la vitesse angulaire. Ainsi, on prendra comme vecteur d'état du système:
\begin{align}
    \left[\boldsymbol{p},~ \boldsymbol{v},~ \boldsymbol{q},~  \boldsymbol{\omega}\right]
\end{align}
Les composantes sont définies de la manière suivante:
\begin{itemize}
    \item $\boldsymbol{p} = {}^{\mathcal{I}}OG$ est la position du centre de gravité du drone par rapport à $\mathcal{I}$, exprimée dans le repère inertiel.
    \item $\boldsymbol{v}={}^{\mathcal{I}}V_{(G/\mathcal{I})}$ est la vitesse du centre de gravité par rapport à $\mathcal{I}$, exprimée dans le repère inertiel.
    \item $\boldsymbol{q} = \left [ \eta ~ \epsilon^\top \right]^\top$ est l'orientation du drone exprimée dans le repère corps.
    \item $\boldsymbol{\omega} = [p,~q,~r]^\top = {}^{\mathcal{B}} \Omega_{(G/\mathcal{I})}$ est le vecteur vitesse instantanée de rotation entre le repère corps et le repère inertiel, exprimé dans le repère corps.
\end{itemize}
\subsubsection{Paramètres cinétiques}
Les paramètres cinétiques caractéristiques du drone sont sa masse, son centrage et sa matrice d'inertie. Le véhicule étant à propulsion électrique, il n'y a pas de variation de masse au cours du temps. Ainsi, tous les paramètres cinétiques sont constants au cours du vol.\\
On définit :
\begin{itemize}
    \item La masse du drone, notée $m$.
    \item Le centre de gravité, noté $G$.
    \item La matrice d'inertie du drone, exprimée au point G dans le repère du corps, notée $J = diag(J1,J2,J3)$.
\end{itemize}
\ \\ \noindent
De plus, le drone évolue dans l'atmosphère terrestre, il est donc soumis à la gravité que nous définissons:
\begin{align}
    \boldsymbol{g} = \begin{bmatrix} 0 \\ 0 \\ g \end{bmatrix}
\end{align} avec $g = \SI{9.81}{\meter\per\square\second}$.

\subsubsection{Description du drone}
Le drone étudié est DarkO, drone conçu et fabriqué à l'ENAC. Il est constitué de plusieurs pièces imprimées en 3D, avec comme matériau d'impression l'Onyx. Ce matériau, à base de fibres de carbone omnidirectionnelles, confère une résistance accrue des pièces ainsi qu'une certaine souplesse permettant d'absorber les chocs.\\
\indent
Pour ce qui est de la structure, DarkO est un drone de type \textit{tail-sitter}, avec deux moteurs situés en avant de l'aile et deux élevons surdimensionnés, agissant comme surface de contrôle. 
\begin{figure}[h]
    \centering
    \includegraphics[width=0.5\textwidth]{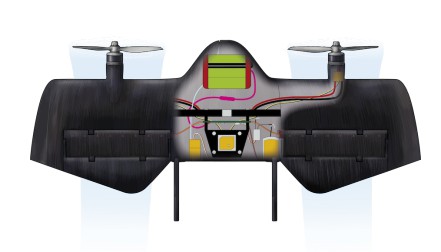}
    \caption{Modélisation de DarkO}
    \label{fig:vue3d}
\end{figure}

\subsubsection{Vecteur de commande}
Le drone est constitué de deux moteurs permettant de générer de la poussée. On peut mettre cet effort généré sous la forme : 
\begin{align}
T_{i} =\begin{bmatrix} k_{f}\omega_{i}^{2} \\ 0 \\ 0 \end{bmatrix}_{b},~i\in [1,2]
\end{align} 
où $\omega_{i}$ représente la vitesse de rotation du moteur en \SI{}{\radian\per\second} et $k_{f}$ est une constante représentant la force générée par l'hélice en \SI{}{\kilogram\meter}.\newline \\ Nous pouvons faire varier la vitesse indépendamment de chacun d'eux, de manière à générer des efforts différentiels. 
\\La rotation des deux moteurs génère aussi des moments:
\begin{align}
    N_{i} =\begin{bmatrix} (-1)^{i}  k_{m} \omega_{i}^{2} \\ 0 \\ 0 \end{bmatrix}_{b}
,~i\in [1,2]
\end{align}
où  $k_{m}$ est une constante représentant le moment généré par l'hélice en \SI{}{\kilogram\square\meter}.\\
On peut exprimer le moment moteur en fonction de la traction engendrée par l'hélice : 
\begin{align}
    N_{i} = (-1)^{i}  \frac{k_{m} }{k_{f}}T_{i}
,~i\in [1,2]
\end{align}  De plus, deux élevons sont présents dans l'aile volante. Ils se situent dans la partie d'aile soufflée par les moteurs. Ils permettent de générer des forces et des moments. Ainsi, nous devons définir deux matrices anti-symétriques, représentant l'efficacité des élevons à générer des forces et des moments: 
\begin{align}
    \Delta_{f,i} =  \skewsym{\xi_{f}} \delta_{i} = \begin{bmatrix} 0 & 0 & \xi_{f}\delta_{i} \\ 0 & 0 & 0 \\ -\xi_{f}\delta_{i} & 0 & 0 \end{bmatrix} 
\end{align}
\begin{align}
    \Delta_{m,i} = \skewsym{\xi_{m}} \delta_{i} = \begin{bmatrix} 0 & 0 & \xi_{m}\delta_{i} \\ 0 & 0 & 0 \\ -\xi_{m}\delta_{i} & 0 & 0 \end{bmatrix} 
\end{align}
La complexité d'un tel système réside dans le fait que l'utilisation des élevons génère en même temps des couples (utilisés pour orienter le drone) et des efforts, qui ont tendance à le déstabiliser.
\newline \\
\noindent
NOTE 1: on définit w1>0 et w2<0 car nous avons des hélices contre-rotatives.\\
NOTE 2: le signe des élevons est pris, par convention, positif pour un moment à cabrer.
\newline \\
\noindent
Il est donc possible de choisir comme vecteur de commande du drone:
\begin{align}
    \left[\omega_{1},~ \omega_{2},~ \delta_{1},~ \delta_{2}\right]
\end{align}
Effectivement, ces commandes représentent les actions possibles sur les moteurs et sur les ailerons. Nous observerons pas la suite qu'il est possible d'obtenir des surfaces de contrôle virtuelles, à partir de ces entrées effectives.

\subsubsection{Saturation des commandes}
Les actionneurs présents sur le drone possèdent des dynamiques qui limitent leurs actions. Chaque actionneur ayant ses caractéristiques propres, nous allons détailler les saturations appliquées, leur sens physique et leur seuil.
\paragraph{}
Pour les moteurs électriques, générant de la traction grâce aux hélices, il est nécessaire de saturer la vitesse de rotation pour avoir un modèle réaliste. Effectivement, saturer la vitesse de rotation permet de générer un modèle à énergie finie. Physiquement, cette saturation provient de la limite de tension que peut accepter le moteur, dans la mesure où la vitesse de rotation est directement liée à la tension présente aux bornes du moteur. Dans notre cas, la saturation permet de maintenir la vitesse de rotation du moteur  $\omega \in [200~; 1000]~\SI{}{\radian\per\second}$. La chaîne d'actionnement des moteurs, constituée notamment de l'ESC,\nomenclature{ESC}{Electronic Speed Control, contrôleur de moteur sans balai.} du moteur et de l'hélice peut être représentée par une dynamique du premier ordre. Cependant, nous avons fait le choix d'utiliser une saturation sur l'accélération du moteur pour représenter le temps de réaction de l'ensemble. En effet, ce choix technique permet de simplifier la mise en place de la simulation. Toutefois, il est possible de montrer que la saturation agit de la même manière que le premier ordre, bien qu'il existe quelques subtilités, en fonction de la fréquence du signal. Ainsi, nous avons $\dot \omega \in [-3000~; 3000]~\SI{}{\radian\per\square\second}$. On observe une grande réactivité dans cette chaîne d'actionnement, notamment par la faible inertie de l'hélice, par la commande exclusivement électrique, et une électronique embarquée à haute fréquence.  \\
\indent 
Les seconds actionneurs sont les élevons qui sont mus par des servomoteurs. Physiquement, leurs positions sont contraintes par la forme du drone et les limites des servomoteurs. Ainsi, nous pouvons observer une limite en position angulaire, qui se traduit dans le modèle par une saturation de la déflexion de gouverne. On maintient donc $\delta \in [-30~; 30]\text{\textdegree}$. Aussi, une limite importante pour les élevons est la saturation en vitesse angulaire. En effet, vu les efforts engendrés par ceux-ci, il est impossible d'atteindre des vitesses d'actionnement très élevées. De plus, avec la technologie utilisée, actionnement par servomoteur, il n'existe aucun réglage de vitesse, celui-ci se déplaçant toujours à 0.2 seconde pour 60~\textdegree. Ainsi $\dot \delta \in \{-5.24, 5.24\} ~\SI{}{\radian\per\square\second}$.

\subsection{Représentation dynamique}

La représentation dynamique du drone est extraite de la modélisation proposée dans \cite{lustosa:hal-03035938}, avec les modifications proposées dans \cite{olszaneckibarth:hal-02542982}, concernant l'utilisation de la formule de Diederich dans les simulations.
\nomenclature{Formule de Diederich}{Formule permettant de déterminer la courbe de portance d'un profil.}
\\
\noindent
On obtient, du théorème fondamental de la dynamique, les équations suivantes:
\begin{align}
    \left\{
        \begin{array}{l}
            \boldsymbol{\dot p} = \boldsymbol{v}\\
            \boldsymbol{\dot v} = \frac{1}{m}R(q)\sum F_{b}(x, u, w) + \boldsymbol{g}\\
            \boldsymbol{\dot q} = \frac{1}{2}\boldsymbol{q} \otimes \boldsymbol{\omega_{b}}\\
            \boldsymbol{\dot \omega_{b}} = J^{-1} \sum M_{b}(x,u,w) - J^{-1}\skewsym{\boldsymbol{\omega_{b}}} J\boldsymbol{\omega_{b}}
        \end{array}
    \right.
\end{align}
L'expression de l'ensemble des forces sur le drone est définie par:

\begin{multline}
   \sum F_{b} = T_{1}+ T_{2} 
   - \frac{S}{4S_{p}} \Phi^{(fv)} T_{1} - \frac{S}{4S_{p}} \Phi^{(fv)} T_{2} + \frac{S}{4S_{p}} \Phi^{(fv)} \Delta_{f,1} T_{1} + \frac{S}{4S_{p}} \Phi^{(fv)} \Delta_{f,2} T_{2}\\ 
   +\frac{1}{4} \rho S \eta \Phi^{(fv)} (\Delta_{f,1} + \Delta_{f,2})  v_{b} + \frac{1}{4} \rho S \eta \Phi^{(mv)} (\Delta_{f,1} + \Delta_{f,2}) B \omega_{b} 
   -\frac{1}{2} \rho S \eta \Phi^{(fv)} v_{b} - \frac{1}{2} \rho S \eta \Phi^{(mv)}  B \omega_{b}
\end{multline}
Que nous pouvons factoriser en :
\begin{multline}
    \sum F_{b} =  T_{1} + T_{2} + \frac{S}{4S_{p}} \Phi^{(fv)} \Big( \Delta_{f,1} - I\Big) T_{1} + \frac{S}{4S_{p}} \Phi^{(fv)} \Big( \Delta_{f,2} - I\Big) T_{2}
    \\ 
    + \frac{1}{4} \rho S \eta \Phi^{(fv)} \Big(\Delta_{f,1}+ \Delta_{f,2} - 2I \Big)  v_{b} + \frac{1}{4} \rho S \eta \Phi^{(mv)} \Big(\Delta_{f,1} + \Delta_{f,2} - 2I\Big) B \omega_{b}
\end{multline}
avec $\eta = \sqrt{v_{b}^{2} + \mu c^{2} \omega_{b}^{2}}$.\\
\nomenclature{$V_{b}$}{Vitesse du drone exprimée dans le repère $\mathcal{B}$  en \SI{}{\meter\per\second}.}\\
Pour tout $i \in [1,2]$, représentant la partie droite et gauche du drone: 
\begin{itemize}
    \item $T_{i}$ est la poussée du moteur.
    \item $\frac{S}{4S_{p}} \Phi^{(fv)} T_{i}$ est l'effort engendré par le flux d'air de l'hélice sur l'aile.
    \item $\frac{S}{4S_{p}} \Phi^{(fv)} \skewsym{\xi_{f}} \delta_{i} T_{i}$ est l'effort engendré par le flux d'air de l'hélice sur l'aileron.
    \item $\frac{1}{4} \rho S \Phi^{(fv)} \skewsym{\xi_{f}} (\delta_{1} + \delta_{2}) v v_{b}$ est l'effort engendré par la vitesse linéaire du drone sur les ailerons.
    \item $\frac{1}{4} \rho S \Phi^{(mv)} \skewsym{\xi_{f}} (\delta_{1} + \delta_{2}) B v \omega_{b}$ est l'effort engendré par la vitesse angulaire du drone sur les ailerons.
    \item $\frac{1}{2} \rho S \Phi^{(fv)} v v_{b} $ est l'effort engendré par la vitesse linéaire du drone sur l'aile.
    \item $\frac{1}{2} \rho S \Phi^{(mv)} v B \omega_{b} $ est l'effort engendré par la vitesse angulaire du drone sur l'aile.
\end{itemize}
\  \\
L'expression de l'ensemble des moments sur le drone peut s'exprimer de la manière suivante : 

\begin{multline}
   \sum M_{b} = -\frac{1}{2} \rho S B \Phi^{(mv)} v v_{b} -\frac{1}{2} \rho S B \Phi^{(mv)} v B \omega_{b} 
   - \frac{S}{4S_{p}} B \Phi^{(mv)} T_{1} - \frac{S}{4S_{p}} B \Phi^{(mv)} T_{2} 
   + \skewsym{p_{1}} T_{1} + \skewsym{p_{2}} T_{2} \\ 
   - \frac{S}{4S_{p}} \skewsym{a_{1}} \Phi^{(fv)} T_{1} - \frac{S}{4S_{p}} \skewsym{a_{2}} \Phi^{(fv)} T_{2} 
   + \frac{S}{4S_{p}} \skewsym{a_{1}} \Phi^{(fv)} \skewsym{\xi_{f}} \delta_{1} T_{1} + \frac{S}{4S_{p}} \skewsym{a_{2}} \Phi^{(fv)} \skewsym{\xi_{f}} \delta_{2} T_{2} \\ 
   + N_{1} + N_{2} 
   + \frac{1}{4} \rho S \skewsym{a_{1}} \Phi^{(fv)} \skewsym{\xi_{f}} \delta_{1} v v_{b} + \frac{1}{4} \rho S \skewsym{a_{2}} \Phi^{(fv)} \skewsym{\xi_{f}} \delta_{2} v v_{b} \\ 
   +\frac{1}{4} \rho S \skewsym{a_{1}} \Phi^{(mv)} \skewsym{\xi_{f}} \delta_{1} B v \omega_{b} + \frac{1}{4} \rho S \skewsym{a_{2}} \Phi^{(mv)} \skewsym{\xi_{f}} \delta_{2} B v \omega_{b}
   + \frac{1}{4} \rho S B \Phi^{(mv)} \skewsym{\xi_{m}} (\delta_{1} + \delta_{2}) v v_{b} \\ + \frac{1}{4} \rho S B \Phi^{(m\omega)} \skewsym{\xi_{m}} (\delta_{1} + \delta_{2}) v B \omega_{b}
   + \frac{S}{4S_{p}} B \Phi^{(mv)} \skewsym{\xi_{m}} \delta_{1}T_{1} + \frac{S}{4S_{p}} B \Phi^{(mv)} \skewsym{\xi_{m}} \delta_{2}T_{2}
\end{multline}
Que nous pouvons factoriser en :
\begin{multline}
   \sum M_{b} =  - \frac{S}{4S_{p}} \bigg[ B \Phi^{(mv)} \Big(\Delta_{m,1}- I \Big) + \skewsym{a_{1}} \Phi^{(fv)} \Big(I + \Delta_{m,1}\Big) \bigg] T_{1}
   - \frac{S}{4S_{p}} \bigg[ B \Phi^{(mv)} \Big(\Delta_{m,2} - I \Big)\\ +  \skewsym{a_{2}} \Phi^{(fv)} \Big(I + \Delta_{m,2}\Big) \bigg] T_{2} 
     + \frac{1}{4} \rho S \eta \bigg[ \Big(\skewsym{a_{1}} \Phi^{(fv)}  + B \Phi^{(mv)} \Big) \Delta_{m,1} 
    + \Big( \skewsym{a_{2}} \Phi^{(fv)} + B \Phi^{(mv)}  \Big) \Delta_{m,2}\\ - 2 B \Phi^{(mv)}  \bigg] v_{b} 
   +\frac{1}{4} \rho S \eta \bigg[ \Big( \skewsym{a_{1}}  + B \Big) \Delta_{m,1} + \Big ( \skewsym{a_{2}}  + B  \Big) \Delta_{m,2} - 2 B  \bigg] \Phi^{(mv)} B  \omega_{b} 
   + \skewsym{p_{1}} T_{1} + \skewsym{p_{2}} T_{2}\\ + N_{1} + N_{2}
\end{multline}

Pour tout $i \in [1,2]$, représentant la partie droite et gauche du drone: 
\begin{itemize}
    \item $\frac{1}{2} \rho S B \Phi^{(mv)} v v_{b}$ est le moment engendré par l'aile en fonction de la vitesse linéaire.
    \item $\frac{1}{2} \rho S B \Phi^{(mv)} v B \omega_{b}  $ est le moment engendré par l'aile en fonction de la vitesse angulaire.
    \item $\frac{S}{4S_{p}} B \Phi^{(mv)} T_{i}$ est le moment engendré par le flux d'air de l'hélice sur l'aile.
    \item $\skewsym{p_{i}} T_{i}$ est le moment engendré par la poussée du moteur.
    \item $\frac{S}{4S_{p}} \skewsym{a_{i}} \Phi^{(fv)} T_{i}$ est le moment engendré par le flux d'air de l'hélice sur l'aile, dû au décalage du centre aérodynamique par rapport au centre de masse.
    \item $\frac{S}{4S_{p}} \skewsym{a_{i}} \Phi^{(fv)} \skewsym{\xi_{f}} \delta_{i} T_{i}$ est le moment engendré par le flux d'air de l'hélice sur l'aileron, dû au décalage du centre aérodynamique par rapport au centre de masse.
    \item $N_{i}$ est le moment engendré par la réaction du moteur.
    \item $\frac{1}{4} \rho S \skewsym{a_{i}} \Phi^{(fv)} \skewsym{\xi_{f}} \delta_{i} v v_{b}$ est le moment engendré par l'aileron en fonction de la vitesse linéaire du drone.
    \item $\frac{1}{4} \rho S \skewsym{a_{i}} \Phi^{(mv)} \skewsym{\xi_{f}} \delta_{i} B v \omega_{b}$ est le moment engendré par l'aileron en fonction de la vitesse angulaire du drone.
    \item $\frac{1}{4} \rho S B \Phi^{(mv)} \skewsym{\xi_{m}} (\delta_{1} + \delta_{2}) v v_{b}$ est le moment engendré par la vitesse linéaire du drone sur les ailerons.
    \item $\frac{1}{4} \rho S B \Phi^{(mv)} \skewsym{\xi_{m}} (\delta_{1} + \delta_{2}) v B \omega_{b}$ est le moment engendré par la vitesse angulaire du drone sur les ailerons.
    \item $\frac{S}{4S_{p}} B \Phi^{(mv)} \skewsym{\xi_{m}} \delta_{i}T_{i}$est le moment engendré par le flux d'air de l'hélice sur les ailerons.
\end{itemize}
\newpage
\section{Étude des modèles }
Comme nous travaillons sur plusieurs lois de commande agissant dans des conditions de vol différentes, nous avons besoin de décrire au mieux la dynamique du drone dans la phase de vol concernée. L'objectif de cette partie est de développer les hypothèses retenues, ainsi que les équations de dynamique.
\subsection{Modèle non-linéaire simplifié}
Dans cette section, nous prenons comme hypothèse que les efforts et les moments engendrés par la vitesse linéaire et angulaire du drone sont très faibles, devant les efforts engendrés par les hélices et le flux d'air qu'elles génèrent. On suppose que nous sommes dans un environnement sans vent, conformément à ce qui a été discuté avec M. Murat Bronz. Cette supposition s'explique par le fait que l'on cherche à stabiliser le drone à la verticale sans vitesse, donc que tous les effets engendrés par une vitesse air sont négligeables. \\ \indent
Lors de la convergence du drone vers un point, celui-ci se déplace en pseudo-stationnaire donc la vitesse air reste faible et inutilisée pour la stabilisation du drone. Dès lors, on la traite comme une perturbation. Les effets de la vitesse angulaire sont négligés dans le modèle et traités comme une perturbation lors de l'orientation du drone. Cela engendre des transitoires lors de la convergence. Il est donc nécessaire d'avoir des preuves de convergence des lois de commande pour assurer le rejet des effets déstabilisants.
\newline \\
\noindent
On prend ainsi:
\begin{align}
    \left\{
        \begin{array}{l}
            \boldsymbol{\dot p} = \boldsymbol{v}\\
            \boldsymbol{\dot v} = \frac{1}{m}R(q)\sum F_{b}(x, u, w) + \boldsymbol{g}\\
            \boldsymbol{\dot q} = \frac{1}{2}\boldsymbol{q} \otimes \boldsymbol{\omega_{b}}\\
            \boldsymbol{\dot \omega_{b}} = J^{-1} \sum M_{b}(x,u,w) - J^{-1}\skewsym{\boldsymbol{\omega_{b}}} J\boldsymbol{\omega_{b}}
        \end{array}
    \right.
\end{align}
Avec l'expression des forces définie par : 

    \begin{flalign}
        \sum F_{b} &=  \Big(I - \frac{S}{4S_{p}} \Phi^{(fv)} \Big) T_{1} + T_{2}) + \frac{S}{4S_{p}} \Phi^{(fv)} (\Delta_{f,1} T_{1} + \Delta_{f,2} T_{2})
    \end{flalign}
Et l'expression des moments :   

\begin{multline}
   \sum M_{b} = - \frac{S}{4S_{p}} B \Phi^{(mv)} (T_{1} + T_{2} )
   + \frac{S}{4S_{p}} B \Phi^{(mv)} (\Delta_{m,1}T_{1} + \Delta_{m,2}T_{2}) 
   - \frac{S}{4S_{p}} \skewsym{a_{1}} \Phi^{(fv)} (T_{1} - T_{2} )\\
   + \frac{S}{4S_{p}} \skewsym{a_{1}} \Phi^{(fv)} (\Delta_{f,1} T_{1}- \Delta_{f,2} T_{2}) 
   +\frac{k_{m} }{k_{f}}(T_{1}- T_{2} ) + \skewsym{p_{1}} T_{1} + \skewsym{p_{2}} T_{2}
 \end{multline}
On peut exprimer cette expression sous la forme:
\begin{flalign}
    \sum F_{b} &= \begin{bmatrix} 1-\frac{S}{4S_{p}} C_{d0} \\ 0  \\ 0 \end{bmatrix}_{b}(T_{1,x} + T_{2,x}) + \begin{bmatrix} 0  \\ 0   \\ -\frac{S}{4S_{p}}(2\pi + C_{d0})\xi_{f}  \end{bmatrix}_{b} (\delta_{1}T_{1,x} + \delta_{2}T_{2,x})
\end{flalign}
 
\begin{multline}
   \sum M_{b} =  \begin{bmatrix} \frac{k_{m} }{k_{f}} \\  0\\ p_{y}+\frac{S}{4S_{p}} a_{y} C_{d0} \end{bmatrix} (T_{1,x} - T_{2,x}) +\begin{bmatrix} \frac{S}{4S_{p}}a_{y}( 2\pi + C_{d0})\xi_{f} \\ 0 \\ 0\end{bmatrix}(\delta_{1}T_{1,x} - \delta_{2}T_{2,x})\\
   + \begin{bmatrix} 0 \\ \frac{S}{4S_{p}} \Delta_{r}( 2\pi + C_{d0})\xi_{m}  \\ 0\end{bmatrix}(\delta_{1}T_{1,x} + \delta_{2}T_{2,x})
\end{multline}

\noindent
On voit apparaître les surfaces de contrôle du drone. Effectivement, la manipulation des équations nous a montré que le drone possède un couplage de la vitesse de rotation de l'hélice avec la déflexion de gouverne soufflée par cette dernière. Ainsi, on préférera utiliser le produit des deux comme commande fictive, pour simplifier la description de la dynamique. Aussi, on observe que l'on n'a pas de découplage entre la force et le moment exercés sur le drone. En d'autres termes, une action quelconque sur les commandes du drone ne permet pas de générer une force sans créer de moment. En réalité, nous observerons plus loin que nous allons prendre des hypothèses pour approcher le découplage. (cf chapitre \ref{chap:hover})
\newline \\
\noindent
Mettons les équations sous la forme matricielle, avec un nouveau vecteur de commande basé sur les commandes virtuelles obtenues précédemment  :
\begin{align}
    u = \begin{bmatrix}T_{1,x}  & T_{2,x}  & \delta_{1}T_{1,x} & \delta_{2}T_{2,x} \end{bmatrix} ^\top
\end{align}

\begin{subequations}
    \begin{flalign}\label{eq:force_matrice}
        \sum F_{b} &= \begin{bmatrix} ( 1-\frac{S_{wet}}{4S_{p}} C_{d0}) & ( 1-\frac{S_{wet}}{4S_{p}} C_{d0}) & 0 & 0 \\  0 & 0 & 0 & 0 \\  0 & 0 & -(\frac{S_{wet}}{4S_{p}}(2\pi + C_{d0})\xi_{f}) & -(\frac{S_{wet}}{4S_{p}}(2\pi + C_{d0})\xi_{f})  \end{bmatrix}
        \begin{bmatrix}T_{1,x} \\  T_{2,x} \\ \delta_{1}T_{1,x} \\ \delta_{2}T_{2,x} \end{bmatrix}\\
        &=F_{b}u
    \end{flalign}
    
    \begin{flalign}\label{eq:moment_matrice}
       \sum M_{b}  &= \begin{bmatrix} (\frac{k_{m} }{k_{f}}) & -(\frac{k_{m} }{k_{f}}) & \frac{S_{wet}}{4S_{p}}a_{y}( 2\pi + C_{d0})\xi_{f} & -\frac{S_{wet}}{4S_{p}}a_{y}( 2\pi + C_{d0})\xi_{f} \\  0 & 0 & \frac{S_{wet}}{4S_{p}} \Delta_{r}( 2\pi + C_{d0})\xi_{m} & \frac{S_{wet}}{4S_{p}} \Delta_{r}( 2\pi + C_{d0})\xi_{m} \\  (p_{y}+\frac{S_{wet}}{4S_{p}} a_{y} C_{d0}) & -(p_{y}+\frac{S_{wet}}{4S_{p}} a_{y} C_{d0}) & 0 & 0  \end{bmatrix}
        \begin{bmatrix}T_{1,x} \\  T_{2,x} \\ \delta_{1}T_{1,x} \\ \delta_{2}T_{2,x} \end{bmatrix}\\
        &=M_{b}u
    \end{flalign}
\end{subequations}
avec $F_{b}$, $M_{b} \in \mathbb{R}^{3x4}$, respectivement, les matrices d'entrée de force et de moment.\newline \\
On note que la deuxième ligne de la matrice de force est nulle, donc il n'est pas possible de créer de forces sur l'axe $y$ avec les commandes présentes sur le drone. Aussi, aucune ligne n'est nulle dans la matrice de moment, ce qui permet de générer des moments sur tous les axes et donc d'orienter de drone. De plus, nous avons $\rank(F_{b}) = 2$ et $\rank(M_{b}) = 3$.
\\
Pour la suite, nous utilisons le modèle suivant pour la commande en vol pseudo-stationnaire (hovering):
\begin{align}
    \left\{
        \begin{array}{l}
            \boldsymbol{\dot p} = \boldsymbol{v}\\
            m\boldsymbol{\dot v} = R(q)F_{b}u - m\boldsymbol{g}\\
            \boldsymbol{\dot q} = \frac{1}{2}\boldsymbol{q} \otimes \boldsymbol{\omega_{b}}\\
            J \boldsymbol{\dot \omega_{b}} =  M_{b}u - \skewsym{\boldsymbol{\omega_{b}}}J\boldsymbol{\omega_{b}}
        \end{array}
    \right.
\end{align}
\subsection{Modèle non-linéaire augmenté par la vitesse air}
Nous allons augmenter le système simplifié précédent en y ajoutant les termes perturbatifs dépendants de $v_{b}$, $\delta_{1}$ et $\delta_{2}$. 
\begin{align}
    \left\{
        \begin{array}{l}
            \boldsymbol{\dot p} = \boldsymbol{v}\\
            m\boldsymbol{\dot v} = R(q)F_{b}u - m\boldsymbol{g} + R(q)F_{p}(v_{b},~u)\\
            \boldsymbol{\dot q} = \frac{1}{2}\boldsymbol{q} \otimes \boldsymbol{\omega_{b}}\\
            J \boldsymbol{\dot \omega_{b}} =  M_{b}u - \skewsym{\boldsymbol{\omega_{b}}}J\boldsymbol{\omega_{b}} + M_{p}(v_{b},~u) 
        \end{array}
    \right.
\end{align}
Nous pouvons définir:

\begin{multline}
    F_{p}(v_{b},~u) =  \frac{1}{4} \rho S \eta \Phi^{(fv)} \Big(\Delta_{f,1} + \Delta_{f,2} - 2I \Big)  v_{b} \\
    = \frac{1}{2} \rho S \eta \begin{bmatrix} -C_{d0}v_{b,x}  \\ 0  \\ -(2\pi +C_{d0})v_{b,z}\end{bmatrix} + \frac{1}{4} \rho S \eta \xi_{f} \begin{bmatrix} C_{d0} v_{b,z}  \\ 0  \\ -(2\pi +C_{d0}) v_{b,x}\end{bmatrix}(\delta_{1}+\delta_{2})
\end{multline}
aussi:
\begin{multline}
    M_{p}(v_{b},~u) = -\frac{1}{2} \rho S \eta B \Phi^{(mv)} v_{b} + \frac{1}{4} \rho S \eta \bigg[ \skewsym{a_{1}} \Phi^{(fv)} (\Delta_{m,1} -\Delta_{m,2}) + B \Phi^{(mv)} (\Delta_{m,1} +\Delta_{m,2}) \bigg] v_{b} \\
         = \frac{1}{2} \rho S \eta  \begin{bmatrix} 0  \\ \Delta_{r}(2\pi +C_{d0}) v_{b,z}  \\ 0\end{bmatrix}+ \frac{1}{4} \rho S \eta \left( \begin{bmatrix} 0  \\ \Delta_{r}(2\pi +C_{d0})\xi_{m} v_{b,x}  \\ 0\end{bmatrix}(\delta_{1}+\delta_{2})\right.\\\left. + \begin{bmatrix} a_{y}(2\pi +C_{d0})\xi_{f} v_{b,x}    \\  0  \\ a_{y}C_{d0}\xi_{f} v_{b,z} \end{bmatrix} (\delta_{1}-\delta_{2})\right)
\end{multline}
Pour cette nouvelle représentation, nous allons définir un nouveau vecteur de commande :
\begin{align}
    u' = \begin{bmatrix}T1+T2  & T1-T2  & \delta_{1}+\delta_{2} & \delta_{1}-\delta_{2} \end{bmatrix} ^\top
\end{align}
Ainsi, on va pouvoir combiner $ F_{b}u + F_{p}(v_{b},~u) = F(u')$, ce qui nous donne :
\begin{subequations} \label{eq:F}
    \begin{align}\label{eq:Fx}
        F_{x} = ( 1-\frac{S_{wet}}{4S_{p}} C_{d0})(T1+T2) - \frac{1}{2} \rho S \eta C_{d0}v_{b,x} + \frac{1}{4} \rho S \eta \xi_{f} C_{d0}v_{b,z} (\delta_{1}+\delta_{2})
    \end{align}
    \begin{align}
        F_{y} = 0
    \end{align}
    \begin{multline}
        F_{z} = -\frac{S_{wet}}{8S_{p}}(2\pi + C_{d0})\xi_{f}\left[ (\delta_{1}+\delta_{2})(T1+T2) + (\delta_{1}-\delta_{2})(T1-T2) \right] - \frac{1}{2} \rho S \eta (2\pi +C_{d0})v_{b,z}\\- \frac{1}{4} \rho S \eta \xi_{f} (2\pi +C_{d0}) v_{b,x}(\delta_{1}+\delta_{2})
    \end{multline}
\end{subequations}
et combiner $M_{b}u + M_{p}(v_{b},~u) = M(u')$, ce qui nous donne :
\begin{subequations} \label{eq:M}
    \begin{multline}
        M_{x} = \left(\frac{k_{m}}{k_{f}} + \frac{S_{wet}}{8S_{p}} a_{y}( 2\pi + C_{d0})\xi_{f} (\delta_{1}+\delta_{2}) \right)(T1-T2)+ \bigg(\frac{S_{wet}}{8S_{p}}a_{y}( 2\pi  + C_{d0})\xi_{f} (T1+T2)\\ + \frac{1}{4} \rho S \eta a_{y}(2\pi +C_{d0})\xi_{f} v_{b,x} \bigg)(\delta_{1}-\delta_{2})
    \end{multline}
    \begin{multline}
        M_{y} = \left(\frac{S_{wet}}{8S_{p}} \Delta_{r}( 2\pi + C_{d0})\xi_{m}(T1-T2) + \frac{1}{4} \rho S \eta  \Delta_{r}(2\pi +C_{d0})\xi_{m} v_{b,x}  \right)(\delta_{1}+\delta_{2})\\ + \frac{S_{wet}}{8S_{p}} \Delta_{r}( 2\pi + C_{d0})\xi_{m}(T1+T2)(\delta_{1}-\delta_{2}) + \frac{1}{2} \rho S \eta \Delta_{r}(2\pi +C_{d0}) v_{b,z}
    \end{multline}
    \begin{align}
        M_{z} = \left((p_{y}+\frac{S_{wet}}{4S_{p}} a_{y} C_{d0}) \right)(T1-T2) + \frac{1}{4} \rho S \eta a_{y}C_{d0}\xi_{f} v_{b,z} (\delta_{1}-\delta_{2})
    \end{align}
\end{subequations}
Nous utiliserons cette modélisation pour effectuer la commande en vol horizontal du drone.

\clearpage\null\newpage
\chapter{Commande hovering}\label{chap:hover}
Dans cette section, nous allons développer des lois de commande conçues pour stabiliser le drone à la verticale. La convergence amènera le drone à s'orienter de manière à annuler la somme du poids du drone et de la traction générée par les hélices. 
\section{Étude du modèle linéarisé}
La première étude porte sur la linéarisation du modèle sur un point d'équilibre. 
Considérons comme modèle non-linéaire :
\begin{align}\label{eq:dyna} 
    \left\{
        \begin{array}{l}
        \boldsymbol{\dot p} = \boldsymbol{v}\\
        m\boldsymbol{\dot v} = R(q)F_{b}u - m\boldsymbol{g}\\
        \boldsymbol{\dot q} = \frac{1}{2}\boldsymbol{q} \otimes \boldsymbol{\omega_{b}}\\
        J \boldsymbol{\dot \omega_{b}} =  M_{b}u - \skewsym{\boldsymbol{\omega_{b}}}J\boldsymbol{\omega_{b}} 
        \end{array}
    \right.
\end{align}
avec comme vecteur d'état $x = \begin{bmatrix} \boldsymbol{p} & \boldsymbol{v} &  \boldsymbol{q} & \boldsymbol{\omega_{b}}  \end{bmatrix}^\top$.\\
\subsection{Caractérisation du point d'équilibre}
\noindent
Pour définir le point d'équilibre, on commence par annuler toutes les dérivées.\\
La première égalité de (\ref{eq:dyna}) nous donne:
\begin{align}
    \boldsymbol{\dot p} = \boldsymbol{v} = 0 \implies \boldsymbol{v} = 0
\end{align}
La troisième égalité de (\ref{eq:dyna}) nous donne:
\begin{align}
    \boldsymbol{\dot q} = \frac{1}{2}\boldsymbol{q} \otimes \boldsymbol{\omega_{b}} = 0 \implies \boldsymbol{\omega_{b}} = 0
\end{align}
Cette implication s'explique par le fait que la norme du quaternion est 1 donc il ne s'annule jamais. On en déduit donc que $\overline{\boldsymbol{\omega_{b}}} = 0$.\\
À partir de l'égalité précédente,
\begin{align}
     J \boldsymbol{\dot \omega_{b}} =  M_{b}\overline{u} - \skewsym{\boldsymbol{\omega_{b}}}J\boldsymbol{\omega_{b}} = 0 \implies  M_{b}\overline{u} = 0
\end{align}
Ainsi, on sait que l'équilibre est atteint quand $\overline{u} \in \ker M_{b}$.\\
La base générant le noyau de $M_{b}$ est de la forme $[1~1~0~0]^\top$ et, par conséquent, $\overline{u} = \lambda [1~1~0~0]^\top ,\text{ avec } \lambda \in \mathds{R}_{+}^{*}$. \\
La seconde égalité de (\ref{eq:dyna}) nous donne:
\begin{align}
     m\boldsymbol{\dot v} = R(q)F_{b}\overline{u} - m\boldsymbol{g} = 0 \implies R(q)F_{b}\overline{u} = m\boldsymbol{g} 
\end{align}
On définit: 
\begin{align}
    \begin{pmatrix}T\\0\\0\end{pmatrix} = F_{b}u = F_{b}\lambda \begin{pmatrix}1\\1\\0\\0\end{pmatrix}
\end{align}
L'objectif étant de déterminer l'orientation à l'équilibre, nous devons déterminer le quaternion solution de:
\begin{align}
    R(q) \begin{pmatrix}T\\0\\0\end{pmatrix} + \begin{pmatrix}0\\0\\ mg\end{pmatrix} = \begin{pmatrix}0\\0\\ 0\end{pmatrix}
\end{align}
Cela correspond à l'ensemble de rotation projetant l'axe $e_{1}$ vers l'axe $e_{3}$.\\
Il est possible d'exprimer l'ensemble de la matrice de rotation en fonction des éléments du quaternion ainsi:
\begin{align}
    \begin{pmatrix}
    \eta^{2} + \epsilon_{1}^{2} - \epsilon_{2}^{2} - \epsilon_{3}^{2} & 2\epsilon_{1} \epsilon_{2} - 2\eta\epsilon_{3} & 2\eta \epsilon_{2} + 2\epsilon_{1}\epsilon_{3}\\
    2\eta \epsilon_{3} + 2\epsilon_{1}\epsilon_{2} & \eta^{2} - \epsilon_{1}^{2} + \epsilon_{2}^{2} - \epsilon_{3}^{2} & 2\epsilon_{2} \epsilon_{3} - 2\eta\epsilon_{1} \\
    2\epsilon_{1} \epsilon_{3} - 2\eta\epsilon_{2} & 2\eta\epsilon_{1} + 2\epsilon_{2} \epsilon_{3}   &\eta^{2} - \epsilon_{1}^{2} - \epsilon_{2}^{2} + \epsilon_{3}^{2}
    \end{pmatrix}\begin{pmatrix}T\\0\\0\end{pmatrix} = \begin{pmatrix}0\\0\\ mg\end{pmatrix}
\end{align}
En observant attentivement l'égalité ci-dessus, on observe que la matrice multiplie un vecteur dont seule sa première composante est non nulle. De ce fait, uniquement la première colonne de la matrice de rotation est multipliée à un terme différent de zéro. Ainsi, on peut simplifier l'équation matricielle précédente en un système de trois équations scalaires: 
\begin{align}
    \left\{
        \begin{array}{l}
        (\eta^{2} + \epsilon_{1}^{2} - \epsilon_{2}^{2} - \epsilon_{3}^{2}) T = 0\\
        (2\eta \epsilon_{3} + 2\epsilon_{1}\epsilon_{2}) T = 0\\
        ( 2\epsilon_{1} \epsilon_{3} - 2\eta\epsilon_{2})T = mg
        \end{array}
    \right.
    \implies
    \left\{
        \begin{array}{l}
        \eta^{2} + \epsilon_{1}^{2} - \epsilon_{2}^{2} - \epsilon_{3}^{2} = 0\\
        2\eta \epsilon_{3} + 2\epsilon_{1}\epsilon_{2} = 0\\
        ( 2\epsilon_{1} \epsilon_{3} - 2\eta\epsilon_{2})T = mg
        \end{array}
    \right.
\end{align}
Une solution évidente de ce système est $\eta =\epsilon_{2} = \frac{1}{\sqrt{2}} \text{ et }  \epsilon_{1} = \epsilon_{3} = 0$. Le quaternion associé est $\overline{q}= [\frac{1}{\sqrt{2}}~0~\frac{1}{\sqrt{2}}~0]^\top$. (Il serait cependant possible de déterminer l'ensemble des orientations du drone qui satisferait l'égalité. En pratique, il existe une multitude de quaternions solutions de cette égalité car il est possible de stabiliser le drone à la verticale avec n'importe quelle rotation, selon son axe longitudinal). 
\newline \\
On peut déterminer la valeur de $\lambda$ permettant d'obtenir cet équilibre:
\begin{align}
    ( 2\epsilon_{1} \epsilon_{3} + 2\eta\epsilon_{2})T = mg \implies 2\eta\epsilon_{2}T = mg \implies T = \frac{mg}{2\eta\epsilon_{2}} = mg
\end{align}
\begin{align}
    \lambda = \frac{mg}{( 1-\frac{S_{wet}}{4S_{p}} C_{d0})}
\end{align}
On observe que l'on applique un effort supérieur au poids du drone pour s'opposer aux pertes aérodynamiques.\\
On obtient ainsi un point d'équilibre pour notre système qui est défini par:

\begin{align}
     \left\{
        \begin{array}{l}
        \boldsymbol{\overline{p}}= \boldsymbol{p_{e}}, \text{ avec } p_{e} \in \mathds{R}^{3} \\
        \boldsymbol{\overline{v}} = 0\\
        \boldsymbol{\overline{q}} = [\frac{1}{\sqrt{2}}~0~\frac{1}{\sqrt{2}}~0]^\top\\
        \boldsymbol{\overline{\omega_{b}}} = 0\\
        \boldsymbol{\overline{u}} = \lambda [1~1~0~0]^\top = \frac{mg}{( 1-\frac{S_{wet}}{4S_{p}} C_{d0})} [1~1~0~0]^\top
        \end{array}
    \right.
\end{align}
\subsection{Linéarisation sur un point d'équilibre}
On linéarise autour de la position d'équilibre ci-dessus ; on définit chaque variable comme la somme de la position d'équilibre et une déviation (variable avec une tilde) : $p = \overline{p} + \tilde{p}$.\\

La linéarisation d'un quaternion nécessite une attention particulière. Nous nous sommes inspirés de \cite[Proof Lemma 1]{tregouet:hal-01760720} pour la réaliser. Effectivement, nous avons $\boldsymbol{q} = \left [ \eta ~ \epsilon^\top \right]^\top$, mais lors de la linéarisation, il est nécessaire de remplacer $\eta$ par sa valeur "contrainte", en raison de la norme unitaire du quaternion. Ainsi, $\eta = (1- \epsilon^\top \epsilon)^\frac{1}{2} = \sqrt{1 - \epsilon_{1}^2- \epsilon_{2}^2- \epsilon_{3}^2}$. L'équation du mouvement linéarisée a pour vecteur d'état $x = \begin{bmatrix} \boldsymbol{\tilde{p}} & \boldsymbol{\tilde{v}} & \boldsymbol{\tilde{\epsilon}} & \boldsymbol{\tilde{\omega_{b}}}  \end{bmatrix}^\top$.
\begin{align}
 \boldsymbol{\dot x} = A \boldsymbol{x} + B \boldsymbol{u} 
\end{align}
avec :
\begin{align}
     A_{ij} &= \frac{\partial f_{i}}{\partial x_{j}} \text{ et } B_{ij} = \frac{\partial f_{i}}{\partial u_{j}}
\end{align}
Le point de départ est l'équation de dynamique suivante:
\begin{align}
    f = \begin{pmatrix}
             v\\
             \boldsymbol{g} + \frac{1}{m}R(q)F_{b}u\\
            \frac{1}{2}q \otimes \omega_{b} \\
            J^{-1}M_{b}u-J^{-1}\skewsym{\omega_{b}}J\omega_{b} 
         \end{pmatrix}
\end{align}
Nous allons effectuer un développement en série de Taylor, que l'on arrête au premier ordre de chaque équation. La première équation ne dépend que de la vitesse du drone. Ainsi, la dérivée sera nulle sauf par rapport à la vitesse et le résultat sera égal à l'identité. La seconde équation dépend de l'orientation du drone. Pour calculer la dérivé par rapport à $\epsilon$, nous avons besoin de $F_{b}u$.
\nomenclature{Série de Taylor}{Série de termes qui coïncide avec la fonction au voisinage d'un point.}
\begin{subequations}
    \begin{align}
        F_{b}u &= \begin{bmatrix} ( 1-\frac{S_{wet}}{4S_{p}} C_{d0}) & ( 1-\frac{S_{wet}}{4S_{p}} C_{d0}) & 0 & 0 \\  0 & 0 & 0 & 0 \\  0 & 0 & -(\frac{S_{wet}}{4S_{p}}(2\pi + C_{d0})\xi_{f}) & -(\frac{S_{wet}}{4S_{p}}(2\pi + C_{d0})\xi_{f})  \end{bmatrix}
            \begin{bmatrix}T_{1,x} \\  T_{2,x} \\ \delta_{1}T_{1,x} \\ \delta_{2}T_{2,x} \end{bmatrix}\\
        &=\begin{bmatrix} ( 1-\frac{S_{wet}}{4S_{p}} C_{d0}) (T_{1,x} +T_{2,x})\\  0  \\ -(\frac{S_{wet}}{4S_{p}}(2\pi + C_{d0})\xi_{f})(\delta_{1}T_{1,x} + \delta_{2}T_{2,x})  \end{bmatrix}
    \end{align}
\end{subequations}
Plaçons nous à $\overline{u} = \frac{mg}{2( 1-\frac{S_{wet}}{4S_{p}} C_{d0})} [1~1~0~0]^\top$, ainsi: 

\begin{subequations}
    \begin{align}
        \frac{1}{m}F_{b}u &= \frac{1}{m}\begin{bmatrix} ( 1-\frac{S_{wet}}{4S_{p}} C_{d0}) (T_{1,x} +T_{2,x})\\
        0  \\
        -(\frac{S_{wet}}{4S_{p}}(2\pi + C_{d0})\xi_{f})(\delta_{1}T_{1,x} + \delta_{2}T_{2,x})  \end{bmatrix} \\
        &= \begin{bmatrix} g\\
        0  \\
        0  \end{bmatrix}
    \end{align} 
\end{subequations}
À partir de ce calcul, nous avons:
\begin{subequations}
    \begin{align}
        \frac{\partial\frac{1}{m}R(q)F_{b}u}{\partial \epsilon} &= 
         \begin{bmatrix} 
        0 & -4\epsilon_{2}g & 0\\
        2\epsilon_{2}g & 0 & 2\sqrt{1  - \epsilon_{2}^{2}}g  \\ 
         0 & (2\sqrt{1 - \epsilon_{2}^{2}} - \frac{2\epsilon_{2}^{2}}{\sqrt{1 - \epsilon_{2}^{2}}})g & 0 \end{bmatrix}\\
         &= \begin{bmatrix} 
        0 & 2\sqrt{2}g & 0\\
        -\sqrt{2}g & 0 & \sqrt{2}g  \\ 
         0 & -\sqrt{2}g & 0 \end{bmatrix}\\
         &=A_{\dot v ,\epsilon} 
    \end{align}
\end{subequations}
La troisième équation dépend de l'orientation et de la vitesse angulaire. Commençons par expliciter l'expression de cette équation.
\begin{align}
    \dot q =\frac{1}{2}\begin{bmatrix} 
    -\epsilon_{1}\omega_{1} -\epsilon_{2}\omega_{2} -\epsilon_{3}\omega_{3}\\
    \eta -\epsilon_{3}\omega_{2} +\epsilon_{2}\omega_{3}  \\
    \epsilon_{3}\omega_{1} + \eta - \epsilon_{1}\omega_{3} \\
    -\epsilon_{2}\omega_{1} + \epsilon_{1}\omega_{2} + \eta
    \end{bmatrix}
    = \frac{1}{2}\begin{bmatrix} 
    -\epsilon_{1}\omega_{1} -\epsilon_{2}\omega_{2} -\epsilon_{3}\omega_{3}\\
    \sqrt{1 - \epsilon_{1}^{2} - \epsilon_{2}^{2} - \epsilon_{3}^{2}} -\epsilon_{3}\omega_{2} +\epsilon_{2}\omega_{3}  \\
    \epsilon_{3}\omega_{1} + \sqrt{1 - \epsilon_{1}^{2} - \epsilon_{2}^{2} - \epsilon_{3}^{2}} - \epsilon_{1}\omega_{3} \\
    -\epsilon_{2}\omega_{1} + \epsilon_{1}\omega_{2} + \sqrt{1 - \epsilon_{1}^{2} - \epsilon_{2}^{2} - \epsilon_{3}^{2}}
    \end{bmatrix}
\end{align} 
Seule la partie vectorielle nous intéresse pour la linéarisation. Ainsi, il est possible de calculer l'expression de la linéarisation de l'équation d'orientation, par rapport à la partie vectorielle du quaternion.  
\begin{align}
\frac{\partial\dot q_{\epsilon}}{\partial \epsilon} &= 
\frac{1}{2}\begin{bmatrix} 
    \frac{-2\epsilon_{1}}{2\sqrt{1 - \epsilon_{1}^{2} - \epsilon_{2}^{2} - \epsilon_{3}^{2}}}\omega_{1} & \omega_{3}-\frac{2\epsilon_{2}}{2\sqrt{1 - \epsilon_{1}^{2} - \epsilon_{2}^{2} - \epsilon_{3}^{2}}}\omega_{1} & -\omega_{2}- \frac{2\epsilon_{3}}{2\sqrt{1 - \epsilon_{1}^{2} - \epsilon_{2}^{2} - \epsilon_{3}^{2}}}\omega_{1}\\ 
    -\omega_{3}-\frac{2\epsilon_{1}}{2\sqrt{1 - \epsilon_{1}^{2} - \epsilon_{2}^{2} - \epsilon_{3}^{2}}}\omega_{2} &  \frac{-2\epsilon_{2}}{2\sqrt{1 - \epsilon_{1}^{2} - \epsilon_{2}^{2} - \epsilon_{3}^{2}}}\omega_{2} & \omega_{1}-\frac{2\epsilon_{3}}{2\sqrt{1 - \epsilon_{1}^{2} - \epsilon_{2}^{2} - \epsilon_{3}^{2}}}\omega_{2} \\
    \omega_{2}-\frac{2\epsilon_{1}}{2\sqrt{1 - \epsilon_{1}^{2} - \epsilon_{2}^{2} - \epsilon_{3}^{2}}}\omega_{3} & -\omega_{1}-\frac{2\epsilon_{2}}{2\sqrt{1 - \epsilon_{1}^{2} - \epsilon_{2}^{2} - \epsilon_{3}^{2}}}\omega_{3} & \frac{-2\epsilon_{3}}{2\sqrt{1 - \epsilon_{1}^{2} - \epsilon_{2}^{2} - \epsilon_{3}^{2}}}\omega_{3}
\end{bmatrix} = 
\begin{bmatrix} 
    0 & 0 & 0 \\ 
    0 & 0& 0  \\
    0 & 0 & 0 
\end{bmatrix}
\end{align} 
De la même manière, il est possible de décrire l'expression de la linéarisation de l'équation de rotation, par rapport à la vitesse angulaire. 
\begin{subequations}
    \begin{align}
    \frac{\partial\dot q_{\epsilon}}{\partial \omega} &= 
    \frac{1}{2}\begin{bmatrix} 
        \sqrt{1 - \epsilon_{1}^{2} - \epsilon_{2}^{2} - \epsilon_{3}^{2}} & -\epsilon_{3} & \epsilon_{2}\\ 
       \epsilon_{3} &  \sqrt{1 - \epsilon_{1}^{2} - \epsilon_{2}^{2} - \epsilon_{3}^{2}}  & -\epsilon_{1} \\
        -\epsilon_{2} & \epsilon_{1} & \sqrt{1 - \epsilon_{1}^{2} - \epsilon_{2}^{2} - \epsilon_{3}^{2}}
    \end{bmatrix} = 
    \frac{1}{2}\begin{bmatrix} 
        \frac{\sqrt{2}}{2} & 0 & -\frac{\sqrt{2}}{2} \\ 
        0 & \frac{\sqrt{2}}{2}& 0  \\
        \frac{\sqrt{2}}{2} & 0 & \frac{\sqrt{2}}{2}
    \end{bmatrix}\\
    &=A_{\dot q ,\omega}
    \end{align} 
\end{subequations}
La dernière équation ne dépend que de la vitesse angulaire mais le résultat de la linéarisation est nul. 
Ainsi, on obtient la matrice d'état suivante: 
\begin{subequations}
\begin{align}
    A &=\begin{bmatrix}
    0 & \mathds{1}_{3} & 0 & 0 \\
    0 & 0 &  A_{\dot v ,\epsilon} & 0 \\
    0 & 0 & 0 & A_{\dot q ,\omega} \\
    0 & 0 & 0 & 0
    \end{bmatrix}\\
    &=\begin{bmatrix}
    0 & 0 & 0 & 1 & 0 & 0 & 0 & 0 & 0 & 0 & 0 & 0\\
    0 & 0 & 0 & 0 & 1 & 0 & 0 & 0 & 0 & 0 & 0 & 0\\
    0 & 0 & 0 & 0 & 0 & 1 & 0 & 0 & 0 & 0 & 0 & 0\\
    0 & 0 & 0 & 0 & 0 & 0 & 0 & 2\sqrt{2}g & 0 & 0 & 0 & 0\\
    0 & 0 & 0 & 0 & 0 & 0 & -\sqrt{2}g & 0 & \sqrt{2}g &  0 & 0 & 0\\
    0 & 0 & 0 & 0 & 0 & 0 & 0 & -\sqrt{2}g & 0 & 0 & 0 & 0\\
    0 & 0 & 0 & 0 & 0 & 0 & 0 & 0 & 0 & \frac{\sqrt{2}}{4} & 0 & -\frac{\sqrt{2}}{4}\\
    0 & 0 & 0 & 0 & 0 & 0 & 0 & 0 & 0 & 0 & \frac{\sqrt{2}}{4} & 0\\
    0 & 0 & 0 & 0 & 0 & 0 & 0 & 0 & 0 & \frac{\sqrt{2}}{4} & 0 & \frac{\sqrt{2}}{4}\\
    0 & 0 & 0 & 0 & 0 & 0 & 0 & 0 & 0 & 0 & 0 & 0\\
    0 & 0 & 0 & 0 & 0 & 0 & 0 & 0 & 0 & 0 & 0 & 0\\
    0 & 0 & 0 & 0 & 0 & 0 & 0 & 0 & 0 & 0 & 0 & 0\\
    \end{bmatrix}
\end{align}
\end{subequations}
Pour déterminer la matrice de commande, nous allons calculer la matrice jacobienne de l'équation de dynamique, par rapport à la commande exprimée au point d'équilibre.
\ \\
\noindent
La première et la troisième équation ne font pas apparaître de terme de commande; ainsi les dérivées partielles seront nulles.\\
La dérivée partielle de la seconde équation nous donne:
\begin{align}
\frac{\partial\frac{1}{m}R(q)F_{b}u}{\partial u} &= 
\frac{1}{m}\begin{bmatrix} 
0 & 0 & (\frac{S_{wet}}{4S_{p}}(2\pi + C_{d0})\xi_{f}) & (\frac{S_{wet}}{4S_{p}}(2\pi + C_{d0})\xi_{f})\\
0 & 0 & 0 & 0 \\ 
( 1-\frac{S_{wet}}{4S_{p}} C_{d0}) & ( 1-\frac{S_{wet}}{4S_{p}} C_{d0}) & 0 & 0 \end{bmatrix}
\end{align}
On observe que seul le terme $J^{-1}M_{b}u$ de la quatrième équation dépend de la commande. Pour déterminer la dérivée partielle, nous allons commencer par calculer l'expression de ce terme.
\begin{subequations}
    \begin{align}
        &J^{-1}M_{b}u =\begin{bmatrix} 
        \frac{1}{J_{1}}(\frac{k_{m} }{k_{f}}(T_{1,x}-T_{2,x}) + a_{y}( 2\pi + C_{d0})\xi_{m}(\delta_{1}T_{1,x}- \delta_{2}T_{2,x}))\\
        \frac{1}{J_{2}}\frac{S_{wet}}{4S_{p}} \Delta_{r}( 2\pi + C_{d0})\xi_{m}(\delta_{1}T_{1,x}+ \delta_{2}T_{2,x})\\
        \frac{1}{J_{3}}(p_{y}+\frac{S_{wet}}{4S_{p}} a_{y} C_{d0})(T_{1,x}-T_{2,x})
        \end{bmatrix}
    \end{align}
\end{subequations}
Il est maintenant possible d'obtenir la dérivée partielle par rapport à la commande de la dernière équation. 

\begin{align}
    \frac{\partial \dot \omega}{\partial u} &= \begin{bmatrix} 
    \frac{1}{J_{1}}(\frac{k_{m}}{k_{f}}) & -\frac{1}{J_{1}}(\frac{k_{m} }{k_{f}}) & \frac{1}{J_{1}} a_{y}( 2\pi + C_{d0})\xi_{m} & -\frac{1}{J_{1}} a_{y}( 2\pi + C_{d0})\xi_{m} \\
    0 & 0 & \frac{1}{J_{2}}\frac{S_{wet}}{4S_{p}} \Delta_{r}( 2\pi + C_{d0})\xi_{m} & \frac{1}{J_{2}}\frac{S_{wet}}{4S_{p}} \Delta_{r}( 2\pi + C_{d0})\xi_{m} \\ 
    \frac{1}{J_{3}}(p_{y}+\frac{S_{wet}}{4S_{p}} a_{y} C_{d0}) & -\frac{1}{J_{3}}(p_{y}+\frac{S_{wet}}{4S_{p}} a_{y} C_{d0}) & 0 & 0 \end{bmatrix}
\end{align}

\begin{align}
    B 
    &=\begin{bmatrix}
    0 & 0 & 0 & 0\\
    0 & 0 & 0 & 0\\
    0 & 0 & 0 & 0\\
    0 & 0 & \frac{1}{m}(\frac{S_{wet}}{4S_{p}}(2\pi + C_{d0})\xi_{f}) & \frac{1}{m}(\frac{S_{wet}}{4S_{p}}(2\pi + C_{d0})\xi_{f})\\
    0 & 0 & 0 & 0\\
    \frac{1}{m}( 1-\frac{S_{wet}}{4S_{p}} C_{d0}) & \frac{1}{m}( 1-\frac{S_{wet}}{4S_{p}} C_{d0})  & 0 & 0\\
    0 & 0 & 0 & 0\\
    0 & 0 & 0 & 0\\
    0 & 0 & 0 & 0\\
    \frac{1}{J_{1}}(\frac{k_{m} }{k_{f}}) & -\frac{1}{J_{1}}(\frac{k_{m} }{k_{f}}) & \frac{1}{J_{1}} a_{y}( 2\pi + C_{d0})\xi_{m} & -\frac{1}{J_{1}} a_{y}( 2\pi + C_{d0})\xi_{m}\\
    0 & 0 & \frac{1}{J_{2}}\frac{S_{wet}}{4S_{p}} \Delta_{r}( 2\pi + C_{d0})\xi_{m} & \frac{1}{J_{2}}\frac{S_{wet}}{4S_{p}} \Delta_{r}( 2\pi + C_{d0})\xi_{m}\\
    \frac{1}{J_{3}}(p_{y}+\frac{S_{wet}}{4S_{p}} a_{y} C_{d0}) & -\frac{1}{J_{3}}(p_{y}+\frac{S_{wet}}{4S_{p}} a_{y} C_{d0}) & 0 & 0\\
    \end{bmatrix}
\end{align}
On observe que l'on obtient une dynamique linéarisée contrôlable.
\subsection{Commande par retour d'état LQR}\label{control_lin}

À partir de la modélisation linéarisée précédente, nous pouvons déterminer un retour d'état. Nous avons choisi de déterminer les gains du retour d'état avec un calcul de type LQR, l'objectif étant de minimiser une fonction de coût quadratique :
\begin{align}
    J(u) =  \int_{0}^{\infty} (x^\top Q x + u^\top R u + 2x^\top N u) \,dt 
\end{align}
avec $Q = \mathbb{1}_{12x12}$, $R = \mathbb{1}_{4x4}$ et $N = \mathbb{0}_{12x4}$.
\newline \\
La matrice calculée par la fonction implémentée dans Matlab nous donne: 
\begin{align}
    \begin{bmatrix}
    0.08&-0.71&0.70&0.14&-1.04&0.92&0.00&5.75&-1.29&-5.69&0.05&0.00&-0.81\\
    0.08&0.71&0.70&0.14&1.04&0.92&0.00&-5.75&-1.29&5.69&-0.05&0.00&0.81\\
    -0.70&-0.05&0.08&-1.13&-0.07&0.09&0.00&-0.11&-15.32&-0.88&-0.71&-1.41&-0.06\\
    -0.70&0.05&0.08&-1.13&0.07&0.09&0.00&0.11&-15.32&0.88&0.71&-1.41&0.06\\\end{bmatrix}
\end{align}
\indent
On observe que cette matrice est de dimension 4 par 13. 
La septième colonne (composée de zéros) s'explique par le fait que la première composante du quaternion n'est pas contrôlée car c'est une variable fixée par la partie vectorielle du quaternion.\\
\newpage
\indent
Il est possible de faire évoluer cette commande vers une commande LQR, avec intégrateur. Ainsi, seraient annulées toutes les erreurs statiques qui proviendraient d'une perturbation telle qu'un vent constant face au drone. Cependant, l'ajout d'un intégrateur peut déstabiliser la dynamique de l'aéronef. Cette proposition a été élaborée sur la base d'un projet de commande réalisé à l'ENAC, où ce mécanisme avait été utilisé.
\begin{figure}[h]
\centering
\includegraphics[width=0.5\textwidth]{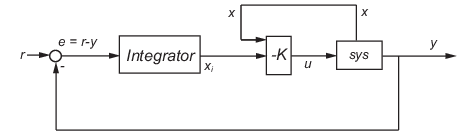}
\caption{Schéma représentant le fonctionnement d'un contrôle LQR avec intégrateur.}
\label{fig:lqi}
\end{figure}\\
On cherche à stabiliser une position donc on sélectionne uniquement les trois premières composantes du vecteur d'état pour les intégrer.
\newline \\
On définit le retour d'état comme : 
\begin{align}
    u = -K \begin{bmatrix}
    x\\
    x_{i}
    \end{bmatrix}
\end{align}
avec $x = \begin{bmatrix} \boldsymbol{p} & \boldsymbol{v} &  \boldsymbol{q} & \boldsymbol{\omega_{b}}  \end{bmatrix}^\top$ et $x_{i} =  \int_{ }^{ } (\boldsymbol{p_{e}} - \boldsymbol{p})^\top  \, \mathrm{d}t $.
\subsection{Simulation}
\begin{figure}[h]
\centering
\includegraphics[width=0.69\textwidth]{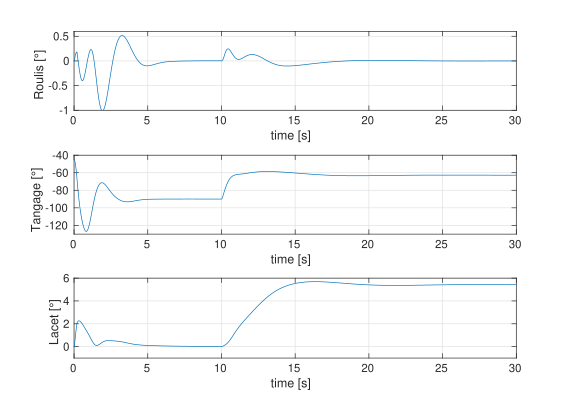}
\caption{Angles d'orientation du drone avec le contrôleur LQR et un intégrateur.}
\label{fig:orient_lqi}
\end{figure}

\begin{figure}[h]
\centering
\includegraphics[width=0.6\textwidth]{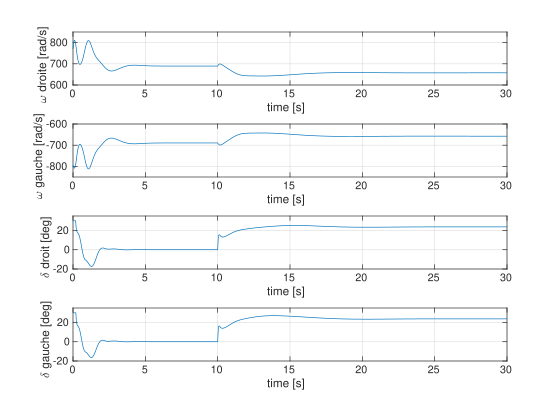}
\caption{Commande générée par le contrôleur LQR et un intégrateur.}
\label{fig:cmd_lqi}
\end{figure}

La position cible est le point d'origine $p=[0,~0,~0]^\top$, maintenu malgré un vent contant de \SI{5}{\meter\per\second}, débutant à dix secondes, face au drone.
\begin{figure}[h]
\centering
\includegraphics[width=0.7\textwidth]{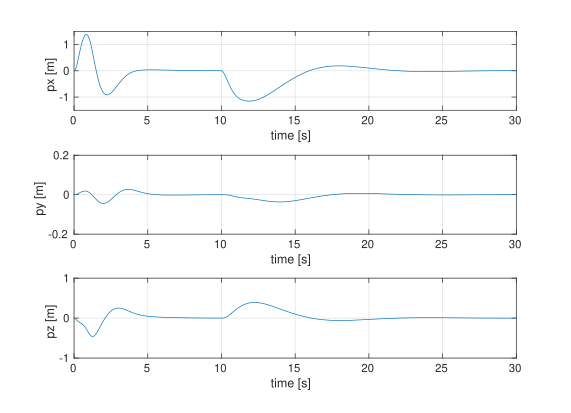}
\caption{Position du drone avec un contrôleur LQR et un intégrateur.}
\label{fig:pos_lqi}
\end{figure}
\newpage
On observe que le drone s'oriente dans le lit du vent, de manière à minimiser son impact et surtout à utiliser le vent pour générer de la portance, comme observé sur l'axe de tangage dans la figure \ref{fig:orient_lqi}
\\
\indent
Dès lors, on observe une diminution de la vitesse de rotation des moteurs dans la figure \ref{fig:cmd_lqi} et donc une consommation électrique plus faible. On en déduit l'importance d'orienter le drone en présence de vent. De plus, les élevons maintiennent une déflexion constante : cela créé un couple qui stabilise le drone à une incidence de -60°.
\\

\newpage
\subsection{Stabilité de la linéarisation}\label{stab_lin}
\begin{figure}[h]
\centering
\includegraphics[width=0.5\textwidth]{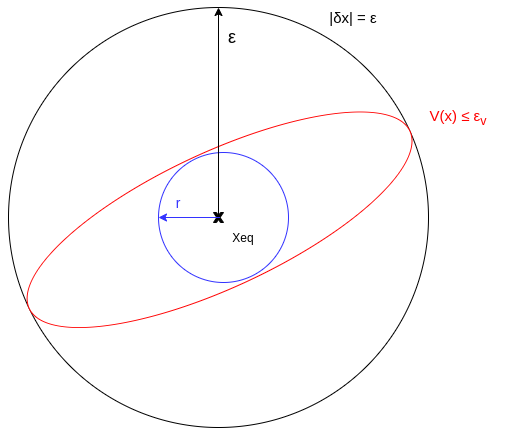}
\caption{Schéma représentant la stabilité d'une linéarisation autour d'un point d'équilibre.}
\label{fig:stab}
\end{figure}

L'intérêt d'une linéarisation a été discuté précédemment mais l'information importante est son domaine de stabilité. On le définit par l'ensemble des points initiaux pour lesquels le système converge vers l'équilibre. Dans notre cas, on cherche à déterminer le sous-espace rouge de la figure \ref{fig:stab}, qui représente l'ensemble de convergence basé sur la fonction de Luyapunov. L'intégralité du domaine noir est définie par la propriété $\dot V(x) < 0 $. L'ensemble bleu est la plus grande boule incluse dans l'ensemble rouge qui vérifie la propriété de convergence. Cette condition est extrêmement restrictive et complexe à obtenir mathématiquement puisque sa démonstration repose sur la majoration des erreurs quadratiques. Nous avons fait le choix, au vu du temps disponible, d'obtenir la borne supérieure de la fonction de Lyapunov, par simulation. Cette méthode empirique repose sur un échantillonage de trajectoires, leur simulation et l'analyse de convergence. On peut effectuer cette analyse grâce à la monotonie de la décroissance de la fonction de Lyapunov, prouvant ainsi la convergence.\\
\\
L'intégralité des contrôleurs proposés dans ce mémoire repose sur cette analyse. De ce fait, on effectue le passage hybride sur le critère de validité de la linéarisation pour s'assurer d'être, dans toutes les situations de vol, piloté par un contrôleur qui assure la convergence.

\section{Commande non-linéaire basée sur la commande \textit{zero-moment}}\label{cmd_zero_moment}
La commande proposée dans \citetitle{2020e-MicCenZacFra} \cite{2020e-MicCenZacFra} est une commande hiérarchique non-linéaire, basée sur la possibilité de générer, indépendamment, un effort longitudinal au drone et des moments dans les trois directions.
Pour utiliser cette loi de commande dans le contrôle de DarkO, nous avons commencé par l'implémenter telle qu'elle est décrite dans l'article \cite{2020e-MicCenZacFra} pour commander un modèle de drone de type quadricoptère. Lors de cette implémentation, nous nous sommes rendus compte que les équations proposées dans le papier étaient incorrectes et ne permettaient pas de stabiliser la position du drone. Un travail important a été fourni pour calculer l'intégralité de la commande et ainsi obtenir les équations de commande. Se trouve dans l'annexe \ref{sec.appendix_mic} une synthèse des équations corrigées de l'article. Dès que les équations ont été obtenues, une simulation a été menée de manière à s'assurer de la qualité des formules. Celle-ci a été suivie d'une implémentation, avec comme système dynamique, un modèle de DarkO. 
\newpage
\begin{figure}[h]
    \centering
    \includegraphics[width=1\textwidth]{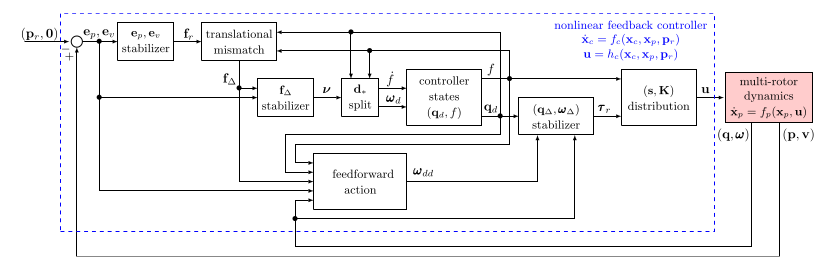}
    \caption{Diagramme représentant la stratégie de contrôle \cite[Fig.1]{2020e-MicCenZacFra}.}
    \label{fig:diagram_miche}
\end{figure}
\noindent
La première équation qui entre en jeu est la stabilisation en fonction de l'erreur de position et de l'erreur de vitesse : 
\begin{align} \label{eq:stab_ep_ev}
    f_{r} = m\boldsymbol{g} - k_{pp} \boldsymbol{e_{p}} - k_{pd}\boldsymbol{e_{v}}
\end{align}
avec $\boldsymbol{e_{p}} = \boldsymbol{p} - \boldsymbol{p_{r}} $ et $\boldsymbol{e_{v}} = \boldsymbol{v}$.
\newline \\
L'équation de distribution est définie par: 
\begin{align}
    \boldsymbol{u} = M_{b}^{\#}\boldsymbol{\tau_{r}} + \boldsymbol{\overline{u}}f
\end{align}
avec $\tau_{r}$ et $f$, respectivement, le couple  et la force nécessaires pour orienter et stabiliser le drone. $\overline{u}$ est défini dans l'équation \ref{eq:u_barre}.
\paragraph{}
L'hypothèse de génération de force sans moment n'est pas exactement vérifiée dans notre cas mais plusieurs articles utilisent cette même approximation \cite{hamel_minhduc}. Le point important de cette commande est l'inversion du modèle. Il est nécessaire d'inverser la matrice des coefficients des moments pour obtenir la commande à appliquer sur les actionneurs, de manière à avoir les moments souhaités. Dans notre cas, nous utilisons la pseudo-inverse de \ref{eq:moment_matrice}. 
\nomenclature{$M_{b}^{\#}$}{Pseudo inverse de $M_{b}$ \nomunit{}.}
\nomenclature{$\Gamma_{c}$}{Moment généré par la loi de commande   \nomunit{}.}
\begin{subequations}
\begin{align}
    \begin{bmatrix}T_{1,x} \\  T_{2,x} \\ \delta_{1}T_{1,x} \\ \delta_{2}T_{2,x} \end{bmatrix} &= M_{b}^{\#} \Gamma_{c}\\
    &=\begin{bmatrix}   0&0&-3.23\\
                        0&0&3.23\\
                        -4.51&-27.75&-1.49\\
                        4.51&-27.75&1.49\\
\end{bmatrix}\begin{bmatrix}\Gamma_{x} \\  \Gamma_{y} \\ \Gamma_{z}  \end{bmatrix}
\end{align}
\end{subequations}
Trois constats s'imposent. En effet, on observe que pour générer du moment, selon l'axe z, on utilise majoritairement le différentiel de poussée ; que le moment en y sera généré par une utilisation symétrique des ailerons avec une grande efficacité, et que le moment selon l'axe x provient d'une utilisation différentielle des ailerons.   
\paragraph{}
Le fonctionnement de cette loi de commande repose sur sa capacité à générer une force qui vienne s'opposer à la gravité, sans avoir de moment. L'étude attentive de \ref{eq:force_matrice} et de \ref{eq:moment_matrice} nous permet d'extraire l'espace du vecteur de commande, validant l'exigence précédemment exposée. On décide de normaliser ce vecteur, pour simplifier la commande. On observe qu'aucun moment n'est obtenu dès lors que l'on maintient les ailerons au neutre et que l'on utilise les vitesses de rotation des hélices de manière symétrique. Cette configuration génère ainsi une force dirigée selon l'axe x.
L'espace du vecteur de commande est donc de la forme :  
\begin{align} \label{eq:u_barre}
    \overline{}{u} = \kappa* \begin{bmatrix}1 \\  1 \\ 0 \\ 0 \end{bmatrix}
\end{align}
avec $\kappa = \lVert F_{b}*[1,1,0,0]^\top \rVert$ un gain permettant de normaliser le vecteur.

\subsection{Saturation de la commande}
La problématique principale de cette commande est la fonction de stabilisation de l'erreur de position et de vitesse présentée dans l'équation \ref{eq:stab_ep_ev}. Effectivement, cette fonction est linéaire donc lors de grandes consignes de déplacement, la loi de commande sature les actionneurs, ce qui empêche le trajet. Lors du stage, nous avons étudié deux manières de résoudre ce problème.\\ \indent La première est extraite de la théorie \textit{reférence governor}, qui s'appuie sur une saturation entre les couches de hauts niveaux et les couches de bas niveaux du contrôleur. Dans notre cas, nous saturons $f_r$ avec comme limite la capacité du drone à générer de la force, une fois la gravité compensée. On utilisera la forme du contrôleur stabilisant quasi temps-optimal proposée dans l'article \cite{journals/csysl/InvernizziLZ20} : 
\begin{align}
    \gamma(e_p, e_v) = -\sigma_{M_{i}}\left(k_p \left( e_p + e_v\mu\left(\frac{\lVert e_v \rVert}{2M_i}, \frac{k_v}{k_p}\right)\right)\right)
\end{align}
avec $\mu$ une fonction continue approximant le maximum de deux nombres positifs et $\sigma_{M_{i}}$ une fonction de saturation continue et différentiable.
\paragraph{}
Un choix courant est la fonction \textit{arctan(x)} mais il existe d'autres types de fonctions sigmoïdes. Nous avons fait le choix d'utiliser la fonction de Gompertz : 
\begin{align}
    \sigma_{M_{i}}(x) = 2M_{i} e^{-e^{b-cx}}-M_{i}
\end{align}
avec 
$b = -\ln(\frac{1}{2})$ et $c = 1$.
\newline \\
L'avantage de cette fonction est présenté dans le graphique suivant. En effet, elle s'approche plus de la fonction de saturation discontinue que la fonction \textit{arctan}.
\begin{figure}[h]
    \centering
    \includegraphics[width=0.6\textwidth]{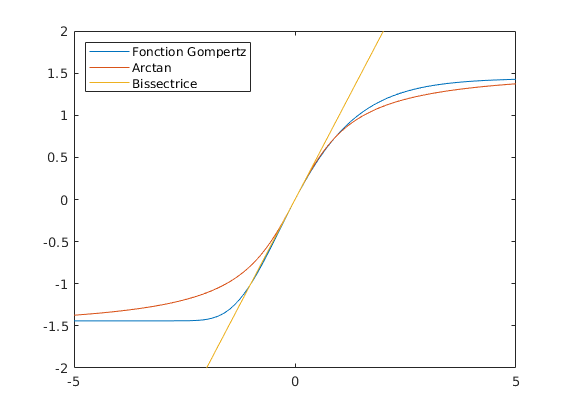}
    \caption{Comparaison de deux fonctions sigmoïdes}
    \label{fig:sigmoide}
\end{figure}

La seconde est basée sur la théorie \textit{error governor}, c'est-à-dire que nous allons saturer l'erreur pour limiter sa croissance et ainsi rester dans le domaine de stabilité de la commande proportionnelle. Dans notre cas, on vient limiter la croissance de $e_p \text{ et } e_v$ (erreurs de position et de vitesse décrites dans l'équation \ref{eq:stab_ep_ev}) à une valeur de 1m et 1 m/s. Ces valeurs ont été trouvées empiriquement avec les simulations, mais permettent à la commande de rester conservative. 
\subsubsection{Rotation selon l'axe de roulis}
Une fois le drone équilibré à la verticale, il maintient sa position. Cependant, il reste un degré de liberté : la rotation sur son axe de roulis, correspondant à une rotation autour de l'axe vertical du référentiel terrestre. Il est ainsi possible d'orienter le drone dans toutes les directions autour de cet axe. Cette capacité est très intéressante car elle permet de diriger le drone vers la cible, avant de passer en vol horizontal. La loi de commande hiérarchique permet ainsi d'ajouter, avec une priorité faible, la convergence de l'orientation du drone, vers un repère cible. La théorie est décrite dans la remarque 3, page 8 de \cite{2020e-MicCenZacFra}. Il est cependant nécessaire d'ajouter le terme de convergence d'orientation dans l'action du \textit{feedforward}. Ce terme est défini dans l'annexe \ref{sec.appendix_res_orient}.

\subsection{Simulation}\label{sim_hover_nl}
\begin{figure}[h]
\centering
\includegraphics[width=0.7\textwidth]{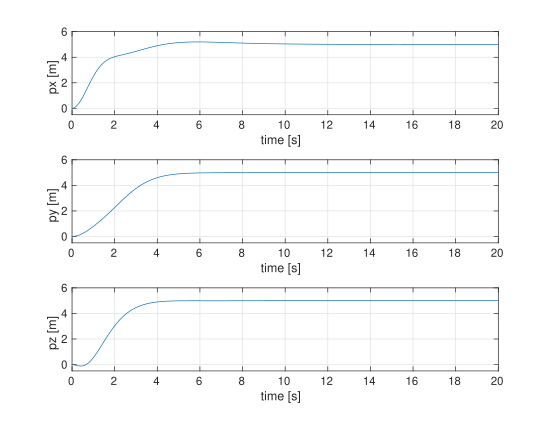}
\caption{Position du drone avec une commande hiérarchique.}
\label{fig:pos_hier}
\end{figure}

\begin{figure}[h]
\centering
\includegraphics[width=0.69\textwidth]{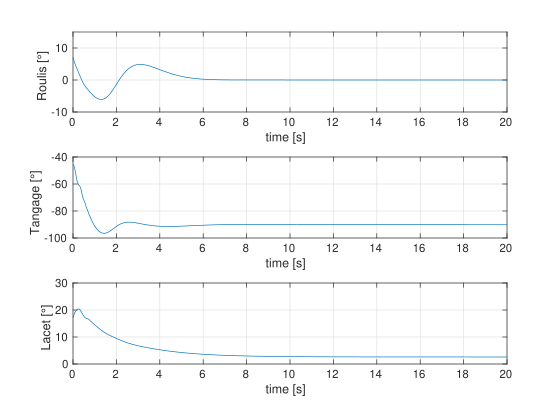}
\caption{Angle d’orientation du drone avec une commande hiérarchique.}
\label{fig:orient_hier}
\end{figure}

\begin{figure}[h]
\centering
\includegraphics[width=0.95\textwidth]{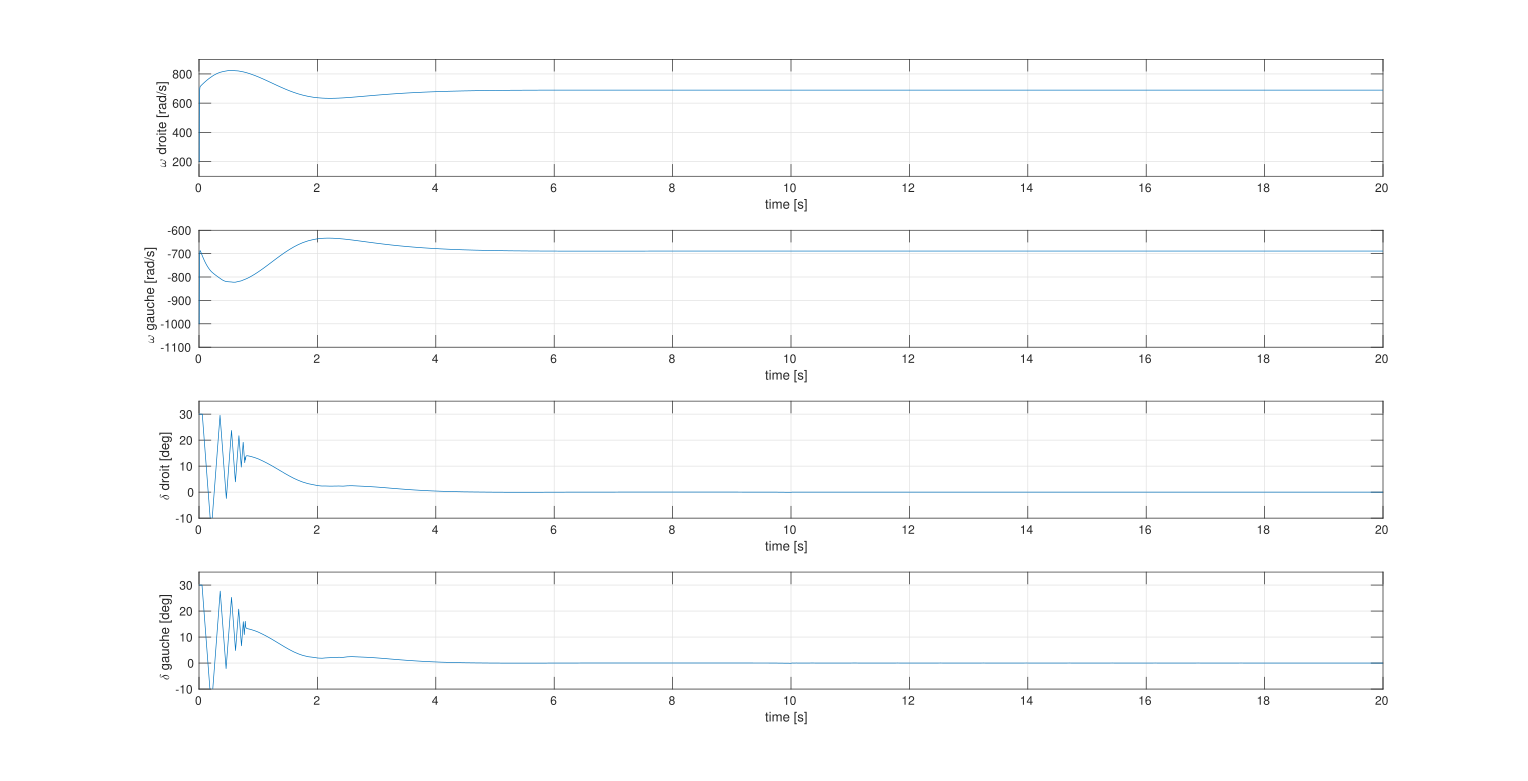}
\caption{Commande générée par le contrôleur hiérarchique.}
\label{fig:cmd_hier}
\end{figure}

La position cible est le point $p=[5,~5,~5]^\top$, atteint et maintenu, comme on peut l'observer sur la figure \ref{fig:pos_hier}.
\\
Nous avons initialisé le drone à une orientation relativement éloignée de son équilibre pour prouver la convergence de la commande ; les angles sont initialisés à roulis égal à 10°, tangage égal à -45° et lacet égal à 25°. Cette orientation correspond à un quaternion initial $q=[0.90,~0.19,~0.36,~0.15]^\top$. Toutefois, on observe sur la figure \ref{fig:orient_hier} que le drone s'aligne avec l'axe vertical, c'est-à-dire en vol pseudo-stationnaire.
\newpage
Le drone maintient une orientation très proche de la verticale, son point d'équilibre. Ainsi, il est maintenu en l'air par la traction des hélices, ce qui explique la vitesse de rotation élevée dans la figure \ref{fig:cmd_hier}. De plus, on peut aussi observer que nous touchons les saturations, au début, quand le drone cherche à converger vers sa position d'équilibre.
\\
Les résultats de cette commande, avec la saturation, sont présentés sur la figure \ref{fig:position_switch_hovering}. Effectivement, sur cette dernière, on observera que le contrôleur est en mesure de stabiliser un vol pseudo-stationnaire avec une cible relativement éloignée, sans diverger.

\chapter{Commande en vol}
Lors d'un vol horizontal, l'avion se maintient dans les airs grâce à la portance de l'aile, portance générée par l'écoulement de l'air sur cette dernière. Ainsi, c'est la vitesse de l'avion qui permet de compenser la gravité. La traction générée par l'hélice permet d'accélérer le drone de manière à atteindre la vitesse de vol, mais aussi de s'opposer à la traînée induite par l'aile. 
\begin{figure}[h]
    \centering
    \includegraphics[width=0.6\textwidth]{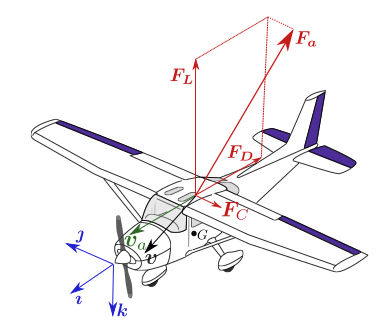}
    \caption{Représentation des forces sur un drone}
    \label{fig:vue_force}
\end{figure}
\section{Commande sur modèle linéarisé}
Nous allons décrire la méthode utilisée pour effectuer la linéarisation de la dynamique du drone à l'aide des fonctions Matlab. \\
La première action est d'obtenir le point de trim. Celui-ci est défini par l'ensemble des valeurs du vecteur d'état et du vecteur de commande qui annule l'ensemble des dérivées de la dynamique du drone. 
La linéarisation est effectuée pour une vitesse de $v = \SI{5}{\meter\per\second}$.
\begin{align}
    u_{eq} = \begin{bmatrix}
        2.18&
        2.18&
        0.92&
        0.92
    \end{bmatrix}
\end{align}
\begin{align}
    x_{eq} = \begin{bmatrix}
        0&
        0 &
        0 &
        5 &
        0 &
        0 & 
        0.85 & 
        0 & 
        -0.52 & 
        0 & 
        0 & 
        0. & 
        0 & 
    \end{bmatrix}
\end{align}
On observe dans le vecteur d'état que la vitesse, selon l'axe x, est fixée à la vitesse de trim et que le drone prend de l'incidence pour maintenir le vol horizontal. Effectivement, le quaternion est égal à $q_{eq} = \begin{bmatrix}0.85 & 0 & -0.52 & 0 \end{bmatrix}$, ce qui correspond à une rotation  autour de l'axe y de 62.67 \textdegree.
Pour la suite, on linéarise la dynamique de notre drone en prenant comme vecteur d'état :
\begin{align}
    x = \left[p_{z},~ \boldsymbol{v},~ \boldsymbol{\epsilon},~  \boldsymbol{\omega}\right]
\end{align}

\begin{align}
    A = \begin{bmatrix}
0&0&0&1&0&0&0&0&0&0\\
0&-2.56&0&1.58&0&53.84&0&0&0&0\\
0&0&0&0&-16.76&0&10.20&-0.05&0&-0.12\\
0&1.30&0&-0.83&0&-20.06&0&0&0&0\\
0&0&0&0&-0&0&0&0.43&0&-0.26\\
0&0&0&0&0&-0.27&0&0&0.43&0\\
0&0&0&0&0&0&0&0.26&0&0.43\\
0&0&0&0&0&0&0&-3.42&0&0.17\\
0&4.77&0&-5.87&0&-90.65&0&0&-7.77&0\\
0&0&0&0&0&0&0&-0.98&0&0.41\\
    \end{bmatrix}
\end{align}

\begin{align}
    B = \begin{bmatrix}
    0&0&0&0\\
    0.92&0.92&2.31&2.31\\
    0&0&0&0\\
    1.79&1.79&-1.21&-1.21\\
    0&0&0&0\\
    0&0&0&0\\
    0&0&0&0\\
    7.33&-7.33&-68.36&68.36\\
    0&0&-11.60&-11.60\\
    -25.58&25.58&-0.18&0.18\\
    \end{bmatrix}
\end{align}
Comme nous l'attendions, la linéarisation montre les mêmes caractéristiques que le modèle non-linéaire. Effectivement, on observe, dans la matrice de commande, que nous ne pouvons que commander l'accélération linéaire et angulaire du drone.

\section{Commande non-linéaire}
De la même manière que nous avons utilisé une commande hiérarchique pour contrôler le drone en hovering, il est possible de concevoir une commande permettant de maintenir un vol horizontal, à une vitesse fixée. 
\subsection{Commande de type "ligne de mire"}
Nous nous somme inspirés de la commande proposée dans \cite{DBLP:conf/isie/AndersenK17}. 
Définissons d'abord le vecteur cible, avec $\boldsymbol{p_{c}}$ le vecteur représentant les coordonnées du point cible : 
\begin{align}
    \boldsymbol{u_c} = \frac{\boldsymbol{p_{c}} - \boldsymbol{p}}{\lVert \boldsymbol{p_{c}} - \boldsymbol{p}\rVert} = [v_{px}~v_{py}~v_{pz} ]^{\top}
\end{align}
On définit le vecteur unitaire portant la vitesse du drone. 
\begin{align}
    \boldsymbol{u_{v}} = \frac{\boldsymbol{v}}{\lVert \boldsymbol{v}\rVert} 
\end{align}
L'objectif est de faire converger le vecteur vitesse vers le vecteur cible et ainsi diriger le drone vers le point $\boldsymbol{p_{c}}$.

On peut caractériser la différence d'orientation comme une rotation portée par l'axe défini par le produit vectoriel des deux vecteurs $\boldsymbol{u_{e}} = \boldsymbol{u_c} \wedge \boldsymbol{u_{v}}$ et de valeur égale au produit scalaire $\cos{\theta} = \langle \boldsymbol{u_c} | \boldsymbol{u_{v}} \rangle $.
On obtient ainsi le quaternion erreur suivant:
\begin{align}
    q_{e} = \begin{bmatrix}
    \sqrt{\frac{1+  \langle \boldsymbol{u_c} | \boldsymbol{u_{v}} \rangle}{2}}
   ,~ 
   \frac{ \boldsymbol{u_c} \wedge \boldsymbol{u_{v}}}{\lVert  \boldsymbol{u_c} \wedge \boldsymbol{u_{v}}\rVert}  \sqrt{\frac{1  \langle \boldsymbol{u_c} | \boldsymbol{u_{v}} \rangle}{2}}
    \end{bmatrix}^\top
\end{align}
Ce quaternion est valable tant que  $\lVert  \boldsymbol{u_c} \wedge \boldsymbol{u_{v}}\rVert \ne 0$. Si cette égalité est vérifiée, deux cas sont possible : soit les deux vecteurs sont alignés, ce qui correspond à $\langle \boldsymbol{u_c} | \boldsymbol{u_{v}} \rangle = 1$  et il n'est plus nécessaire d'effectuer une rotation donc $ q_{e} = [1,~0,~0,~0]^\top$, soit les deux vecteurs sont opposés, ce qui correspond à $\langle \boldsymbol{u_c} | \boldsymbol{u_{v}} \rangle = -1$ et il est nécessaire d'effectuer une rotation de 180°.
\newline \\
\noindent
On obtient le couple à appliquer sur le drone avec :
\begin{align}
    \boldsymbol{\Gamma_{c}} = -k_{c} \sign(\eta_{e}) \boldsymbol{\epsilon_{e}} + \skewsym{\boldsymbol{\omega_{b}}}J\boldsymbol{\omega_{b}} - k_{d} \boldsymbol{\omega_{b}}
\end{align}
avec $q_{e} = [\eta_{e},~\epsilon_{e}]^\top$.\newline \\Nous avons choisi d'utiliser un retour discontinu, basé sur le $\sign(\eta_{e})$ pour assurer la convergence, même dans le cas où $\eta_{e} \to 0$.
\subsection{Allocation des commandes}
La problématique principale de toutes les lois de commande sur modèle non-linéaire est l'allocation des commandes, pour déterminer la quantité à appliquer sur les actionneurs, et ce dans le but d'obtenir l'effet désiré (génération de forces ou de moments). La commande hiérarchique proposée s'appuie sur un suivi de vitesse agissant sur les entrées virtuelles. On suppose que la vitesse de convergence de l'orientation du drone est plus rapide que le suivi de vitesse. Ainsi, on propose un contrôleur de type PID pour le suivi de la vitesse cible.
 
\begin{align}
    (T1+T2) = k_{p}(v_{c} - v_{b,x}) + k_{i}\int_{}^{} v_{b,x} \,dt + k_{d}*\dot{v}_{b,x}
\end{align}
avec $k_{p}$, $k_{i}$ et $k_{d}$, respectivement, le gain proportionnel, intégral et dérivé, assurant le suivi de vitesse. De plus, nous avons $v_{c}$ la vitesse cible  et $v_{b,x}$ la vitesse selon l'axe $x$ du drone et donc exprimée dans le repère du corps $\mathcal{B}$.
\newline \\
Ainsi, comme nous avons fixé $T1+T2$, il nous reste les trois commandes $T1-T2$, $(\delta_{1}+\delta_{2})$ et $(\delta_{1}-\delta_{2})$ pour générer les moments nécessaires à l'orientation du drone. 
Ainsi, on définit $\bar{u} =  \begin{bmatrix} T1-T2  & \delta_{1}+\delta_{2} & \delta_{1}-\delta_{2} \end{bmatrix} ^\top $ comme vecteur de commande pour l'orientation.
Le modèle étant relativement complexe, nous allons utiliser un état interne au contrôleur, qui permettra la convergence des moments générés sur le drone vers les moments commandés. La dynamique interne est de la forme:
\begin{align}
    \dot{\bar{u}} = - k_{u} \frac{\partial M}{\partial \bar{u}}^{-1} 
    ( M(\bar{u}) - \Gamma_{c} )
\end{align}
avec $M$ une fonction non-linéaire exprimant le lien entre les commandes et les moments générés sur le drone, définie en \ref{eq:M} et $\frac{\partial M}{\partial \bar{u}}$ une matrice 3x3 représentant la jacobienne de M par rapport à $\bar{u}$.
\newline \\
On définit $C_{l} = ( 2\pi + C_{d0})$ :

\begin{subequations}
    \begin{align}
        \frac{\partial M}{\partial \bar{u}} = \begin{bmatrix}
         \frac{k_{m}}{k_{f}} + \frac{S_{wet}}{8S_{p}} a_{y}C_{l}\xi_{f} (\delta_{1}+\delta_{2}) & \frac{S_{wet}}{8S_{p}} a_{y}C_{l}\xi_{f} (T1-T2)& \partial M_{1,3} \\
         \frac{S_{wet}}{8S_{p}} \Delta_{r}C_{l}\xi_{m}(\delta_{1}+\delta_{2})& \partial M_{2,2} & \frac{S_{wet}}{8S_{p}} \Delta_{r}C_{l}\xi_{m}(T1+T2)\\
        \left((p_{y}+\frac{S_{wet}}{4S_{p}} a_{y} C_{d0}) \right) & 0 & \frac{1}{4} \rho S \eta a_{y}C_{d0}\xi_{f} v_{b,z}
        \end{bmatrix}
    \end{align}
    avec :
    \begin{align}
        \partial M_{1,3}= \frac{S_{wet}}{8S_{p}}a_{y}( 2\pi  + C_{d0})\xi_{f} (T1+T2) + \frac{1}{4} \rho S \eta a_{y}(2\pi +C_{d0})\xi_{f} v_{b,x}
    \end{align}
    \begin{align}
        \partial M_{2,2} = \frac{S_{wet}}{8S_{p}} \Delta_{r}C_{l}\xi_{m}(T1-T2) + \frac{1}{4} \rho S \eta  \Delta_{r}(2\pi +C_{d0})\xi_{m} v_{b,x}
    \end{align}
\end{subequations}
\newpage

Les simulations ont montré que notre système est mal conditionné. Cela veut dire qu'il est impossible de déterminer l'inverse de la matrice jacobienne. Cette condition est représentée par un \textit{condition number}, défini par $\kappa = \frac{\lVert \lambda_{max}\rVert}{\lVert \lambda_{min}\rVert}$. Il est admis, lors de l'utilisation d'une précision double flottant, qu'un \textit{condition number} supérieur à $10^{14}$ ne permet pas d'inverser efficacement une matrice \cite{pillow_2018}. Nous avons quand même obtenu des résultats lors d'un vol dans un plan vertical. Pour ce faire, nous avons utilisé une décomposition en valeurs singulières de la matrice, avant d'utiliser cette forme pour l'inverser. L'avantage de cela est la possibilité de contraindre les valeurs singulières pour obtenir un approximation de l'inverse.
\subsection{Simulation}\label{sim_vol_nl}
La dynamique présentée précédemment pour l'allocation des commandes a été implémentée pour un vol à la vitesse $V_{b,x} = \SI{20}{\meter\per\second}$, en direction du point $p=[1000,~0,~1000]^\top$, contenu dans un plan vertical, et ce pour s'affranchir du problème cité précédemment. On initialise le modèle à une vitesse, dans le repère inertiel $\mathcal{I}$, de $V = [10,~0,~0]^\top$ et à une orientation nulle, c'est-à-dire $q=[1,~0,~0,~0]^\top$.
\begin{figure}[h]
\centering
\includegraphics[width=0.7\textwidth]{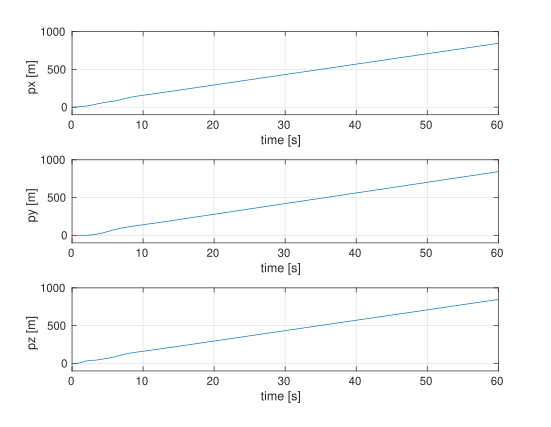}
\caption{Position du drone avec une commande non-linéaire de vol.}
\label{fig:pos_vol_nl}
\end{figure}\\
La figure \ref{fig:pos_vol_nl} indique une légère oscillation de la position. Ce problème devra être résolu pour envisager une implémentation.
\newpage
\begin{figure}[h]
\centering
\includegraphics[width=0.6\textwidth]{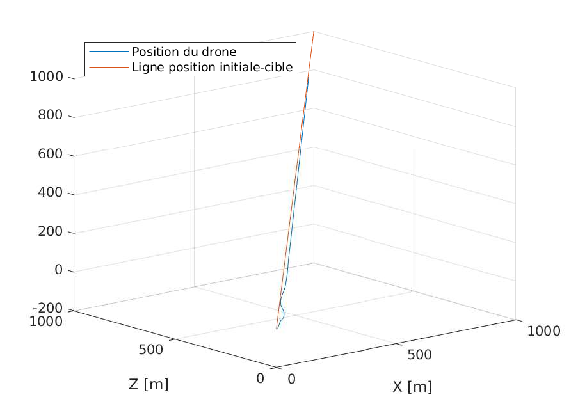}
\caption{Position, dans le plan, du drone avec une commande non-linéaire de vol.}
\label{fig:pos_plan_vol_nl}
\end{figure}
On observe, sur la figure \ref{fig:pos_plan_vol_nl}, que le drone ne suit pas une trajectoire rectiligne : cela est dû au type de commande "ligne de mire". Effectivement, on n'a pas d'exigence, comme lors d'un suivi de trajectoire. Ce dernier impliquerait une définition d'une erreur de \textit{cross track} et une commande plus gourmande énergiquement. Notre commande permet uniquement d'atteindre un point d'intérêt.  

\begin{figure}[h]
\centering
\includegraphics[width=0.69\textwidth]{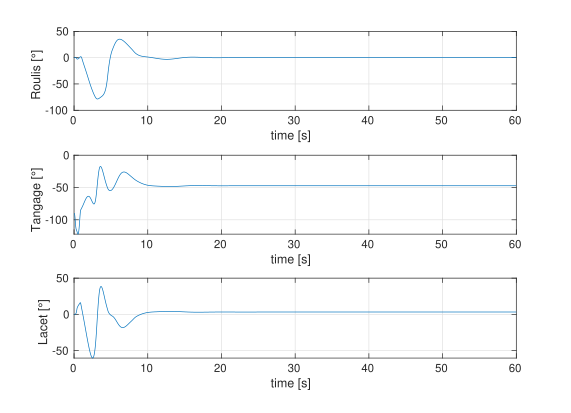}
\caption{Angle d’orientation du drone avec une commande non-linéaire de vol.}
\label{fig:orient_vol_nl}
\end{figure}
La figure \ref{fig:orient_vol_nl} montre que l'on stabilise l'assiette du drone vers -54° alors que notre trajectoire décrit un angle avec l'axe horizontal de 45°. La différence entre les deux est l'angle d'incidence. Effectivement, la loi de commande fait converger l'orientation de la vitesse du drone vers le point cible.
\newpage
\begin{figure}[h]
\centering
\includegraphics[width=0.7\textwidth]{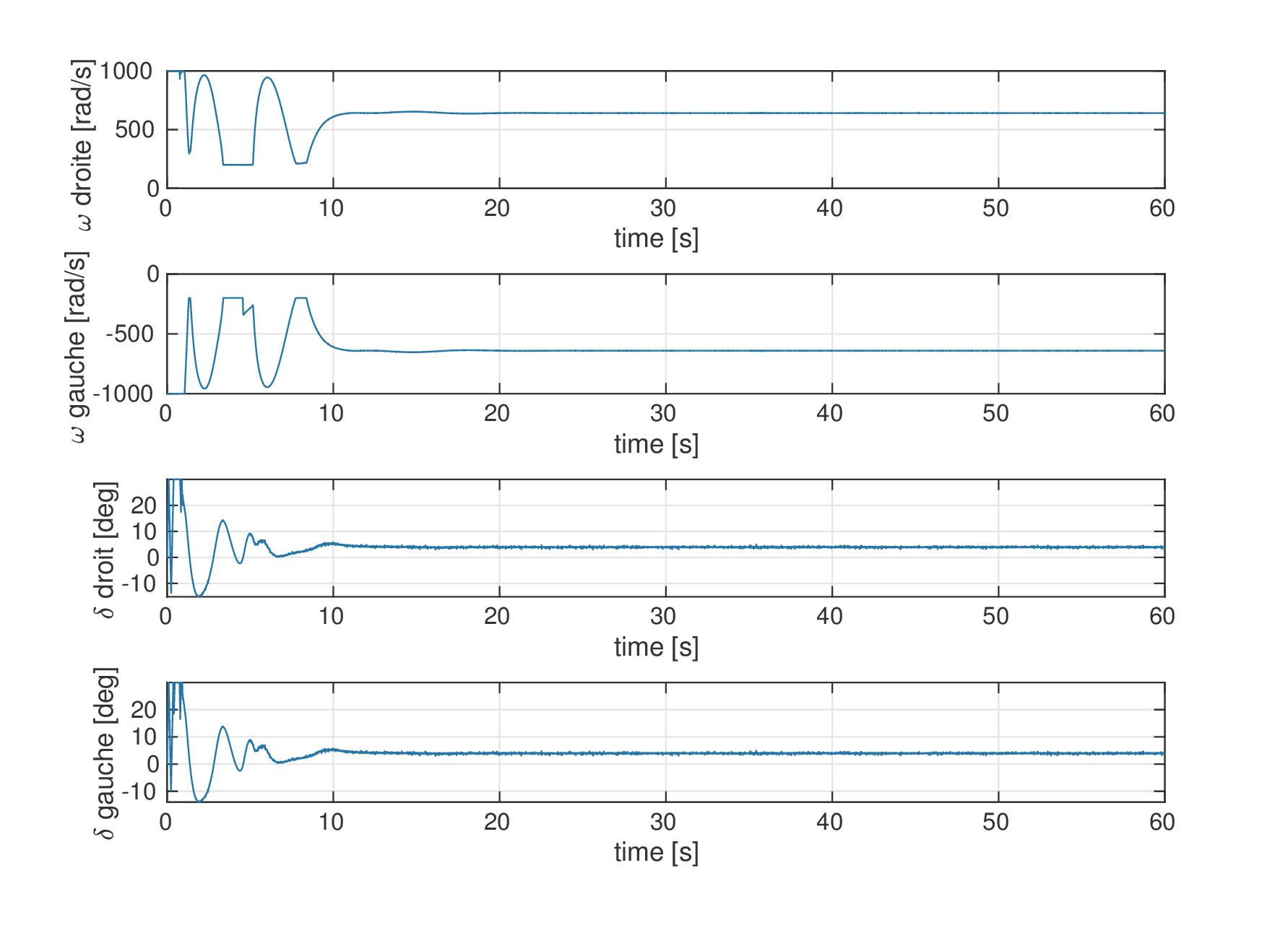}
\caption{Commande générée par la commande non-linéaire de vol.}
\label{fig:cmd_vol_nl}
\end{figure}
Les commandes nécessaires à la convergence du drone vers le point cible sont exposées dans la figure \ref{fig:cmd_vol_nl}. Les moteurs stabilisent leur vitesse à $\lvert\omega \rvert = \SI{661}{\radian\per\second}$. Cette vitesse est relativement proche de celle obtenue dans la commande présentée dans \ref{cmd_zero_moment}. Les deux simulations \ref{sim_hover_nl} et \ref{sim_vol_nl} présentent les mêmes caractéristiques, en ce qu'elles proposent d'amener le drone vers un point formant un angle de 45° avec l'horizontal. Cependant, la commande non-linéaire basée sur la commande \textit{zero-moment} propose une convergence en cinq secondes pour effectuer un déplacement de 8.6 mètres, ce qui nous donne une vitesse air de \SI{1.72}{\meter\per\second}. Quant à la commande ci-dessus, nous arrivons à stabiliser une vitesse air de \SI{20}{\meter\per\second}. Ainsi, on observe l'avantage énergétique du vol horizontal par rapport au vol en pseudo-stationnaire.
\chapter{Transitions}
L'obtention des lois de commande précédentes permet d'utiliser la théorie hybride pour effectuer un passage d'une loi de commande à l'autre.

\section{Vol en hovering, du non-linéaire au linéaire}

Contrairement aux commandes non-linéaires, les commandes linéaires ont des outils et des méthodes d'optimisation basés sur divers critères, rendant possible l'optimisation des vitesses de convergence, tels que les LMIs. La problématique d'une commande linéarisée réside dans la validité du domaine. En effet, comme cette commande est basée sur une linéarisation du modèle en un point d'équilibre, dès que le drone s'éloigne trop de ce point, le modèle est invalide et la commande n'est plus stabilisante. L'idée est donc de passer d'une commande à l'autre, sur la base de l'évaluation d'une fonction de Lyapunov, celle-ci décrivant le domaine de stabilité de la commande linéaire. Ainsi, le drone sera contrôlé avec la commande non-linéaire, basée \textit{zero-moment}, à l'extérieur du domaine de stabilité et avec la commande linéaire, à l'intérieur de celui-ci. \\

\begin{figure}[h]
    \centering
    \includegraphics[width=0.59\textwidth]{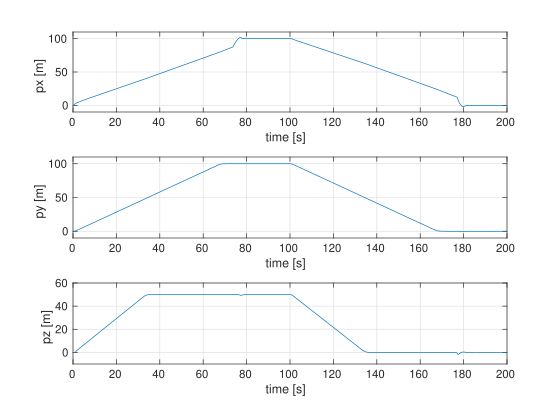}
    \caption{Évolution de la position du drone en hovering}
    \label{fig:position_switch_hovering}
\end{figure}

\begin{figure}[h]
    \centering
    \includegraphics[width=0.6\textwidth]{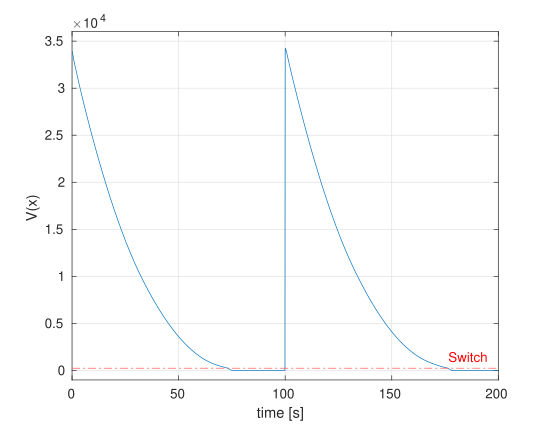}
    \caption{Évolution de la  valeur de la fonction de Lyapunov}
    \label{fig:Vx_switch_hovering}
\end{figure}

\begin{figure}[h]
    \centering
    \includegraphics[width=0.6\textwidth]{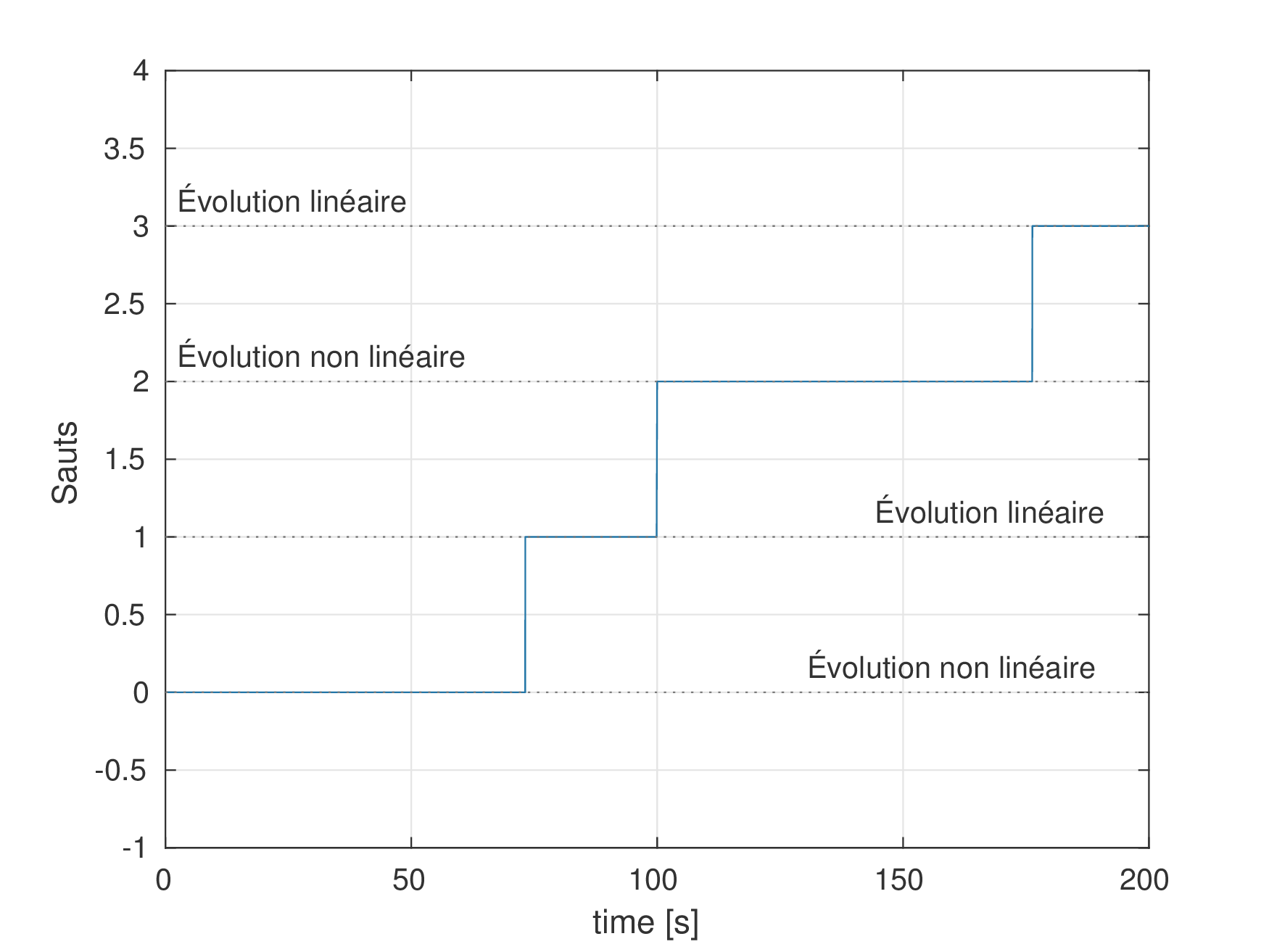}
    \caption{Évolution des sauts effectués}
    \label{fig:saut_switch_hovering}
\end{figure}

La simulation présentée ci-dessous, dans la figure \ref{fig:position_switch_hovering} montre l'évolution de la position du drone pour une cible définie par: 
\begin{align}
    p_{cible} = \left\{
    \begin{array}{ll}
    \left [ 100, 100, 50\right], & \forall t \in \left[0;100\right[ \\
     \left[0, 0, 0\right],& \forall t \in [100, \infty[ 
    \end{array}
\right.
\end{align}
\\
\noindent
On définit la fonction de Lyapunov par :
\begin{align}
    V(\delta x) = \delta x^\top S \delta x
\end{align}
avec S la solution de l'équation de Riccati donnée par la commande LQR de Matlab. Ainsi, S permet de s'assurer que le bouclage $A-BK$ est Hurwitz et que V(x) est bien une fonction de Lyapunov.\\ La figure \ref{fig:Vx_switch_hovering} présente son évolution ainsi que les conditions de passage.
\nomenclature{Équation de Riccati}{Équation différentielle utilisée en commande optimale, assurant les propriétés d'optimalité.}
\nomenclature{Hurwitz}{Une matrice carrée est appelée matrice de Hurwitz si toutes les valeurs propres de cette dernière ont une partie réelle strictement négative.}
\\
L'avantage de ce type de commande est la possibilité d'ajouter un mécanisme d'hystéresis, ce qui évite de multiplier les sauts. Celui-ci repose sur deux valeurs: dans notre cas, on passe de la commande non-linéaire à la commande linéaire si $V(x) < 250$ mais on n'effectuera le passage de la commande linéaire vers la commande non-linéaire que si $V(x) < 400$.\\
Il est possible, en simulation, de suivre les sauts effectués, présentés dans la figure \ref{fig:saut_switch_hovering}
\newpage
\section{D'un vol horizontal à un hovering}
\begin{figure}[h]
    \centering
    \includegraphics[width=0.6\textwidth]{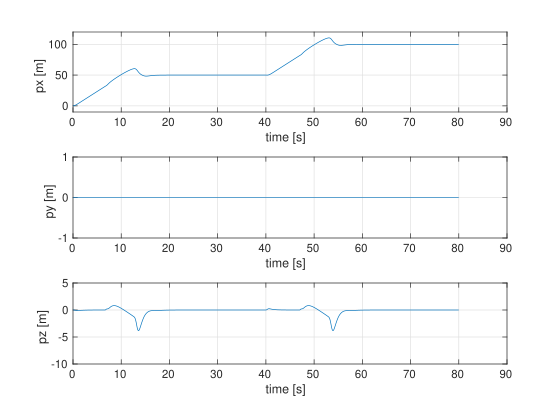}
    \caption{Évolution de la position du drone vol et hovering}
    \label{fig:position_switch_vol_hovering}
\end{figure}

\begin{figure}[h]
    \centering
    \includegraphics[width=0.6\textwidth]{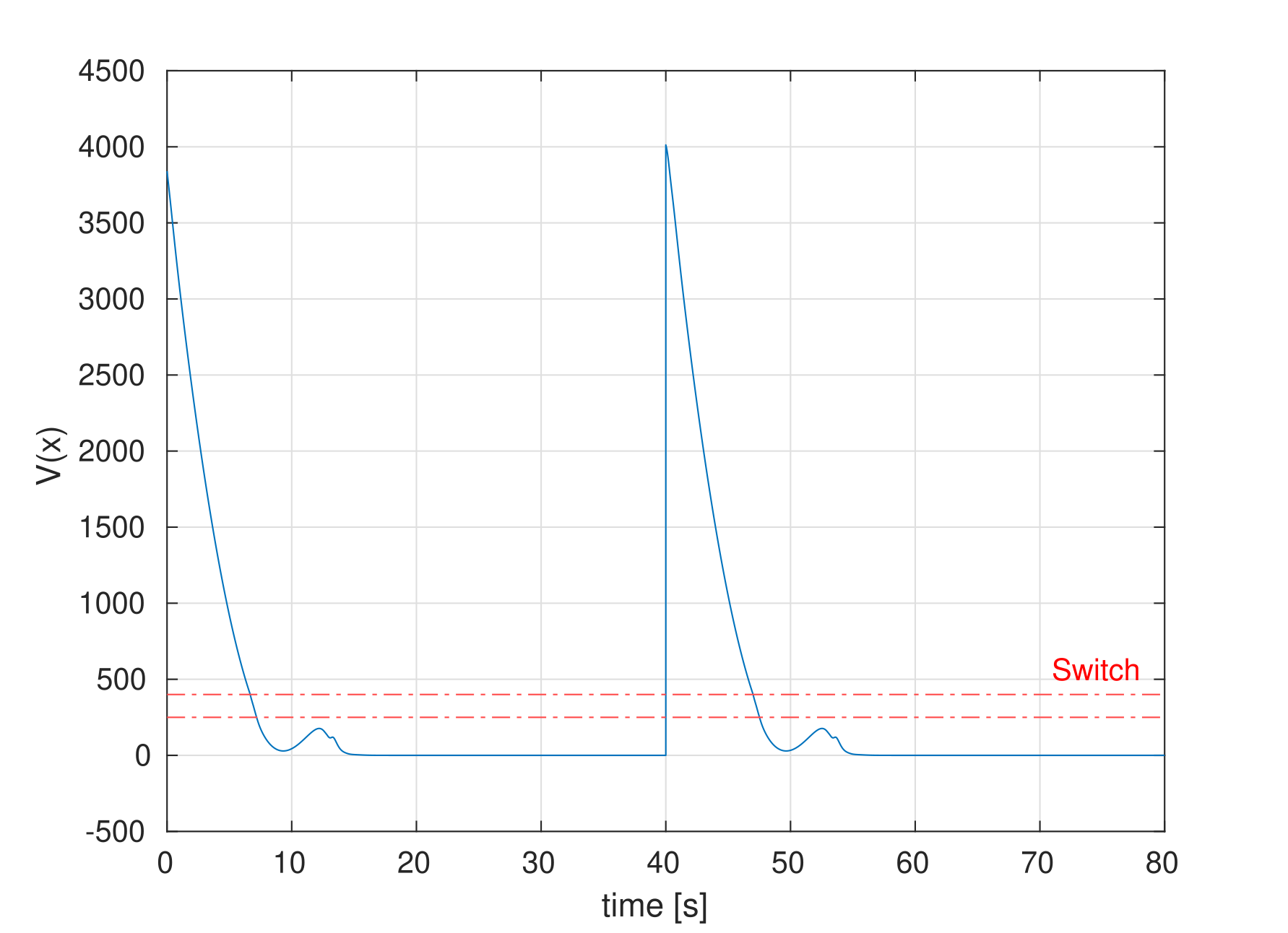}
    \caption{Évolution de la  valeur de la fonction de Lyapunov, lors d'un vol en hovering}
    \label{fig:Vx_switch_vol_hovering}
\end{figure}

\begin{figure}[h]
    \centering
    \includegraphics[width=0.6\textwidth]{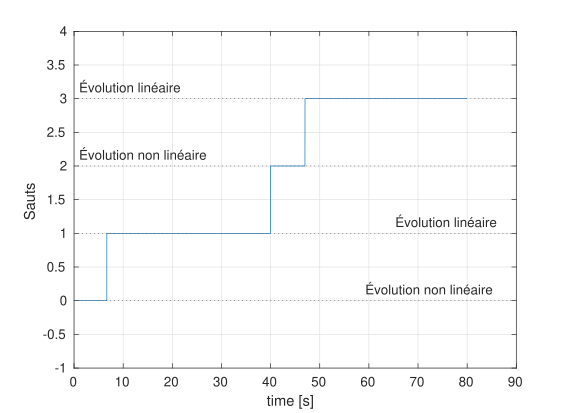}
    \caption{Évolution des sauts effectués lors d'un vol en hovering}
    \label{fig:saut_switch_vol_hovering}
\end{figure}

L'obtention d'une commande de vol horizontal, sur modèle linéarisé, nous permet d'effectuer une simulation entre un vol horizontal et une stabilisation en hovering.
La simulation présentée ci-dessous (figure \ref{fig:position_switch_vol_hovering}) montre l'évolution de la position du drone pour une cible définie par: 
\begin{align}
    p_{cible} = \left\{
    \begin{array}{ll}
    \left [ 50, 0, 0\right], & \forall t \in \left[0;40\right[ \\
     \left[100, 0, 0\right],& \forall t \in [40, \infty[ 
    \end{array}
\right.
\end{align}
\\
On observe un changement de consigne au bout de quarante secondes, après une stabilisation du drone en hovering. Ce changement correspond à une nouvelle commande opérateur, qui souhaite stabiliser le drone sur un autre point. Cette évolution de commande s'observe aisément sur la figure \ref{fig:Vx_switch_vol_hovering}, qui représente l'évolution de la fonction de Lyapunov et ainsi un saut lors de la nouvelle cible. 
\newpage
On suit l'évolution du contrôleur ainsi que la sélection des lois de commande avec le diagramme de sauts de la figure \ref{fig:saut_switch_vol_hovering}. 
\\
Comme nous l'avons vu, il est possible de faire tourner le drone lors d'un hovering ; ainsi il est en capacité, en vol horizontal, d'aller dans toutes les directions.  

\section{Implémentations}
L'ENAC dispose d'une volière où il est possible de réaliser des expérimentations. Dans celle-ci, pour contrôler les drones, nous utilisons la platerforme Paparazzi. Il s'agit d'un projet \textit{open-source} de développement matériel et logiciel d'un système de pilote automatique et de station sol pour la commande des drones. La création d'un autopilote est réalisée à l'aide de fichiers descriptifs en XML ; l'implémention de lois de commande est en C. Il a donc été nécessaire de coder la loi de commande à l'aide des outils proposés par Paparazi et notamment la bibliothèque \textit{math} et l'estimation de l'état du drone. L'intégralité du code source est disponible \cite{sansou_2021}.

Si nous avions eu assez de temps, nous aurions poursuivi l'intégration de la loi de commande. Pour celle-ci, nous aurions simulé le drone avec sa commande codée en C, grâce aux outils de Paparazzi. Cela permet de détecter d'éventuelles erreurs. Par la suite, nous aurions téléversé l'auto-pilote dans le drone cible. Dans notre cas, il aurait été opportun d'utiliser, en premier lieux, un quadricoptère, cible initiale de la commande \textit{zero-moment} \cite{2020e-MicCenZacFra}. Par la suite, nous aurions testé notre implémentation sur DarkO, dans une atmosphère contrôlée, comme celle de la volière. Enfin, il aurait été possible d'utiliser un module de génération de vent et de rafales pour perturber le vol du drone et tester la robustesse des lois de commande.

\newpage
\section*{Conclusion}\label{sec:conclusion}
\addcontentsline{toc}{chapter}{\nameref{sec:conclusion}}
\vspace{1cm}

L'objectif principal de ce mémoire était d'obtenir un contrôleur en mesure d'utiliser les deux modes de vol d'un drone convertible, pour relier deux points, en fonction des exigences de la mission. Au moment où cette conclusion est rédigée, si de nombreux axes de recherche ont été explorés, cette problématique n'a pas entièrement trouvé réponse.

\paragraph{}
En premier lieu, nous avons entrepris l'étude du modèle de DarkO. La manipulation des équations, avec l'exhibition des surfaces de contrôle virtuelles et l'obtention de la dynamique de l'appareil, m'a permis de mieux saisir les enjeux et les spécificités techniques du drone. Aussi, une parfaite compréhension du contexte opérationnel, permettant la soumission d'hypothèses simplificatrices, engendre la clarification du modèle, qui devient la base préalable à toutes lois de commande. Dès lors, il convient de souligner qu'une attention particulière a été apportée à cette tâche, notre asservissement découlant de cette dernière.

À partir de cette première étape, nous avons proposé quatre lois de commande, deux par mode de vol, basées sur une linéarisation du modèle et sur une commande non-linéaire.
Pour le vol en hovering, nous avons conçu une linéarisation autour d'un point d'équilibre, pour obtenir une loi de commande linéaire par retour d'état. L'avantage de cette dernière réside dans sa possibilité d'optimisation du bouclage et ainsi du réglage de sa vitesse de convergence.  Toutefois, elle possède un domaine de stabilité réduit, en ce qu'il s'agit d'une loi de commande locale, valable au voisinage du point d'équilibre. Au contraire, la seconde loi proposée, non-linéaire, est globale. Celle-ci, stable sur tout le domaine de vol, ne permet pas, toutefois, d'optimiser la vitesse de convergence, étant donné sa complexité.
La même logique a été utilisée pour le vol horizontal.

Au vu des avantages et des inconvénients observés sur les lois de commande, il nous a semblé opportun d'envisager un passage hybride sur la base de la stabilité des bouclages. Autrement dit, on utilise le contrôleur non-linéaire là où le contrôleur linéaire est instable, de manière à faire converger l'aéronef vers la cible. Dès que le drone entre dans le domaine de stabilité du contrôleur linéaire, le mécanisme hybride se charge du changement de loi, utilisant la vitesse de convergence maximale permise par l'optimisation linéaire. L'avantage de ce mécanisme réside dans l'obtention d'un contrôleur ayant l'autorité sur tout le domaine de vol, assurant la convergence, tout en optimisant les capacités du drone. Bien que les premiers résultats ne soient obtenus que par simulation, ils semblent prometteurs pour une intégration en volière.

\paragraph{}
Toutefois, il convient de noter que, pour répondre pleinement à notre question initiale, d'autres points restent à aborder. En effet, évaluer la pertinence de l'utilisation d'une commande vis-à-vis de l'autre ne peut se faire sans une étude énergétique des manoeuvres. Celle-ci permettrait d'obtenir les conditions de passage d'un contrôleur à l'autre, sur la base de la consommation.

\vspace{1cm}
\section*{Perspectives}\label{sec:perspectives}
\addcontentsline{toc}{chapter}{\nameref{sec:perspectives}}
\vspace{1cm}
Comme cela vient d'être évoqué, les limites de ce travail semblent claires, dans la mesure où il est nécessaire d'obtenir des résultats d'essais en vol, suite aux simulations effectuées. Nonobstant, plusieurs points collatéraux ont été soulevés lors de ce stage et mériteraient une étude plus approfondie.

Premièrement, il serait intéressant de mener une analyse plus poussée des saturations, notamment sur l'utilisation d'un modèle saturé ou du premier ordre, pour la limite en accélération angulaire des moteurs. Celle-ci permettrait de mettre en lumière et peut-être de simuler l'apparition du cycle limite, observé lors des campagnes de test en volière antérieures à ce stage. Ce phénomène a été évoqué lors d'une réunion de stage mais son étude n'a pas été menée, par manque de temps.\\

Aussi, il semble pertinent d'utiliser des LMIs pour optimiser le contrôleur linéaire proposé dans \ref{control_lin}. Effectivement, si celui-ci s'appuie sur la minimisation d'une erreur quadratique, sa convergence pourrait être optimisée grâce aux LMIs.\\

Enfin, a été décrit dans le paragraphe \ref{stab_lin} la manière que nous avons utilisée pour obtenir le domaine de stabilité de la linéarisation. Cependant, il serait beaucoup plus rigoureux d'effectuer une analyse mathématique de la stabilité de la linéarisation, basée sur la majoration des termes quadratiques.

\vspace{4cm}
\paragraph{}
Pour conclure cet écrit et ce d'un point de vue plus personnel, je souhaiterais souligner que ce mémoire est l'aboutissement de mon projet de fin d'études mais aussi la conclusion de mon diplôme d'ingénieur. Cependant, il n'apparaît pas comme une fin en soit car il s'inscrit dans une démarche de recherche qui dépasse ce stage. Effectivement, les résultats obtenus sont les prémisses d'un doctorat, conjointement mené avec l'ENAC et l'ONERA, qui doit débuter en octobre. Le double objectif de cette expérience, c'est-à-dire celui de s'initier aux modèles de drone et d'acquérir une méthodologie de recherche, a été entièrement rempli. Ce stage a été, pour moi, une grande opportunité, en ce qu'il m'a initié à de nouvelles méthodes de commande, m'a permis de débuter une recherche bibliographique et a engagé un questionnement sur les problématiques de commande des convertibles. 
\newpage
\addcontentsline{toc}{chapter}{Glossaire}
\printnomenclature[4cm] 
\clearpage\null\newpage
\addcontentsline{toc}{chapter}{Bibliographie}
\printbibliography[title=Bibliographie]

\newpage
\addcontentsline{toc}{chapter}{Annexes}
\vspace*{7cm}
\begin{center}
  \Huge\bfseries 
  Annexes
\end{center}

\clearpage\null\newpage
\appendix 
\renewcommand{\thesection}{\Alph{section}}
\renewcommand{\thefigure}{\Alph{section}.\arabic{figure}}
\renewcommand{\theequation}{\Alph{section}.\arabic{equation}}
\section{Paramètres DarkO}
\begin{table}[h]
   \centering
    \begin{tabular}{|l|c|c|}
      \hline
      Paramètres & Valeurs & Unités  \\
      \hline
       Masse ($m$)  & 0.492 & \SI{}{\kilogram} \\
      \hline
      Corde moyenne ($c$) & 0.13 & \SI{}{\meter}\\
      \hline
      Envergure ($b$) & 0.55 & \SI{}{\meter}\\
      \hline
      Surface alaire ($S$) & 0.0743 & \SI{}{\square\meter}\\
      \hline
      Surface hélice ($S_{f}$) & 0.3989 & \SI{}{\square\meter}\\
      \hline
      $J_{xx}$ & 0.0070 & \SI{}{\kilogram\square\meter}\\
      \hline
      $J_{yy}$ & 0.0028 & \SI{}{\kilogram\square\meter}\\
      \hline
      $J_{zz}$ & 0.0061 & \SI{}{\kilogram\square\meter}\\
      \hline
      $J_{p}$ & 5.1116e-6 & \SI{}{\kilogram\square\meter}\\
      \hline
      $k_{f}$ & 5.13e-6 & \SI{}{\kilogram\meter}\\
      \hline
      $k_{m}$ & 2.64e-7 & \SI{}{\kilogram\square\meter}\\
      \hline
      $C_{d0}$ & 0.025 & Pas d'unité\\
      \hline
      $C_{y0}$ & 0 & Pas d'unité\\
      \hline
      $C_{lp}$ & 0.2792 & Pas d'unité\\
      \hline
      $C_{lq}$ & 0.0 & Pas d'unité\\
      \hline
      $C_{lr}$ & 0.1145 & Pas d'unité\\
      \hline
      $C_{mp}$ & 0.0 & Pas d'unité\\
      \hline
      $C_{mq}$ & 1.2715 & Pas d'unité\\
      \hline
      $C_{mr}$ & 0.0 & Pas d'unité\\
      \hline
      $C_{np}$ & 0.081 & Pas d'unité\\
      \hline
      $C_{nq}$ & 0.0 & Pas d'unité\\
      \hline
      $C_{nr}$ & 0.0039 & Pas d'unité\\
      \hline
      $p_{x}$ & 0.065 & \SI{}{\meter}\\
      \hline
      $p_{y}$ & 0.155 & \SI{}{\meter}\\
      \hline
      $p_{z}$ & 0.0 & \SI{}{\meter}\\
      \hline
      $a_{x}$ & 0.0 & \SI{}{\meter}\\
      \hline
      $a_{y}$ & 0.155 & \SI{}{\meter}\\
      \hline
      $a_{z}$ & 0.0 & \SI{}{\meter}\\
      \hline
      $\xi_{f}$ & 0.85 & Pas d'unité\\
      \hline
      $\xi_{m}$ & 0.55 & Pas d'unité\\
      \hline
    \end{tabular}
    \label{table:Param}
    \caption{Paramètres DarkO}
\end{table}

\newpage
\section{Obtention des matrices du modèle non-linéaire}
\label{matrice_droit}
\setcounter{figure}{0} 
\setcounter{equation}{0} 
\noindent
Définissons et simplifions les équations : 
\begin{align}
    \skewsym{a_{1}} &= \begin{bmatrix} 0 & -a_{z} & a_{y} \\ a_{z} & 0 & -a_{x} \\ -a_{y} & a_{x} & 0 \end{bmatrix}=\begin{bmatrix} 0 & 0 & -a_{y} \\ 0 & 0 & 0 \\ a_{y} & 0 & 0 \end{bmatrix}, \text{ car } a_{x} \text{ et } a_{z} \text{ sont nuls}.
\end{align}

\begin{align}
    \skewsym{a_{2}} &= \begin{bmatrix} 0 & -a_{z} & a_{y} \\ a_{z} & 0 & -a_{x} \\ -a_{y} & a_{x} & 0 \end{bmatrix}=\begin{bmatrix} 0 & 0 & a_{y} \\ 0 & 0 & 0 \\ -a_{y} & 0 & 0 \end{bmatrix}, \text{ car } a_{x} \text{ et } a_{z} \text{ sont nuls}.
\end{align}
On observe que $\skewsym{a_{2}} = -\skewsym{a_{1}}$.
\begin{align}
    \skewsym{p_{1}} &= \begin{bmatrix} 0 & -p_{z} & p_{y} \\ p_{z} & 0 & -p_{x} \\ -p_{y} & p_{x} & 0 \end{bmatrix}=\begin{bmatrix} 0 & 0 & -p_{y} \\ 0 & 0 & -p_{x} \\ p_{y} & p_{x} & 0 \end{bmatrix}, \text{ car } p_{z} \text{ est nul}.
\end{align}

\begin{align}
    \skewsym{p_{2}} &= \begin{bmatrix} 0 & -p_{z} & p_{y} \\ p_{z} & 0 & -p_{x} \\ -p_{y} & p_{x} & 0 \end{bmatrix}=\begin{bmatrix} 0 & 0 & p_{y} \\ 0 & 0 & -p_{x} \\ -p_{y} & p_{x} & 0 \end{bmatrix}, \text{ car } p_{z} \text{ est nul}.
\end{align}
\begin{align}
    \Phi^{(fv)} = \begin{bmatrix} C_{d0} & 0 & 0 \\ 0 & C_{y0} & 0 \\ 0 & 0 & 2\pi + C_{d0} \end{bmatrix} = \begin{bmatrix} C_{d0} & 0 & 0 \\ 0 & 0 & 0 \\ 0 & 0 & 2\pi + C_{d0} \end{bmatrix}, \text{ car } C_{y0} \text{ est nul}.
\end{align}

\begin{align}
    \Phi^{(mv)} = \begin{bmatrix} 0 & 0 & 0 \\ 0 & 0 & -c^{-1} \Delta_{r}(2\pi +C_{d0})  \\ 0 & c^{-1} \Delta_{r}C_{y0} & 0 \end{bmatrix} = \begin{bmatrix} 0 & 0 & 0 \\ 0 & 0 & -c^{-1} \Delta_{r}(2\pi +C_{d0})  \\ 0 & 0 & 0 \end{bmatrix} \text{ car } C_{y0} \text{ est nul}.
\end{align}
Expression de la force due à la traction des hélices et à la traînée :
\begin{subequations}
    \begin{align}
        \Big(I - \frac{S}{4S_{p}} \Phi^{(fv)} \Big) (T_{1} + T_{2}) &= \Bigg(\begin{bmatrix} 1 & 0 & 0 \\ 0 & 1 & 0 \\ 0 & 0 & 1 \end{bmatrix} - \frac{S}{4S_{p}} \begin{bmatrix} C_{d0} & 0 & 0 \\ 0 & 0 & 0 \\ 0 & 0 & 2\pi + C_{d0} \end{bmatrix}\Bigg) \begin{bmatrix} T_{1,x} + T_{2,x}  \\ 0  \\ 0 \end{bmatrix} \\
        & = \begin{bmatrix} (1-\frac{S}{4S_{p}} C_{d0})(T_{1,x} + T_{2,x})  \\ 0  \\ 0 \end{bmatrix}
    \end{align}
\end{subequations}
On observe que le premier terme de la force n'agit que sur l'axe $x_{b}$.\\
Pour la suite, calculons dans un premier temps:
\begin{subequations}
    \begin{align}
        \Delta_{f,i} T_{i} &=  \begin{bmatrix} 0 & 0 &  \xi_{f}\delta_{i}\\ 0 & 0 & 0 \\ -\xi_{f}\delta_{i} & 0 & 0 \end{bmatrix}\begin{bmatrix} T_{i,x}  \\ 0  \\ 0 \end{bmatrix}\\
        & = \begin{bmatrix} 0  \\ 0  \\ -\xi_{f}\delta_{i}T_{i,x} \end{bmatrix}
    \end{align}
\end{subequations}
On observe que ce produit est une projection de l'effort de traction vers les deux autres axes. 
Expression de la force de portance :
\begin{subequations}
    \begin{align}
        \frac{S}{4S_{p}} \Phi^{(fv)} (\Delta_{f,1} T_{1} + \Delta_{f,2} T_{2}) &= \frac{S}{4S_{p}} \begin{bmatrix} C_{d0} & 0 & 0 \\ 0 & 0 & 0 \\ 0 & 0 & 2\pi + C_{d0} \end{bmatrix} \begin{bmatrix} 0  \\ 0  \\ -\xi_{f}(\delta_{1}T_{1,x} + \delta_{2}T_{2,x})  \end{bmatrix}\\
        &= \begin{bmatrix} 0  \\ 0   \\ -\frac{S}{4S_{p}}(2\pi + C_{d0})\xi_{f}(\delta_{1}T_{1,x} + \delta_{2}T_{2,x})  \end{bmatrix}
    \end{align}
\end{subequations}
Comme prévu, on observe un effort sur les axes $y_{b}$ et $z{b}$.
\noindent
Pour le moment:
\begin{subequations}
    \begin{align}
        - \frac{S}{4S_{p}} B \Phi^{(mv)} (T_{1} + T_{2} ) &= -\frac{S}{4S_{p}} \begin{bmatrix} b & 0 & 0 \\ 0 & c & 0 \\ 0 & 0 & b \end{bmatrix}
        \begin{bmatrix} 0 & 0 & 0 \\ 0 & 0 & -c^{-1} \Delta_{r}(2\pi +C_{d0})  \\ 0 & 0 & 0 \end{bmatrix}\begin{bmatrix} T_{1,x} + T_{2,x}  \\ 0  \\ 0 \end{bmatrix} \\
        & = \begin{bmatrix} 0  \\ 0  \\ 0 \end{bmatrix} 
    \end{align}
\end{subequations}
\begin{subequations}
    \begin{align}
        \frac{S}{4S_{p}} B \Phi^{(mv)} (\Delta_{m,1}T_{1} + \Delta_{m,2}T_{2}) &= \frac{S}{4S_{p}} \begin{bmatrix} b & 0 & 0 \\ 0 & c & 0 \\ 0 & 0 & b \end{bmatrix}
        \begin{bmatrix} 0 & 0 & 0 \\ 0 & 0 & -c^{-1} \Delta_{r}(2\pi +C_{d0})  \\ 0 & 0 & 0 \end{bmatrix}\begin{bmatrix} 0  \\ 0  \\ -\xi_{m}(\delta_{1}T_{1,x} + \delta_{2}T_{2,x})  \end{bmatrix}\\
        &=\begin{bmatrix} 0  \\ \frac{S}{4S_{p}} \Delta_{r}(2\pi +C_{d0}) \xi_{m}(\delta_{1}T_{1,x} + \delta_{2}T_{2,x})\\ 0\end{bmatrix}
    \end{align}
\end{subequations}
Expression du moment dû au différentiel de vitesse sur les ailes : 
\begin{subequations}
    \begin{align}
        \frac{S}{4S_{p}} \skewsym{a_{1}} \Phi^{(fv)} T_{1} &= -\frac{S}{4S_{p}}  \begin{bmatrix} 0 & 0 & -a_{y} \\ 0 & 0 & 0 \\ a_{y} & 0 & 0 \end{bmatrix}\begin{bmatrix} C_{d0} & 0 & 0 \\ 0 & 0 & 0 \\ 0 & 0 & 2\pi + C_{d0} \end{bmatrix} \begin{bmatrix} T_{1,x}  \\ 0  \\ 0 \end{bmatrix}\\
        &=\begin{bmatrix} 0  \\  0 \\ \frac{S}{4S_{p}} a_{y} C_{d0}T_{1,x}   \end{bmatrix} 
    \end{align}
\end{subequations}
\begin{subequations}
    \begin{align}
        \frac{S}{4S_{p}} \skewsym{a_{2}} \Phi^{(fv)} T_{2} &= -\frac{S}{4S_{p}}  \begin{bmatrix} 0 & 0 & a_{y} \\ 0 & 0 & 0 \\ -a_{y} & 0 & 0 \end{bmatrix}\begin{bmatrix} C_{d0} & 0 & 0 \\ 0 & 0 & 0 \\ 0 & 0 & 2\pi + C_{d0} \end{bmatrix} \begin{bmatrix} T_{2,x}  \\ 0  \\ 0 \end{bmatrix}\\
        &=\begin{bmatrix} 0  \\  0 \\ -\frac{S}{4S_{p}} a_{y} C_{d0}T_{2,x}   \end{bmatrix} 
    \end{align}
\end{subequations}
On obtient ainsi :
\begin{align}
        \frac{S}{4S_{p}} \skewsym{a_{1}} \Phi^{(fv)} T_{1} + \frac{S}{4S_{p}} \skewsym{a_{2}} \Phi^{(fv)} T_{2} = \begin{bmatrix} 0  \\  0 \\ \frac{S}{4S_{p}} a_{y} C_{d0}(T_{1,x}-T_{2,x} )  \end{bmatrix} 
\end{align}
Expression du moment dû au différentiel de vitesse sur les élevons :
\begin{subequations}
    \begin{align}
        \frac{S}{4S_{p}} \skewsym{a_{1}} \Phi^{(fv)} \Delta_{f,1} T_{1} &= \frac{S}{4S_{p}}  \begin{bmatrix} 0 & 0 & -a_{y} \\ 0 & 0 & 0 \\ a_{y} & 0 & 0 \end{bmatrix} \begin{bmatrix} C_{d0} & 0 & 0 \\ 0 & 0 & 0 \\ 0 & 0 & 2\pi + C_{d0} \end{bmatrix} \begin{bmatrix} 0  \\ 0   \\ -\xi_{f}\delta_{1}T_{1,x}   \end{bmatrix}\\
        & = \frac{S}{4S_{p}}\begin{bmatrix}  a_{y}C_{l} \xi_{f}\delta_{1}T_{1,x}  \\ 0 \\ 0\end{bmatrix} 
    \end{align}
\end{subequations}
\begin{subequations}
    \begin{align}
        \frac{S}{4S_{p}} \skewsym{a_{2}} \Phi^{(fv)} \Delta_{f,2} T_{2} &= \frac{S}{4S_{p}}  \begin{bmatrix} 0 & 0 & a_{y} \\ 0 & 0 & 0 \\ -a_{y} & 0 & 0 \end{bmatrix} \begin{bmatrix} C_{d0} & 0 & 0 \\ 0 & 0 & 0 \\ 0 & 0 & 2\pi + C_{d0} \end{bmatrix} \begin{bmatrix} 0  \\ 0   \\ -\xi_{f}\delta_{2}T_{2,x}   \end{bmatrix}\\
        & = \frac{S}{4S_{p}}\begin{bmatrix}  -a_{y}( 2\pi + C_{d0}) \xi_{f}\delta_{2}T_{2,x}  \\ 0 \\ 0\end{bmatrix} 
    \end{align}
\end{subequations}
On obtient ainsi :
\begin{align}
       \frac{S}{4S_{p}} \skewsym{a_{1}} \Phi^{(fv)} \Delta_{f,1} T_{1} + \frac{S}{4S_{p}} \skewsym{a_{2}} \Phi^{(fv)} \Delta_{f,2} T_{2} = \begin{bmatrix} \frac{S}{4S_{p}} a_{y}( 2\pi + C_{d0}) \xi_{f}(\delta_{1}T_{1,x}-\delta_{2}T_{2,x})  \\ 0 \\ 0\end{bmatrix}  
\end{align}
Expression du moment dû au différentiel de vitesse des hélices :
\begin{align}
    \skewsym{p_{1}} T_{1} &= \begin{bmatrix} 0 & 0 & -p_{y} \\ 0 & 0 & -p_{x} \\ p_{y} & p_{x} & 0 \end{bmatrix} \begin{bmatrix} T_{1,x}  \\ 0  \\ 0 \end{bmatrix} = \begin{bmatrix} 0 \\  0 \\ p_{y}T_{1,x}  \end{bmatrix}
\end{align}
\begin{align}
    \skewsym{p_{2}} T_{2} &= \begin{bmatrix} 0 & 0 & p_{y} \\ 0 & 0 & -p_{x} \\ -p_{y} & p_{x} & 0 \end{bmatrix} \begin{bmatrix} T_{2,x}  \\ 0  \\ 0 \end{bmatrix} = \begin{bmatrix} 0 \\  0 \\ -p_{y}T_{2,x}  \end{bmatrix}
\end{align}
On obtient ainsi :
\begin{align}
       \skewsym{p_{1}} T_{1} + \skewsym{p_{2}} T_{2}= \begin{bmatrix} 0 \\  0 \\ p_{y}(T_{1,x} - T_{2,x})  \end{bmatrix}
\end{align}
Expression du moment dû à la réaction du moteur : 

\begin{align}
    \frac{k_{m} }{k_{f}}(T_{1}- T_{2} ) = \begin{bmatrix} \frac{k_{m} }{k_{f}}(T_{1,x} - T_{2,x})  \\ 0  \\ 0 \end{bmatrix}
\end{align}
Ainsi en combinant les calculs précédents, on obtient :
\begin{flalign}
    \sum F_{b} &= \begin{bmatrix} 1-\frac{S}{4S_{p}} C_{d0} \\ 0  \\ 0 \end{bmatrix}_{b}(T_{1,x} + T_{2,x}) + \begin{bmatrix} 0  \\ 0   \\ -\frac{S}{4S_{p}}(2\pi + C_{d0})\xi_{f}  \end{bmatrix}_{b} (\delta_{1}T_{1,x} + \delta_{2}T_{2,x})
\end{flalign}
\begin{subequations}    
    \begin{flalign}
       \sum M_{b} &=  \begin{bmatrix} \frac{k_{m} }{k_{f}} \\  0\\ p_{y}+\frac{S}{4S_{p}} a_{y} C_{d0} \end{bmatrix} (T_{1,x} - T_{2,x}) +\begin{bmatrix} \frac{S}{4S_{p}}a_{y}( 2\pi + C_{d0})\xi_{f} \\ 0 \\ 0\end{bmatrix}(\delta_{1}T_{1,x} - \delta_{2}T_{2,x})\\
       &+ \begin{bmatrix} 0 \\ \frac{S}{4S_{p}} \Delta_{r}( 2\pi + C_{d0})\xi_{m}  \\ 0\end{bmatrix}(\delta_{1}T_{1,x} + \delta_{2}T_{2,x})
    \end{flalign}
\end{subequations}

\newpage
\section{Comparaison des dynamiques, en fonction de la forme du drone} \label{ann:comp}
\setcounter{figure}{0} 
\setcounter{equation}{0} 
\noindent
Lors du développement de DarkO plusieurs versions ont été réalisées, avec des formes variées.\\
\begin{figure}[h]
    \centering
    \includegraphics[width=0.7\textwidth]{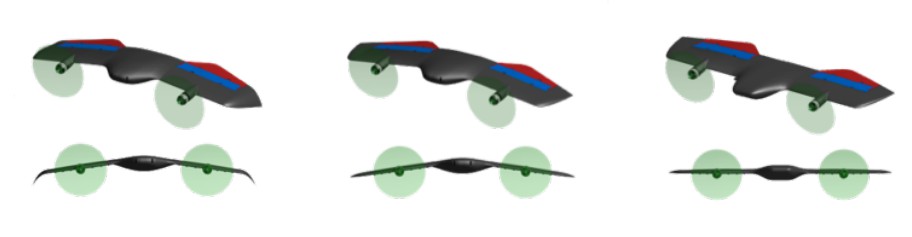}
    \caption{Géométrie du drone}
    \label{fig:geo_drone}
\end{figure}\\
\noindent
Dans l'annexe \ref{matrice_droit} précédente, nous avons déterminé les matrices caractéristiques pour un drone droit. Maintenant, nous allons effectuer les calculs pour un drone en arc.
\begin{align}
    \Phi^{(fv)} = \begin{bmatrix} C_{d0} & 0 & 0 \\ 0 & C_{y0} & 0 \\ 0 & 0 & 2\pi + C_{d0} \end{bmatrix} 
\end{align}

\begin{align}
    \Phi^{(mv)} = \begin{bmatrix} 0 & 0 & 0 \\ 0 & 0 & -c^{-1} \Delta_{r}(2\pi +C_{d0})  \\ 0 & c^{-1} \Delta_{r}C_{y0} & 0 \end{bmatrix} 
\end{align}
Expression de la force due à la traction des hélices et à la traînée :
\begin{subequations}
    \begin{align}
        \Big(I - \frac{S}{4S_{p}} \Phi^{(fv)} \Big) (T_{1} + T_{2}) &= \Bigg(\begin{bmatrix} 1 & 0 & 0 \\ 0 & 1 & 0 \\ 0 & 0 & 1 \end{bmatrix} - \frac{S}{4S_{p}} \begin{bmatrix} C_{d0} & 0 & 0 \\ 0 & C_{y0} & 0 \\ 0 & 0 & 2\pi + C_{d0} \end{bmatrix} \Bigg) \begin{bmatrix} T_{1,x} + T_{2,x}  \\ 0  \\ 0 \end{bmatrix} \\
        & = \begin{bmatrix} (1-\frac{S}{4S_{p}} C_{d0})(T_{1,x} + T_{2,x})  \\ 0  \\ 0 \end{bmatrix}
    \end{align}
\end{subequations}
On observe que le premier terme de la force n'agit que sur l'axe $x_{b}$.\\
Pour la suite, calculons dans un premier temps:
\begin{subequations}
    \begin{align}
        \Delta_{f,i} T_{i} &=  \begin{bmatrix} 0 & -\xi_{m}\delta_{i} &  \xi_{m}\delta_{i}\\ \xi_{m}\delta_{i} & 0 & -\xi_{m}\delta_{i} \\ -\xi_{m}\delta_{i} & \xi_{m}\delta_{i} & 0 \end{bmatrix}\begin{bmatrix} T_{i,x}  \\ 0  \\ 0 \end{bmatrix}\\
        & = \begin{bmatrix} 0  \\ \xi_{m}\delta_{i}T_{i,x}  \\ -\xi_{m}\delta_{i}T_{i,x} \end{bmatrix}
    \end{align}
\end{subequations}
On observe que ce produit est une projection de l'effort de traction vers les deux autres axes. 
Expression de la force de portance :
\begin{subequations}
    \begin{align}
        \frac{S}{4S_{p}} \Phi^{(fv)} (\Delta_{f,1} T_{1} + \Delta_{f,2} T_{2}) &= \frac{S}{4S_{p}} \begin{bmatrix} C_{d0} & 0 & 0 \\ 0 & C_{y0} & 0 \\ 0 & 0 & 2\pi + C_{d0} \end{bmatrix}  \begin{bmatrix} 0  \\ 0   \\ -\xi_{f}(\delta_{1}T_{1,x} + \delta_{2}T_{2,x})  \end{bmatrix}\\
        &= \begin{bmatrix} 0  \\  0  \\ -\frac{S}{4S_{p}}(2\pi + C_{d0})\xi_{f}(\delta_{1}T_{1,x} + \delta_{2}T_{2,x})  \end{bmatrix}
    \end{align}
\end{subequations}
Comme prévu, on observe un effort sur les axes $y_{b}$ et $z{b}$.\\
Pour le moment:
\begin{subequations}
    \begin{align}
        - \frac{S}{4S_{p}} B \Phi^{(mv)} (T_{1} + T_{2} ) &= -\frac{S}{4S_{p}} \begin{bmatrix} b & 0 & 0 \\ 0 & c & 0 \\ 0 & 0 & b \end{bmatrix}
        \begin{bmatrix} 0 & 0 & 0 \\ 0 & 0 & -c^{-1} \Delta_{r}(2\pi +C_{d0})  \\ 0 & c^{-1} \Delta_{r}C_{y0} & 0 \end{bmatrix} \begin{bmatrix} T_{1,x} + T_{2,x}  \\ 0  \\ 0 \end{bmatrix} \\
        & = \begin{bmatrix} 0  \\ 0  \\ 0 \end{bmatrix} 
    \end{align}
\end{subequations}
\begin{subequations}
    \begin{align}
        \frac{S}{4S_{p}} B \Phi^{(mv)} (\Delta_{m,1}T_{1} + \Delta_{m,2}T_{2}) &= \frac{S}{4S_{p}} \begin{bmatrix} b & 0 & 0 \\ 0 & c & 0 \\ 0 & 0 & b \end{bmatrix}
        \begin{bmatrix} 0 & 0 & 0 \\ 0 & 0 & -c^{-1} \Delta_{r}(2\pi +C_{d0})  \\ 0 & c^{-1} \Delta_{r}C_{y0} & 0 \end{bmatrix} \begin{bmatrix} 0  \\0   \\ -\xi_{m}(\delta_{1}T_{1,x} + \delta_{2}T_{2,x})  \end{bmatrix}\\
        &=\begin{bmatrix} 0  \\ 0\\ \frac{S}{4S_{p}} b c^{-1} \Delta_{r}C_{y0} \xi_{m}(\delta_{1}T_{1,x} + \delta_{2}T_{2,x})\end{bmatrix}
    \end{align}
\end{subequations}
Expression du moment dû au différentiel de vitesse sur les ailes :
\begin{subequations}
    \begin{align}
        \frac{S}{4S_{p}} \skewsym{a_{1}} \Phi^{(fv)} T_{1} &= -\frac{S}{4S_{p}}  \begin{bmatrix} 0 & 0 & -a_{y} \\ 0 & 0 & 0 \\ a_{y} & 0 & 0 \end{bmatrix}\begin{bmatrix} C_{d0} & 0 & 0 \\ 0 & C_{y0} & 0 \\ 0 & 0 & 2\pi + C_{d0} \end{bmatrix}\begin{bmatrix} T_{1,x}  \\ 0  \\ 0 \end{bmatrix}\\
        &=\begin{bmatrix} 0  \\  0 \\ \frac{S}{4S_{p}} a_{y} C_{d0}T_{1,x}   \end{bmatrix} 
    \end{align}
\end{subequations}
\begin{subequations}
    \begin{align}
        \frac{S}{4S_{p}} \skewsym{a_{2}} \Phi^{(fv)} T_{2} &= -\frac{S}{4S_{p}}  \begin{bmatrix} 0 & 0 & a_{y} \\ 0 & 0 & 0 \\ -a_{y} & 0 & 0 \end{bmatrix}\begin{bmatrix} C_{d0} & 0 & 0 \\ 0 & C_{y0} & 0 \\ 0 & 0 & 2\pi + C_{d0} \end{bmatrix} \begin{bmatrix} T_{2,x}  \\ 0  \\ 0 \end{bmatrix}\\
        &=\begin{bmatrix} 0  \\  0 \\ -\frac{S}{4S_{p}} a_{y} C_{d0}T_{2,x}   \end{bmatrix} 
    \end{align}
\end{subequations}
On obtient ainsi :
\begin{align}
        \frac{S}{4S_{p}} \skewsym{a_{1}} \Phi^{(fv)} T_{1} + \frac{S}{4S_{p}} \skewsym{a_{2}} \Phi^{(fv)} T_{2} = \begin{bmatrix} 0  \\  0 \\ \frac{S}{4S_{p}} a_{y} C_{d0}(T_{1,x}-T_{2,x} )  \end{bmatrix} 
\end{align}
Expression du moment dû au différentiel de vitesse sur les élevons :
\begin{subequations}
    \begin{align}
        \frac{S}{4S_{p}} \skewsym{a_{1}} \Phi^{(fv)} \Delta_{f,1} T_{1} &= \frac{S}{4S_{p}}  \begin{bmatrix} 0 & 0 & -a_{y} \\ 0 & 0 & 0 \\ a_{y} & 0 & 0 \end{bmatrix} \begin{bmatrix} C_{d0} & 0 & 0 \\ 0 & C_{y0} & 0 \\ 0 & 0 & 2\pi + C_{d0} \end{bmatrix} \begin{bmatrix} 0  \\ 0  \\ -\xi_{f}\delta_{1}T_{1,x}   \end{bmatrix}\\
        & = \begin{bmatrix}  a_{y}( 2\pi + C_{d0}) \xi_{f}\delta_{1}T_{1,x}  \\ 0 \\ 0\end{bmatrix} 
    \end{align}
\end{subequations}
\begin{subequations}
    \begin{align}
        \frac{S}{4S_{p}} \skewsym{a_{2}} \Phi^{(fv)} \Delta_{f,2} T_{2} &= \frac{S}{4S_{p}}  \begin{bmatrix} 0 & 0 & a_{y} \\ 0 & 0 & 0 \\ -a_{y} & 0 & 0 \end{bmatrix} \begin{bmatrix} C_{d0} & 0 & 0 \\ 0 & C_{y0} & 0 \\ 0 & 0 & 2\pi + C_{d0} \end{bmatrix} \begin{bmatrix} 0  \\ 0   \\ -\xi_{f}\delta_{2}T_{2,x}   \end{bmatrix}\\
        & = \begin{bmatrix}  -a_{y}( 2\pi + C_{d0}) \xi_{f}\delta_{2}T_{2,x}  \\ 0 \\ 0\end{bmatrix} 
    \end{align}
\end{subequations}
On obtient ainsi :
\begin{align}
       \frac{S}{4S_{p}} \skewsym{a_{1}} \Phi^{(fv)} \Delta_{f,1} T_{1} + \frac{S}{4S_{p}} \skewsym{a_{2}} \Phi^{(fv)} \Delta_{f,2} T_{2} = \begin{bmatrix}  \frac{S}{4S_{p}}a_{y}( 2\pi + C_{d0}) \xi_{f}(\delta_{1}T_{1,x}-\delta_{2}T_{2,x})  \\ 0 \\ 0\end{bmatrix}  
\end{align}
Expression du moment dû au différentiel de vitesse des hélices :

\begin{align}
       \skewsym{p_{1}} T_{1} + \skewsym{p_{2}} T_{2}= \begin{bmatrix} 0 \\  0 \\ p_{y}(T_{1,x} - T_{2,x})  \end{bmatrix}
\end{align}
Expression du moment dû à la réaction du moteur : 

\begin{align}
    \frac{k_{m} }{k_{f}}(T_{1}- T_{2} ) = \begin{bmatrix} \frac{k_{m} }{k_{f}}(T_{1,x} - T_{2,x})  \\ 0  \\ 0 \end{bmatrix}
\end{align}
On peut exprimer cette expression sous la forme:

    \begin{flalign}
        \sum F_{b} &= \begin{bmatrix} 1-\frac{S}{4S_{p}} C_{d0} \\ 0  \\ 0 \end{bmatrix}_{b}(T_{1,x} + T_{2,x}) + \begin{bmatrix} 0  \\ 0  \\ -\frac{S}{4S_{p}}(2\pi + C_{d0})\xi_{f}  \end{bmatrix}_{b} (\delta_{1}T_{1,x} + \delta_{2}T_{2,x})
    \end{flalign}

\begin{subequations}
    \begin{flalign}
       \sum M_{b} &=  \begin{bmatrix} \frac{k_{m} }{k_{f}} \\  0\\ p_{y}+\frac{S}{4S_{p}} a_{y} C_{d0} \end{bmatrix} (T_{1,x} - T_{2,x}) +\begin{bmatrix} \frac{S}{4S_{p}}a_{y}( 2\pi + C_{d0})\xi_{f} \\ 0 \\ 0\end{bmatrix}(\delta_{1}T_{1,x} - \delta_{2}T_{2,x})\\
       &+ \begin{bmatrix} 0 \\ \frac{S}{4S_{p}} \Delta_{r}( 2\pi + C_{d0})\xi_{m}  \\ 0\end{bmatrix}(\delta_{1}T_{1,x} + \delta_{2}T_{2,x})
    \end{flalign}
\end{subequations}
Mettons les équations sous la forme matricielle, avec $u = \begin{bmatrix}T_{1,x}  & T_{2,x}  & \delta_{1}T_{1,x} & \delta_{2}T_{2,x} \end{bmatrix} ^\top$ :

\begin{subequations}
    \begin{flalign}
        \sum F_{b} &= \begin{bmatrix} ( 1-\frac{S_{wet}}{4S_{p}} C_{d0}) & ( 1-\frac{S_{wet}}{4S_{p}} C_{d0}) & 0 & 0 \\  0 & 0 & 0 & 0 \\  0 & 0 & -(\frac{S_{wet}}{4S_{p}}(2\pi + C_{d0})\xi_{f}) & -(\frac{S_{wet}}{4S_{p}}(2\pi + C_{d0})\xi_{f})  \end{bmatrix}
        \begin{bmatrix}T_{1,x} \\  T_{2,x} \\ \delta_{1}T_{1,x} \\ \delta_{2}T_{2,x} \end{bmatrix}\\
        &=F_{b}u
    \end{flalign}
\end{subequations}
\begin{subequations}
    \begin{flalign}
       \sum M_{b}  &= \begin{bmatrix} (\frac{k_{m} }{k_{f}}) & -(\frac{k_{m} }{k_{f}}) & \frac{S}{4S_{p}}a_{y}( 2\pi + C_{d0})\xi_{f} & -\frac{S}{4S_{p}}a_{y}( 2\pi + C_{d0})\xi_{f} \\  0 & 0 & (\frac{S_{wet}}{4S_{p}} \Delta_{r}( 2\pi + C_{d0})\xi_{m}) & (\frac{S_{wet}}{4S_{p}} \Delta_{r}( 2\pi + C_{d0})\xi_{m}) \\  (p_{y}+\frac{S_{wet}}{4S_{p}} a_{y} C_{d0}) & -(p_{y}+\frac{S_{wet}}{4S_{p}} a_{y} C_{d0}) &0 & 0  \end{bmatrix}
        \begin{bmatrix}T_{1,x} \\  T_{2,x} \\ \delta_{1}T_{1,x} \\ \delta_{2}T_{2,x} \end{bmatrix}\\
        &=M_{b}u
    \end{flalign}
\end{subequations}
On observe que l'on obtient exactement la même matrice que pour le drone droit. Ainsi, la modification de la forme du drone n'a aucune influence sur la commande. 

\newpage
\section{Re-formulation des équations de \texorpdfstring{\citetitle{2020e-MicCenZacFra}}{} \texorpdfstring{\cite{2020e-MicCenZacFra}}{} }\label{sec.appendix_mic}
\setcounter{figure}{0} 
\setcounter{equation}{0} 
\subsection*{Préliminaires et notations} 
\noindent
Nous utilisons les mêmes notations introduites dans le papier \cite{2020e-MicCenZacFra}.\\
Nous ajoutons simplement deux propriétés sur les matrices asymétriques : 
\begin{align}\label{eq:skew_matrix}
    \skewsym{\omega} \skewsym{\omega} = \omega\omega^\top - \omega^\top\omega I_{3}
\end{align}
Soit q un quaternion et $R(q)$ sa matrice de rotation associée : 
\begin{align}\label{eq:der_rot_matrix}
  \dot R(q) = \skewsym{\omega'} R = \skewsym{R\omega}R = R\skewsym{\omega}R^\top R = R\skewsym{\omega}
\end{align}
\subsection*{Analyse de stabilité} 
\noindent
Pour étudier la stabilité de notre bouclage, nous nous intéressons à la dynamique de l'erreur de force $f_{\Delta}$. Pour ce faire, calculons $\dot f_{\Delta}$.
\\À partir de la définition \cite[(24)]{2020e-MicCenZacFra}, on sait que $f_{\Delta} = R(q_{d})d_{*}f - f_{r}$ ; ainsi :
\begin{subequations}
\begin{align}
    \dot f_{\Delta} &= R(q_{d})d_{*}\dot f + \dot R(q_{d})d_{*}f - \dot f_{r}\\
                    &= \dot f_{\Delta, 1} + \dot f_{\Delta, 2} + \dot f_{\Delta, 3}
\end{align}
\end{subequations}
\begin{subequations}
\begin{align}
    \dot f_{\Delta, 1} &= R(q_{d})d_{*}\dot f \\
                    \overset{[1, (26)]}&{=} (R(q_{d})d_{*}) (R(q_{d})d_{*})^\top \nu\\
                    &= R(q_{d})d_{*} d_{*}^\top R(q_{d})^\top \nu
\end{align}
\end{subequations}

\begin{subequations}
\begin{align}
    \dot f_{\Delta, 2} &= \dot R(q_{d})d_{*}f \\
                    \overset{\eqref{eq:der_rot_matrix}}&{=} R(q_{d})\skewsym{\omega_{d}}d_{*} f \\
                    \overset{[1, (25)]}&{=} R(q_{d})\skewsym{\frac{1}{f}\skewsym{d_{*}}R(q_{d})^\top \nu }d_{*} f\\
                    &= -R(q_{d}) \skewsym{d_{*}} \skewsym{d_{*}} R(q_{d})^\top \nu \\
                    \overset{\eqref{eq:skew_matrix}}&{=} -R(q_{d}) ( d_{*}d_{*}^\top - d_{*}^\top d_{*}I_{3})R(q_{d})^\top \nu \\
                    &= -R(q_{d}) d_{*}d_{*}^\top R(q_{d})^\top \nu + R(q_{d})R(q_{d})^\top \nu\\
                    &= -\dot f_{\Delta, 1} + \nu
\end{align}
\end{subequations}
On obtient ainsi:
\begin{align}
    \dot f_{\Delta, 1}  + \dot f_{\Delta, 2} = \nu
\end{align}
Pour le calcul de $\dot f_{\Delta, 1}$, le calcul préliminaire de $\dot e_{v}$ est nécessaire : 
\begin{subequations}
\begin{align}\label{eq:e_v_d}
    \dot e_{v} \overset{[1, (40)]}&{=} -g e_{3} + \frac{1}{m} \Big[\big(R(q) -R(q_{d})\big)f_{c} + f_{r} + f_{\Delta}\Big]\\
    \overset{[1, (20),(22)]}&{=} \frac{1}{m} \Big[\big(R(q) -R(q_{d}) \big)d_{*}f -k_{pp} e_{p} - k_{pd} e_{v} + f_{\Delta} \Big]
\end{align}
\end{subequations}

\begin{subequations}
\begin{align}
    \dot f_{\Delta, 3} &= -\dot f_{r} \\
                    &=k_{pp} \dot e_{p} + k_{pd} \dot e_{v}\\
                    \overset{[1, (39)],\eqref{eq:e_v_d}}&{=} k_{pp} e_{v} + \frac{k_{pd}}{m} \Big[\big(R(q) -R(q_{d}) \big)d_{*}f -k_{pp} e_{p} - k_{pd} e_{v} + f_{\Delta} \Big] \\
                    &= -\frac{k_{pp}k_{pd}}{m}e_{p}  -\left(\frac{k_{pd}^2}{m} -k_{pp}\right) e_{v} + \frac{k_{pd}}{m}\Big(\big(R(q) -R(q_{d}) \big)d_{*}f + f_{\Delta}\Big)
\end{align}
\end{subequations}
On peut donc déterminer $\dot f_{\Delta}$:
\begin{subequations}
\begin{align}\label{eq:f_delta}
    \dot f_{\Delta} &= \dot f_{\Delta, 1} + \dot f_{\Delta, 2} + \dot f_{\Delta, 3}\\
                    &= \nu - \frac{k_{pp}k_{pd}}{m}e_{p}  -\left(\frac{k_{pd}^2}{m} -k_{pp}\right) e_{v} + \frac{k_{pd}}{m}\Big(\big(R(q) -R(q_{d}) \big)d_{*}f + f_{\Delta}\Big)\\
                    \overset{[1, (27)]}&{=} \frac{k_{pp}k_{pd}}{m}e_{p} + \left(\frac{k_{pd}^2}{m} -k_{pp}\right) e_{v} - \left(\frac{k_{pd}}{m}+ k_{\Delta}\right)f_{\Delta} - \frac{k_{pp}k_{pd}}{m}e_{p}  -\left(\frac{k_{pd}^2}{m} -k_{pp}\right) e_{v}\\
                    &+ \frac{k_{pd}}{m}\Big(\big(R(q) -R(q_{d}) \big)d_{*}f + f_{\Delta}\Big)\\
                    &= -k_{\Delta}f_{\Delta} + \frac{k_{pd}}{m}\Big(\big(R(q) -R(q_{d})\big)d_{*}f\Big)
\end{align}
\end{subequations}
On observe que la convergence n'est pas de la forme d'un premier ordre, il y a un terme perturbateur. Cependant, il est intéressant de constater que, dès lors que le drone a aligné sa direction de force avec la direction calculée par le contrôleur, le terme devient nul car $R(q) - R(q_{d}) = 0$.  À partir de cette expression de $\dot f_{\Delta}$, il est maintenant possible de calculer le terme $\dot \omega_{d, 3}$, défini \cite[(A.7)]{2020e-MicCenZacFra}:
\begin{subequations}
\begin{align}
    \dot \omega_{d, 3} &= \frac{1}{f}\skewsym{d_{*}}R(q_{d})^\top \dot \nu\\
    \overset{[1, (27)]}&{=} \frac{1}{f}\skewsym{d_{*}}R(q_{d})^\top \left[ \frac{k_{pp}k_{pd}}{m} \dot e_{p} + \left(\frac{k_{pd}^2}{m} -k_{pp}\right) \dot e_{v} - \left(\frac{k_{pd}}{m}+ k_{\Delta}\right)\dot f_{\Delta} \right]\\
    \overset{\eqref{eq:e_v_d}, \eqref{eq:f_delta}}&{=} \frac{1}{f}\skewsym{d_{*}}R(q_{d})^\top \left[ \frac{k_{pp}k_{pd}}{m} e_{v} + \left(\frac{k_{pd}^2}{m^2} -\frac{k_{pp}}{m}\right)  \Big[\big(R(q) -R(q_{d}) \big)d_{*}f -k_{pp} e_{p} - k_{pd} e_{v} + f_{\Delta} \Big] \right. \\ & \left. + \left(\frac{k_{pd}}{m}+ k_{\Delta}\right) \left(k_{\Delta}f_{\Delta} - \frac{k_{pd}}{m}\Big(\big(R(q) -R(q_{d})\big)d_{*}f\Big) \right) \right]
\end{align}
\end{subequations}
On peut cependant redéfinir $\nu$ de manière à obtenir une convergence de $f_{\Delta}$ sous la forme d'un premier ordre. 
\begin{align}
    \nu' = \frac{k_{pp}k_{pd}}{m}e_{p} + \left(\frac{k_{pd}^2}{m} -k_{pp}\right) e_{v} - \left(\frac{k_{pd}}{m}+ k_{\Delta}\right)f_{\Delta} - \frac{k_{pd}}{m}\big(R(q) -R(q_{d})\big)d_{*}f
\end{align}
Calculons $\dot \nu'$:
\begin{subequations}
\begin{align}
    \dot \nu' &= \frac{k_{pp}k_{pd}}{m}\dot e_{p} + \left(\frac{k_{pd}^2}{m} -k_{pp}\right) \dot e_{v} + \left(\frac{k_{pd}}{m}+ k_{\Delta}\right) \dot f_{\Delta} - \frac{k_{pd}}{m}\big(\dot R(q) - \dot R(q_{d})\big)d_{*}f - \frac{k_{pd}}{m}\big(R(q) - R(q_{d})\big)d_{*}\dot f\\
    &= \frac{k_{pp}k_{pd}}{m} e_{v} \left(\frac{k_{pd}^2}{m^2} - \frac{k_{pp}}{m}\right)\Big(\big(R(q) -R(q_{d})\big)d_{*}f -k_{pp} e_{p} - k_{pd} e_{v} + f_{\Delta}\Big) + \left(\frac{k_{pd}}{m}+ k_{\Delta}\right) k_{\Delta}f_{\Delta} \\&+ \frac{k_{pd}}{m}\big(R(q_{d})\skewsym{\omega_{d}} - R(q)\skewsym{\omega}\big)d_{*}f + \frac{k_{pd}}{m}\big( R(q_{d}) - R(q) \big)d_{*}\big(R(q_{d})d_{*}\big)^\top \nu
    \\ &= -\left(\frac{k_{pd}^2}{m^2} - \frac{k_{pp}}{m}\right)k_{pp} e_{p} + \left[\frac{k_{pp}k_{pd}}{m} - \left(\frac{k_{pd}^2}{m^2} - \frac{k_{pp}}{m}\right)k_{pd}\right] e_{v} + \left[\left(\frac{k_{pd}^2}{m^2} - \frac{k_{pp}}{m}\right) + \left(\frac{k_{pd}}{m}+k_{\Delta}\right)k_{\Delta}\right]f_{\Delta} \\&+ \left[\left(\frac{k_{pd}^2}{m^2} - \frac{k_{pp}}{m}\right)\big(R(q) -R(q_{d})\big) + \frac{k_{pd}}{m}\big(R(q_{d})\skewsym{\omega_{d}} - R(q)\skewsym{\omega}\big) \right]d_{*}f + \frac{k_{pd}}{m}\big( R(q_{d} - R(q) \big)d_{*}\big(R(q_{d})d_{*}\big)^\top \nu
\end{align}
\end{subequations}
Avec cette expression de $\dot \nu'$, on peut déterminer $\dot \omega_{d, 3}' = \frac{1}{f}\skewsym{d_{*}}R(q_{d})^\top \dot \nu'$.\\
On observe, par simulation, que les deux contrôleurs convergent de la même manière. 

\begin{figure}[h]
\centering
\includegraphics[width=1\textwidth]{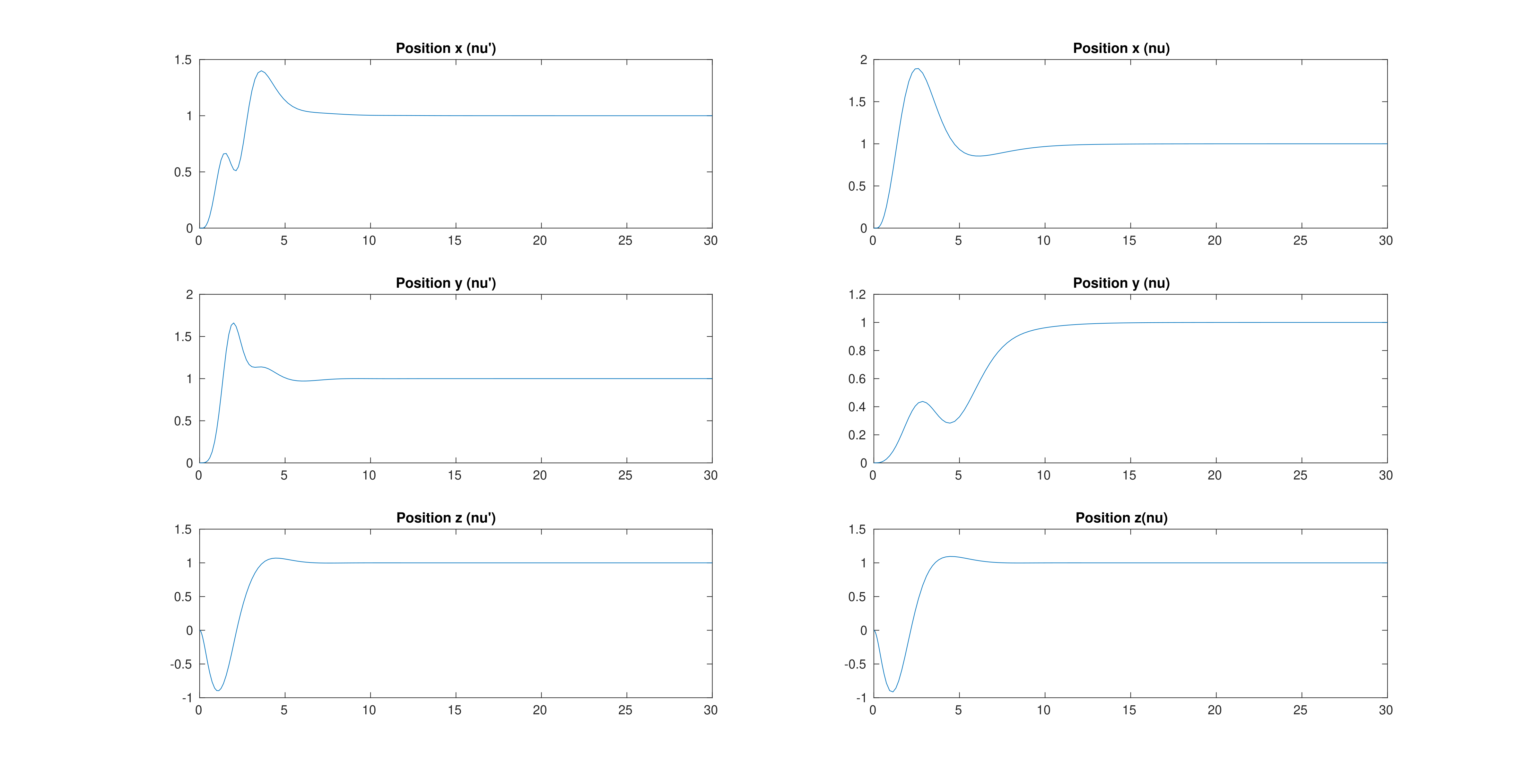}
\caption{Résultats des simulations avec la redéfinition de $\nu$ : position}
\end{figure}
\newpage
\begin{figure}[h]
\centering
\includegraphics[width=1\textwidth]{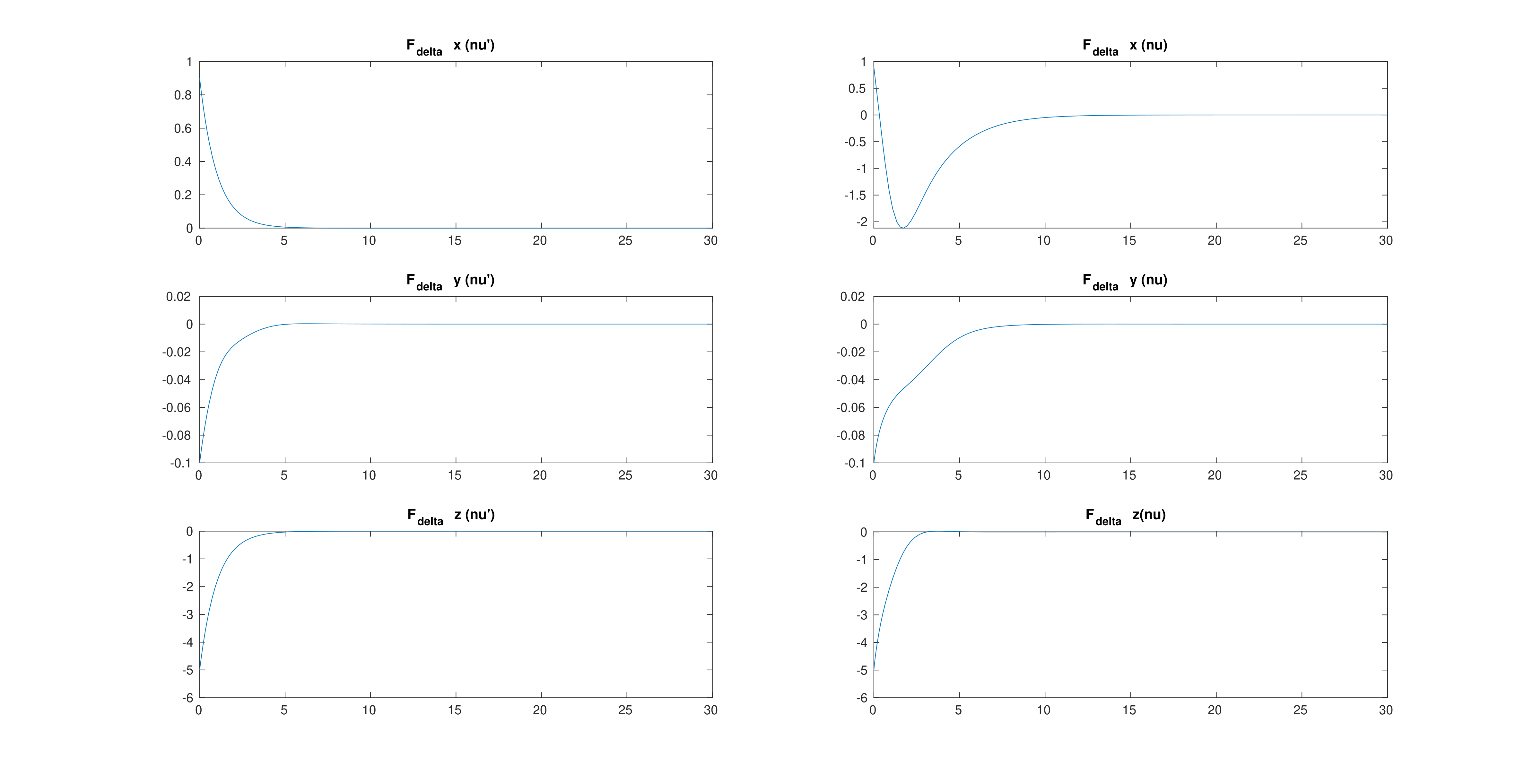}
\caption{Résultats des simulations avec la redéfinition de $\nu$ : dynamique de $f_{delta}$}
\end{figure}
On observe que la nouvelle définition de $\nu$ ($\nu'$) permet d'avoir une dynamique du premier ordre sur $f_{delta}$.
\subsection*{Restricted Orientation Stabilization} \label{sec.appendix_res_orient}

Dès lors que l'on souhaite implémenter la convergence de l'orientation vers une orientation cible, il est nécessaire de faire évoluer l'expression de l'action \textit{feedforward}. 
La première intuition que nous avons eue a été d'effectuer une réinitialisation de $q_{d}$, l'état du contrôleur, vers un quaternion tourné de l'angle de rotation souhaité. Toutefois, cette solution n'était pas satisfaisante car elle implique une discontinuité dans l'état du contrôleur, ce qui engendre des transitoires non souhaités. On préfère utiliser une action hiérarchique pour faire converger le drone vers l'orientation cible. Pour cela, il est nécessaire de faire évoluer l'expression de $\omega_{d}$ en y ajoutant ce terme de convergence :
\begin{align}
    \omega_{d} = \frac{1}{f}*\skewsym{d_{*}}R(q_{d})^{\top} \nu - k_{q}d_{*}d_{*}^{\top} \epsilon_{\Delta}^{'}
\end{align}
avec $k_{q}$ un gain pour assurer la vitesse de convergence et $\epsilon_{\Delta}^{'}$ la partie vectorielle de la différence entre le quaternion de référence de l'orientation et le quaternion représentant l'orientation du drone.
\newline \\
Cette nouvelle expression modifie l'expression de l'action \textit{feedforward} : 
\begin{align}
    \dot \omega_{d,4} = - k_{q}d_{*}d_{*}^{\top} \dot \epsilon_{\Delta}^{'}
\end{align}
\clearpage\null\newpage
\end{document}